\documentclass[twocolumn]{aastex631}
\usepackage{natbib}
\usepackage{url}
\usepackage{colortbl}
\usepackage{float}
\DeclareUnicodeCharacter{2212}{-}

\newcommand{\msun}{\mbox{$\,{\rm M}_\odot$}}

\begin{document}

\title{The Isaac Newton Telescope Monitoring Survey of Local Group Dwarf Galaxies -- VIII. A Census of Long-Period Variable Stars across the Andromeda Dwarf Satellite System}

\correspondingauthor{Atefeh Javadi}
\email{atefeh@ipm.ir}
\email{hedieh.abdollahi@csfk.org}

\author[0000-0002-7823-7169]{Hedieh Abdollahi}
\affiliation{Konkoly Observatory, HUN-REN Research Centre for Astronomy and Earth Sciences, MTA Centre of Excellence, Konkoly-Thege Mikl\'os \'ut 15-17, H-1121, Budapest, Hungary}
\affiliation{School of Astronomy, Institute for Research in Fundamental Sciences (IPM), P.O. Box 1956836613, Tehran, Iran}

\author[0000-0001-8392-6754]{Atefeh Javadi}
\affiliation{School of Astronomy, Institute for Research in Fundamental Sciences (IPM), P.O. Box 1956836613, Tehran, Iran}

\author[0000-0002-1272-3017]{Jacco Th. van Loon}
\affiliation{Lennard-Jones Laboratories, Keele University, ST5 5BG, UK}

\author[0000-0003-0356-0655]{Iain McDonald}
\affiliation{Department of Physical Sciences, The Open University, Walton Hall, Milton Keynes, UK}
\affiliation{Jodrell Bank Centre for Astrophysics, Alan Turing Building, University of Manchester, M13 9PL, UK}

\author[0009-0006-3280-5622]{Mahdi Abdollahi}
\affiliation{School of Astronomy, Institute for Research in Fundamental Sciences (IPM), P.O. Box 1956836613, Tehran, Iran}
\author[0000-0002-5075-1764]{Elham Saremi}
\affiliation{School of Physics \& Astronomy, University of Southampton, Highfield Campus, Southampton SO17 1BJ, UK}

\author[0000-0003-0558-8782]{Habib G. Khosroshahi}
\affiliation{School of Astronomy, Institute for Research in Fundamental Sciences (IPM), P.O. Box 1956836613, Tehran, Iran}
\affiliation{Iranian National Observatory, Institute for Research in Fundamental Sciences (IPM), Tehran, Iran}

\author[0000-0002-8159-1599]{L\'aszl\'o Moln\'ar}
\affiliation{Konkoly Observatory, HUN-REN Research Centre for Astronomy and Earth Sciences, MTA Centre of Excellence, Konkoly-Thege Mikl\'os \'ut 15-17, H-1121, Budapest, Hungary}

\affiliation{ELTE E\"otv\"os Lor\'and University, Institute of Physics and Astronomy, 1117, P\'azm\'any P\'eter s\'et\'any 1/A, Budapest, Hungary}
\author[0009-0002-9957-5818]
{Hamidreza Mahani}
\affiliation{School of Astronomy, Institute for Research in Fundamental Sciences (IPM), P.O. Box 1956836613, Tehran, Iran}



\begin{abstract}
We present a comprehensive catalog, in the Sloan $i$ and Harris $V$ filters, of long-period variable (LPV) stars in the spheroidal dwarf satellites of the Andromeda galaxy, based on a dedicated survey for variable stars in Local Group dwarf systems. Using photometric time-series data obtained with the Wide Field Camera (WFC) on the 2.5 m Isaac Newton Telescope (INT), we identify approximately 2800 LPV candidates across 17 Andromeda satellites, spanning a broad range in luminosity and variability amplitude. This study is accompanied by a public data release that includes two comprehensive catalogs, a catalog of the complete stellar populations for each galaxy and a separate catalog listing all identified LPV candidates. Both are available through CDS/VizieR and provide a valuable resource for investigating quenching timescales, stellar mass distributions, and the effects of mass-loss and dust production in dwarf galaxies. We derive updated structural parameters, including newly measured half-light radii, and determine distance moduli using the Tip of the Red Giant Branch (TRGB) method with Sobel-filter edge detection, yielding values between $23.38\pm0.06$ and $25.35\pm0.06$ mag. 

\end{abstract}



\section{Introduction} \label{sec:intro}

Local Group dwarf galaxies provide valuable laboratories for studying galaxy formation and evolution across a wide range of environments, morphologies, and metallicities \citep{2010AdAst2010E...3C}. Their well-resolved stellar populations and varied star-formation histories (SFHs) make them ideal for examining the connection between stellar populations and galactic evolution. These characteristics also make dwarf galaxies excellent testbeds for studying stellar evolution and the cosmic cycle of matter in environments that differ significantly from those of massive galaxies such as the Milky Way \citep{Dib05}. The generally metal-poor conditions under which many of their stars formed are more representative of the early Universe, where individual stars cannot be directly observed \citep{Tolstoy09}.

Because these systems are typically dominated by dark matter, they also serve as important probes of dark-matter distribution within galaxy clusters \citep{2012AJ....144...76W, 2012AJ....144..183L}. In the context of the $\Lambda$CDM cosmological model, dwarf galaxies, including the ultra-faint dwarfs that represent the smallest dark matter-dominated systems, are understood to constitute one of the smallest bound structures within the dark matter halo hierarchy, thereby offering crucial observational tests of hierarchical galaxy formation scenarios \citep{1978ApJ...225..357S, Moore99}. Moreover, their low stellar densities and limited dynamical evolution imply that their present-day mass functions remain close to their initial forms \citep{2019UFD}.

The contrasting environments of satellite and isolated dwarf galaxies allow the investigation of environmental effects such as tidal interactions and gas removal in shaping their evolution \citep{Mahani25}. In particular, comparisons between the satellite systems of the Milky Way and Andromeda (M31) provide insight into how similar galaxies evolve under different conditions.  

Since Andromeda lies within the northern hemisphere, it can be observed in its entirety using northern facilities, which enables a complete study of its satellite system. In contrast, the Milky Way’s satellite system is distributed across both the northern and southern hemispheres, making it more challenging to obtain homogeneous observational coverage. This accessibility motivated the \textit{Isaac Newton Telescope Monitoring Survey of Dwarf Galaxies in the Local Group} (hereafter the INT Survey), which targets the dwarf satellites of both the Milky Way and M31, along with several isolated dwarfs and globular clusters. The survey was designed to trace the evolutionary histories of these galaxies by monitoring their resolved stellar populations over multiple epochs between June 2015 and February 2018 \citep{2021ApJ...923..164S}.

%\subsection{AGB and LPV Stars}
To explore the late stages of stellar evolution in these galaxies, this study focuses on asymptotic giant branch (AGB) stars, which are luminous tracers that can be detected in distant systems. Although their evolutionary tracks are difficult to model precisely, AGB stars are valuable because of their high luminosities ($L \sim 10^4\,L_\odot$) and their sensitivity to both age and metallicity distributions within stellar populations \citep{Javadi13, 2018A&ARv..26....1H}. Stars with initial masses between about 0.5 and $8\,M_\odot$ pass through the AGB phase during their evolution \citep{2018A&ARv..26....1H}.  

Thermal pulses and associated deep convection transport material from the stellar interior to the surface, enriching the photosphere with heavy elements \citep{Karakas14, Herwig05}. In particular, neutron-capture nucleosynthesis during thermal pulses leads to the production of s-process elements (e.g., Sr, Y, Zr, Ba), which are subsequently brought to the surface by third dredge-up episodes \citep{1999ARA&A..37..239B, Karakas14}. Pulsation-driven shocks then lift material to regions where it can cool and condense into dust, initiating mass-loss \citep{Gail99, McDonald17}. These processes inject newly formed dust and gas, and s-process-enriched material into the interstellar medium, driving the chemical evolution of galaxies.  

At this stage, AGB stars exhibit long-period variability caused by large convection cells that excite resonant oscillations in their extended atmospheres. These long-period variable (LPV) stars typically have pulsation periods of 60--500 days and mass-loss rates of approximately $10^{-7}$ to $10^{-5}\,M_\odot\,{\rm yr^{-1}}$ \citep{vanLoon99, Scicluna22, Mahani25}. Their luminosities depend primarily on the core mass \citep{Bloecker93}, while their time spent on the AGB, and therefore their maximum luminosity, is determined by their initial mass and mass-loss rate. The dust injected by AGB stars into the ISM helps regulate the thermal balance and formation of molecular clouds in which star formation ensues.
  
Here we present a catalog of AGB-type LPVs in the majority of Andromeda's satellite dwarf galaxies, extending our previous work on a limited number of Andromeda's dwarf satellites to its satellite system as a whole \citep{2023ApJ...948...63A, 2023ApJ...942...33P, 2021ApJ...923..164S, 2021ApJ...910..127N, Gholami25, Javadi11}. By identifying and characterizing these evolved, mass-losing stars across diverse environments, we aim to refine the location of the TRGB, determine the half-light radii of the host systems, and establish a foundation for future studies investigating how metallicity and environment influence stellar evolution and dust production in low-mass galaxies.

The structure of this paper is as follows. Section~\ref{sec:data} describes the observations and data. Section~\ref{sec:method} outlines the methods used to identify and analyze variable stars. Section~\ref{sec:parameter} presents the derived stellar parameters. Section~\ref{sec:Discussion} discusses the properties of each Andromeda dwarf satellite, and Section~\ref{sec:Summary} summarizes the main results.

\section{Observation and Data} \label{sec:data}

The observations in this study are part of the INT Survey of Dwarf Galaxies in the Local Group. A full description of the survey strategy, observational setup, and data reduction, as well as the number of observations in each photometric band for each galaxy, is provided in \citet{2020ApJ...894..135S}.

The observations were carried out using the Wide Field Camera (WFC) on the 2.5-m Isaac Newton Telescope at the Observatorio del Roque de los Muchachos, La Palma. The WFC consists of four 2048\,$\times$\,4096 CCDs with a pixel scale of 0.33\,arcsec\,pixel$^{-1}$. Time-series photometry was obtained in the Sloan $i$, Harris $V$, and RGO $I$ filters. The RGO $I$ filter was employed only for the initial epoch of several targets. 

The combination of $V$ and $i$ bands allows the characterization of cool, evolved stars over a wide wavelength range. The $i$ band enhances the detectability of luminous, red variables whose spectral energy distributions peak near 1\,$\mu$m, while the $V$ band provides higher sensitivity to pulsation-induced variations, which are stronger at shorter wavelengths due to molecular absorption changes in stellar atmospheres. Color indices such as $V-i$ were used to estimate effective temperatures and to investigate color–magnitude behavior during variability cycles \citep{McDonald12}.

Among the 24 known satellites of Andromeda, four galaxies (M\,110, M\,32, Pisces\,I, and Pegasus) lacked sufficient observational coverage to allow LPV detection, while three systems have been analyzed in previous works \citep{2023ApJ...948...63A, 2021ApJ...910..127N, 2021ApJ...923..164S}. %The present study therefore, focuses on the remaining 17 galaxies.

%\textcolor{red}{M\,32 and M\,110, however, already have published studies that document their evolved-star content and variable-star populations; M32 hosts numerous variable AGB and RR~Lyrae stars \citep{Davidge2004, Fiorentino2012} and has recently been surveyed for infrared variables with Spitzer \citep{Jones2021}, while M\,110 shows a well-populated AGB and dusty-AGB population in near-IR/optical studies \citep{Davidge2005} and is included among the DUSTiNGS Spitzer targets that yield infrared variable-AGB candidates \citep{Boyer2015}. The existing catalogs offer a basis for extending the LPV census to M\,32 and M\,110 in a future study. For consistency, the present analysis includes only the 17 dwarf satellites with uniform INT coverage.}
%%%%%%%%%%%%%%%%%%%%%%%%%%%%%%%%%

% Comment: None of the above-mentioned studies were conducted in the $i$ and $V$ bands, so it may be better to omit them here. Including them could be viewed as over-citation or off-topic in this context.

%%%%%%%%%%%%%%%%%%%%%%%%%%%%%%%%%

\begin{figure}
\includegraphics [width=0.48\textwidth]{ 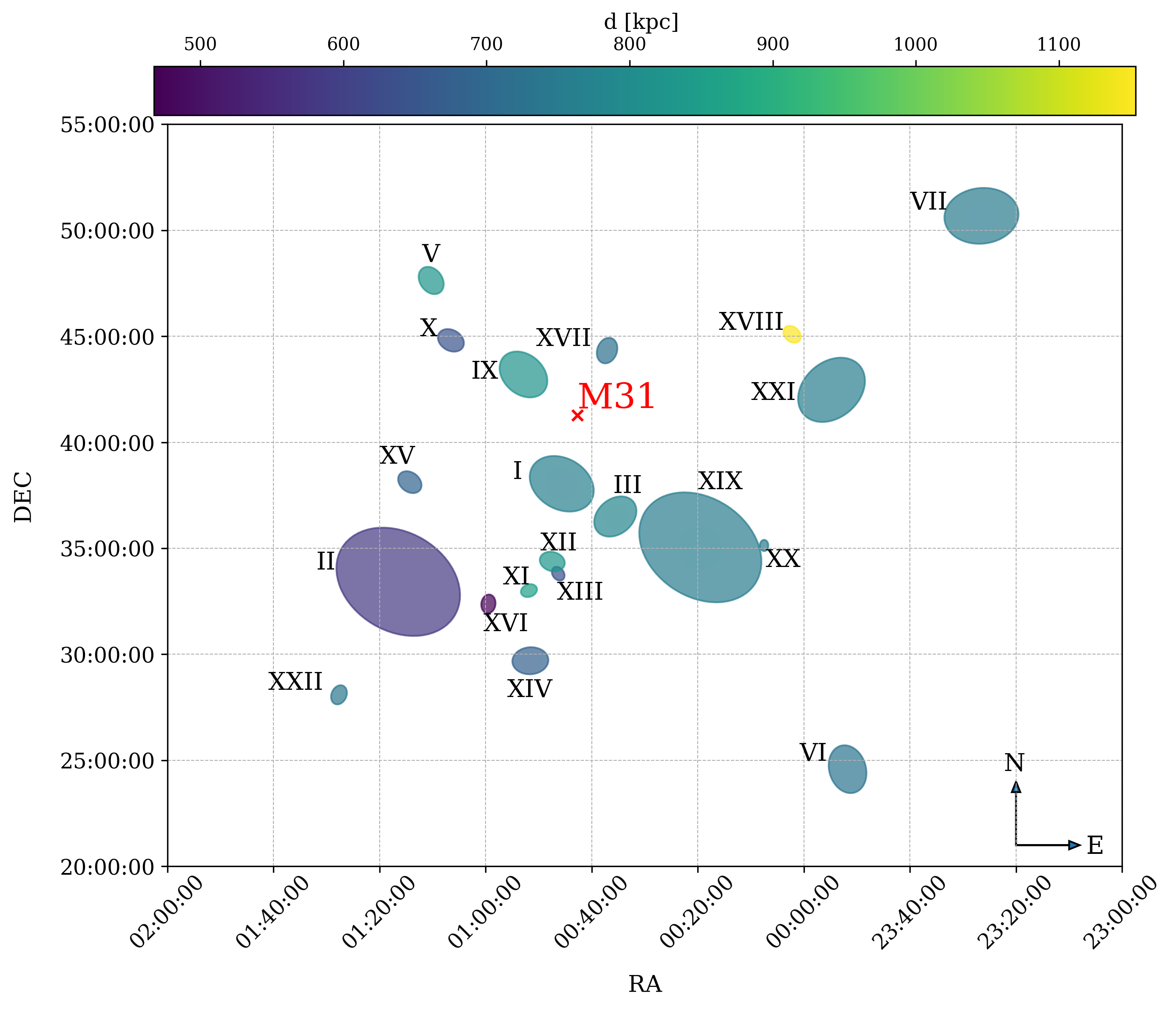}
\caption{Satellite galaxies of Andromeda. The semi-major axis of the galaxies is represented proportionally to their actual sizes, based on data from Table~\ref{table:objects}.}
\label{fig:M31_map}
\end{figure}

Figure~\ref{fig:M31_map} shows the spatial distribution of the observed satellites around M31. The size of each ellipse represents the half-light radius derived in this work (see Section~\ref{Section half-light}), while the color indicates the distance estimated using the TRGB method (see Section~\ref{sec:TRGB}). The position of M31 is marked with a red cross.

\begin{table*}
\caption{Observational properties and half-light radius of targets.}
\footnotesize % Adjust the font size to make the table smaller
\setlength{\tabcolsep}{2pt}
\centering
\begin{tabular}{lllllllllllll}
\hline\hline

 \noalign{\smallskip}
{Galaxy}           &
{R.A.$^a$}             &
{Dec$^a$}              &
{$\epsilon^b$} & 
{[Fe/H]$^c$}        &
{$r_{\rm h}$ $^b$}   &
{$r_{\rm h}$$^{Exponential}$ }    &
{$r_{\rm h}$$^{Plummer}$ }        &
{$r_{\rm h}$$^{S\acute{e}rsic}$ } &
{E$_{i}$}$^h$&
{E$_{V}$}$^i$&
N$_{Total}$ $^j$ &
N$_{LPV}$ $^k$\\

 &
(J2000) &
(J2000) &
 & 
(dex) &
(arcmin) &
(arcmin) &
(arcmin) &
(arcmin) & & & &\\

\hline
\multicolumn{13}{l}{}\\ 
And\,I $^d$ & 00 45 39.8 & $+38$ 02 28 & $0.28\pm0.03$ & $-1.45\pm0.04$ & $3.90\pm0.10$ & $3.20\pm0.30$ & - & - & 8&5 & 10243 & 470\\

And\,II  & 01 16 29.7 & $+33$ 25 08 & $0.16\pm0.02$ & $-1.64\pm0.04$ & $5.30\pm0.10$ & 5.32$^{+0.19}_{-0.02}$ & 5.21$^{+0.09}_{-0.12}$ &5.35$^{+0.16}_{-0.05}$ & 8 & 4 &11230 & 155\\

And\,III & 00 35 33.7 & $+36$ 29 51 &  $0.59\pm0.04$ & $-1.78\pm0.04$ & $2.20\pm0.20$ & $ 2.33\pm 0.20$ & $2.21 \pm 0.19$ & $2.15 \pm 0.25$ & 9 & 4 & 5878  & 185 \\

And\,V & 01 10 17.1 & $+47$ 37 41 &  $0.26 \pm 0.09$ & $-1.60\pm0.30$  & $1.60 \pm 0.20$ &  $1.85^
{+0.07}_{-0.14}$ & $1.82^{+0.10}_{-0.11}$ & $1.81^{+0.11}_{-0.10}$ & 8 & 5 & 7916 & 112\\

And\,VI & 23 51 46.3 & $+24$ 34 57 &  $0.41\pm0.03$ & $-1.30 \pm0.14$ & $2.30\pm0.20$ &$ 2.24\pm 0.15$ & $2.00\pm 0.16$ & $2.27\pm 0.12$ & 8 & 5 & 6342 & 117\\

And\,VII $^e$ & 23 26 31.7 & $+50$ 40 32 &  $0.13\pm0.04$ & $-1.40\pm0.30$ & $3.50\pm0.10$ & $3.80\pm0.30$ & - & - & 9 & 5 & - & 55\\

And\,IX $^f$ & 00 52 53.0 & $+43$ 11 45 &  0.00$^{+0.16}_{-0.00}$ & $-2.20\pm0.20$ & $2.00\pm  0.30$ & $2.50\pm0.26$ & - & - & 8 & 4 & 8653 & 77\\

And\,X & 01 06 39.5 & $+44$ 47 19 &  0.10$^{+0.34}_{-0.10}$ & $-1.93\pm0.11$ & $1.10\pm 0.40$ & $1.64^{+0.46}_{-0.24}$& $1.63^{+0.47}_{-0.23}$  & $1.63^{+0.47}_{-0.23}$ & 8 & 5 & 3502 & 123\\

And\,XI & 00 46 20.0 & $+33$ 48 05 &  0.19$^{+0.28}_{-0.19}$  & $-2.00\pm0.20$ & $0.60\pm0.20$ &  $0.65^{+0.07}_{-0.17}$ & $0.63^{+0.09}_{-0.15}$ & $ 0.60^{+0.12}_{-0.12}$ & 7 & 6 & 5219 & 84 \\

And\,XII & 00 47 27.0 & $+34$ 22 29 &  $0.61 \pm 0.48 $  & $-2.10\pm0.20 $ &$ 1.80\pm 1.20$ & $1.89^{+0.17}_{-0.34}$ &  $1.82^{+0.23}_{-0.28}$ & $1.76^{+0.29}_{-0.22}$ & 8 & 4 & 3552 & 116\\

And\,XIII & 00 51 51.0 & $+33$ 00 16 & $0.61 \pm 0.20$ & $-1.90\pm0.20$ & $0.80\pm 0.40$ & $0.71\pm 0.25$ & $0.69\pm 0.27$ & $0.73\pm 0.23$ & 8 & 4 & 4052 & 111\\

And\,XIV & 00 51 35.0 & $+29$ 41 49 &  $0.17 \pm  0.17 $ & $-2.26\pm0.05$ & $1.50\pm0.20$ & $1.67\pm 0.20$ & $1.66\pm 0.21$   & 1.66$^{+0.21}_{-0.16}$ & 6 & 4 & 7823 & 48\\

And\,XV & 01 14 18.7 & $+38$ 07 03 &  $0.24\pm0.10$ & $-1.80\pm0.20$ & $1.30\pm0.10$ & $1.55\pm 0.25$ & $1.48\pm 0.18$ & $1.46\pm 0.16$ & 7 & 3 & 7228 & 214\\

And\,XVI & 00 59 29.8 & $+32$ 22 36 &  $0.29\pm0.08$ & $-2.10\pm0.20$ & $1.00\pm0.10$ & $1.19\pm 0.19$ & $1.14\pm 0.14$ &$ 1.13\pm 0.17$ & 7 & 3 & 2979 & 78 \\

And\,XVII  & 00 37 07.0 & $+44$ 19 20 &  $0.50\pm0.10$ & $-1.90\pm0.20$ & $1.48\pm0.30$ & $1.47^{+0.02}_{-0.16}$ & $1.46^{+0.03}_{-0.16}$ & $1.47^{+0.02}_{-0.16}$ & 8 & 3 & 11097 & 174 \\

And\,XVIII & 00 02 14.5 & $+45$ 05 20 &  0.03$^{+0.28}_{-0.03}$ & $-1.80\pm0.10$ & $0.80\pm0.10$ & $0.89^{+0.07}_{-0.09}$ & $0.87^{+0.09}_{-0.07}$ & $0.87^{+0.09}_{-0.07}$ & 5 & 2 & 2544 & 123\\

And\,XIX & 00 19 32.1 & $+35$ 02 37 &  $0.58\pm 0.10$ & $-1.90\pm0.10$ &$ 14.20\pm 3.40$ & $14.29^{+0.48}_{-0.09}$ & $14.06^{+0.14}_{-0.42}$ & $14.35^{+0.42}_{-0.15}$ & 8 & 4 & 7470 & 179 \\

And\,XX & 00 07 30.7 & $+35$ 07 56 &  0.11$^{+0.41}_{-0.11}$ & $-1.50\pm0.10$ & $0.40\pm 0.20$ & $0.50^{+0.03}_{-0.10}$ & $0.49^{+0.04}_{-0.09}$ & $0.49^{+0.05}_{-0.09}$ & 8 & 5 & 4065 & 120\\

And\,XXI & 23 54 47.7 & $+42$ 28 15 & $0.36\pm 0.13$ & $-1.80\pm0.20$ & $4.10\pm 0.80$ &$ 3.83\pm 0.55$ & $3.82\pm 0.54$ &$ 3.82\pm 0.54$ & 10 & 6 & 4078 & 52\\

And\,XXII & 01 27 40.0 & $+28$ 05 25 &  $0.61\pm 0.14$ & $-1.80$ & $0.90\pm 0.30$ & $0.94^{+0.14}_{-0.22}$ & $0.90^{+0.18}_{-0.18}$ & $ 0.95^{+0.13}_{-0.23}$ & 4 & 1 & 4651 & 158\\

\hline

\end{tabular}
\tablecomments{
\footnotesize{$^a$ Coordinates inferred from the \cite{NED_Full} portal.}\\
\footnotesize{$^b$ All ellipticities ($\epsilon$) and half-light radius in 6$^{th}$ column ({$r_{\rm h}$}) are referred from \cite{2016ApJ...833..167M}, except for And\,VI which is from \cite{2012AJ....144....4M}.}\\
\footnotesize{$\epsilon$ = $1-b/a$, where $b$ is the semi-minor axis and $a$ is the semi-major axis.}\\
\footnotesize{$^c$ \cite{2012AJ....144....4M}, $^d$ \cite{2020ApJ...894..135S}, $^e$ \cite{2021ApJ...910..127N}, and $^f$ \cite{2023ApJ...948...63A}. }\\
\footnotesize{{$r_{\rm h}$ $^{Exponential}$ }, {$r_{\rm h}$ $^{Plummer}$ }, and {$r_{\rm h}$ $^{S\acute{e}rsic}$} are calculated in this work.}\\
\footnotesize{$^h$ Total number of observations in $i$-band.}\\
\footnotesize{$^i$ Total number of observations in $V$-band.}\\
\footnotesize{$^j$ Total number of stellar population detected in the observed field.}\\
\footnotesize{$^k$ The number of confirmed LPV candidates in the observed field based on the criteria}.}
\label{table:objects}
\end{table*}
%@@@@@@@@@@@@@@@@@@@@@@@@@@@@@@@@@@@@@@@@@@@@@@@@@@@@@@@@@@@@@@@@@@ section 2.1

\section{Methodology} \label{sec:method}

This section outlines the comprehensive methodology employed to determine the photometric variability of stars. The approach involves several critical steps, from initial data reduction to preprocessing of the raw observational images. Subsequent photometric measurements were conducted using specialized software to quantify the stellar magnitudes accurately. Additionally, the survey's completeness was evaluated to ensure the reliability of the detected variable stars. These steps collectively ensure the precision and accuracy necessary for robust photometric analysis and variability assessment.

\subsection{Data reduction}

Before starting the photometry of observational images, initial data reduction steps are necessary. These steps involve preparing the raw images through several calibration processes, including bias correction, flat-fielding, and dark current subtraction, to correct for instrumental and environmental effects. These preprocessing steps are critical to ensure the images accurately reflect the brightness of the sources. Bias correction removes the electronic offset added during image readout, flat-fielding corrects for pixel sensitivity variations using images of a uniform light source, and dark current subtraction removes thermal noise using dark frames taken with the same exposure time as the science images, but with the shutter closed. 
The calibration of astrometry enables the accurate determination of the celestial coordinates of objects captured in images. By utilizing a high signal-to-noise ratio from combined images, a final stacked image is produced. This coaddition process significantly enhances the quality of the data, allowing for more in-depth scientific analysis.

Image reduction and astrometric alignment were performed using the \texttt{THELI} pipeline \citep{Erben05}, which is optimized for wide-field mosaic CCD data. A summary of the observational parameters, including epoch, filter, exposure time, airmass, and seeing for each target, is presented in Table~\ref{table:objects} and described in detail by \citet{2020ApJ...894..135S}.
\subsection{Photometry}

The {\sc \texttt{DAOPHOT/ALLSTAR}} package offers comprehensive tools for performing precise photometric analysis in crowded fields \citep{1987PASP...99..191S}. The following sections outline the key steps required to utilize the Stetson package for this purpose.

Photometry in both $i$ and $V$ filters was performed using the {\sc \texttt{DAOPHOT/ALLSTAR}} package. Initially, a selection of approximately 30-40 isolated stars located at various positions in the field was made using the {\sc \texttt{PSF}} routine suitable for a crowded field. The purpose was to construct a point-spread function (PSF) model for each image following the {\sc \texttt{FIND}} and {\sc \texttt{PHOT}} procedures. A master image was then created by combining individual images through the {\sc \texttt{DAOMATCH}}, {\sc \texttt{DAOMASTER}}, and {\sc \texttt{MONTAGE2}} routines. This master image was used to generate a star list with the {\sc \texttt{ALLSTAR}} routine. Subsequently, the {\sc \texttt{ALLFRAME}} routine employed the star list to estimate the instrumental magnitudes of stars by fitting the PSF models to the individual images \citep{Stetson94}.

The transformation of the instrumental magnitudes onto the standard system was accomplished using observations of standard stars \citep{Landolt92} and the {\sc \texttt{NEWTRIAL}} routine \citep{Stetson96}.

\subsection{Calibration and photometry assessment}

The photometric calibration process was conducted in three stages. First, aperture corrections were applied using the {\sc \texttt{DAOGROW}} and {\sc \texttt{COLLECT}} routines to calculate the differences in magnitude between the PSF-fitting and largest aperture photometry of about 40 isolated bright stars in each frame \citep{Stetson90}. The {\sc \texttt{NEWTRIAL}} routine then adjusted these aperture corrections for all stars in each frame. 

Second, the transformation to the standard photometric system was carried out by constructing transformation equations for each frame, which accounted for the zero-point and atmospheric extinction. The mean of other zero-points was used for frames lacking a standard-field observation. The {\sc \texttt{CCDAVE}} routine applied these transformation equations to the program stars for each frame, and the {\sc \texttt{NEWTRIAL}} routine subsequently calibrated all other stars using the program stars as local standards. 

Finally, relative photometry between epochs was conducted to distinguish variable from non-variable sources accurately. Approximately 1000 common stars were selected across all frames within the magnitude interval 18 to 21 mag. Each star's deviation at each epoch was determined relative to the mean magnitude calculated from all epochs. These mean magnitudes were computed by weighting the individual measurements. The resulting corrections were then applied to the frames.

 To evaluate the completeness of the survey, artificial stars were added using the {\sc \texttt{ADDSTAR}} routine \citep{1987PASP...99..191S} in both $i$ and $V$-bands single frames, across discrete 0.50 magnitude bins ranging from 16 to 24.50 mag. The fraction of recovered artificial stars was estimated with the {\sc \texttt{ALLFRAME}} routine. The results indicated that the survey is approximately 90–95\% complete up to 22 mag in the $i$ and $V$-bands, near the TRGB (Section~\ref{sec:TRGB}), and up to 50\% complete for stars with a magnitude of approximately 23 mag in both filters, confirming that nearly the entire AGB populations are detected for the primary research purpose. In the color-magnitude diagram for each dwarf galaxy (Section~\ref{sec:CMD} and Figs.~\ref{fig:CMDs1}--\ref{fig:CMDs3}), the black dashed-line signifies the estimated completeness limit.
 
\subsection{Detection of long-period variables}
%%%%%%here
Identifying LPVs in sparsely sampled observations requires statistical techniques that can reliably distinguish intrinsic stellar variability from random noise. In the present dataset, the temporal coverage often spans only one or roughly 2-3 pulsation cycles, making classical periodogram-based methods unreliable. We therefore adopt the variability indices introduced by \citet{Welch93} and refined by \citet{Stetson96}, which provide a quantitative and statistically robust measure of variability by accounting for photometric uncertainties and correlated brightness changes across different filters. This approach is well-suited for large datasets and remains effective even when periodicity cannot be reliably determined.

The method begins by computing standardized magnitude deviations for each observation, defined as the difference between an individual measurement and the mean magnitude, normalized by the corresponding photometric uncertainty. This procedure places all observations on a common scale, ensuring that variability is measured consistently despite differences in measurement errors. The standardized deviations for the $i$ and $V$ filters are given in Equation~\ref{eq:mag_dev}, where $m_j$ is the magnitude in filter $j$, $\langle m \rangle$ is the mean magnitude, $\sigma_j$ is the measurement uncertainty, and $N$ is the total number of observations.
\begin{equation}
\delta_j={\sqrt{\frac {N}{N-1}}}{\frac{m_j-\langle m\rangle}{\sigma_j}}
\label{eq:mag_dev}
\end{equation}

To assess correlated variability, the method computes pairwise products of the standardized deviations obtained within a time interval shorter than half the minimum expected LPV period (60 days; \citealt{McDonald16}). For variable stars, brightness changes in different filters are expected to occur in the same direction, leading to positive products, whereas random noise produces uncorrelated deviations that tend to cancel out. This behavior is quantified by the Stetson $J$ index (Equation~\ref{eq:J_indx}), which assumes high positive values for variable stars and approaches zero for non-variable sources. The weighting factor $w_k$ accounts for the reliability of each observation pair, while $P_k$ is defined in Equation~\ref{eq:P_K}.
\begin{equation}
J=\frac{\sum_{k=1}^{N} w_k\ {\rm sign}(P_k)\sqrt{|P_k|}}{\sum_{k=1}^{N} w_k}
\label{eq:J_indx}
\end{equation}

\begin{equation}
P_k=\left\{
\begin{array}{lll}
(\delta_i\delta_j)_k & {\rm if} & i\neq j \\
\delta_i^2-1 & {\rm if} & i=j \\
\end{array}
\right.
\label{eq:P_K}
\end{equation}

To further characterize the variability, the kurtosis index $K$ (Equation~\ref{eq:K_indx}) is computed from the distribution of standardized deviations. This index probes the shape of the light-curve distribution by comparing the mean absolute deviation to the root mean square deviation, thereby distinguishing between different variability patterns. For example, $K \approx 0.9$ corresponds to sinusoidal variability, $K \approx 0.798$ reflects a Gaussian distribution dominated by measurement noise, and values approaching zero indicate distributions influenced by isolated outliers.
\begin{equation}
K=\frac{\frac{1}{N} \Sigma_{i=1}^N |\delta_i|}{\sqrt{\frac{1}{N}\Sigma_{i=1}^N\delta_i^2}}
\label{eq:K_indx}
\end{equation}

\begin{figure}
\includegraphics [width=0.48\textwidth]{ 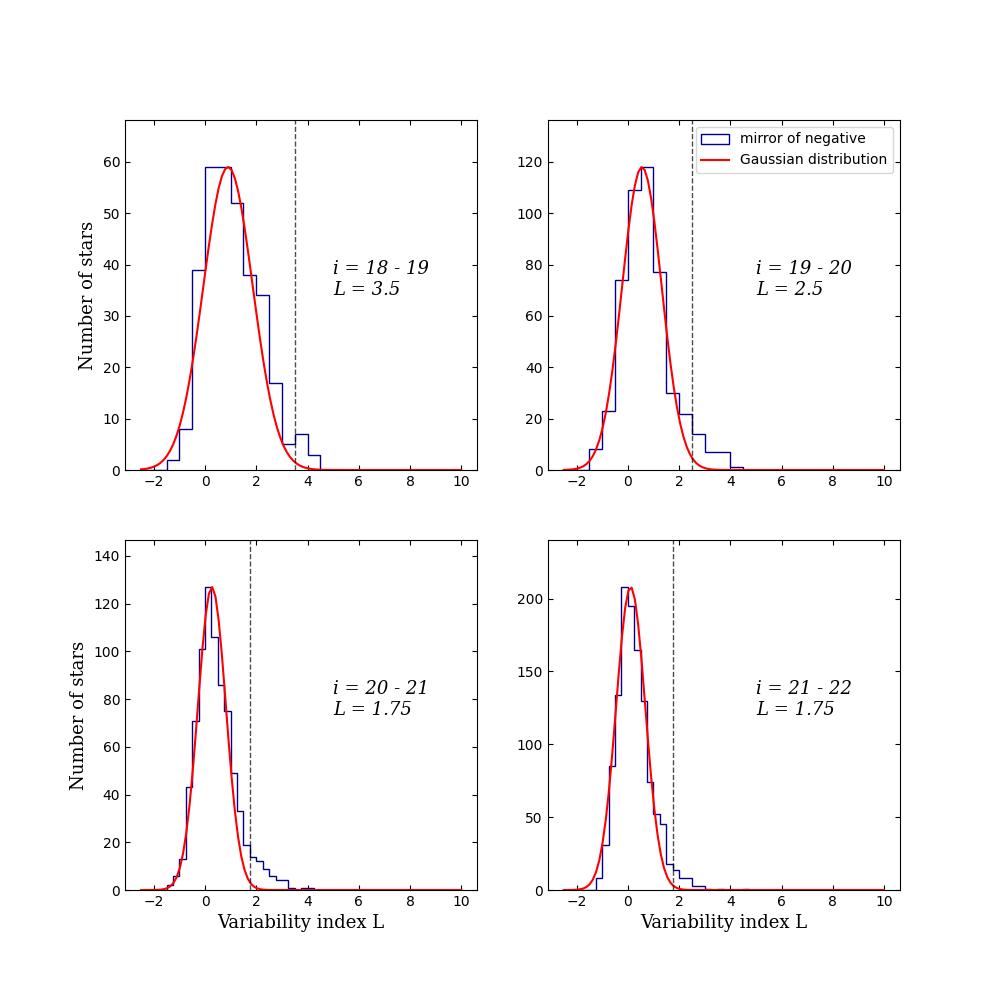}
\caption{The red curves show Gaussian functions fitted to the histograms of the variability index $L$ for the And\,V stellar population. The negative part of the blue histogram ($L<0$) is mirrored about this value. Vertical black dashed-lines indicate the variability index thresholds adopted in each magnitude bin.}
\label{fig:Gaussian_fiting}
\end{figure}

The final variability index, $L$, combines the $J$ and $K$ indices and is normalized by the total observational weight (Equation~\ref{eq:L_indx}). A higher $L$ value indicates a greater likelihood that a star is an LPV. A threshold for identifying candidate variable stars is established from the distribution of the variability index $L$ in each magnitude bin. The negative side of the $L$ index distribution is fitted with a Gaussian function, and stars with $L$ values exceeding this fit by a factor of ten are flagged as potential variables, corresponding to an approximate 90\% confidence level. Fig.~\ref{fig:Gaussian_fiting} provides an example of such a Gaussian fit. Additional details on this procedure can be found in previous INT survey studies \citep{2023ApJ...948...63A, 2021ApJ...923..164S, 2021ApJ...910..127N}.

\begin{equation}
L=\frac{J\times K}{0.798} \frac {\Sigma_{i=1}^N w_i}{w_{\rm all}}.
\label{eq:L_indx}
\end{equation}

The reliability of the method has been validated through simulations using the {\sc \texttt{ADDSTAR}} routine, in which artificial stars with known properties are injected into the data and subsequently recovered. Examples of light-curves for the identified LPV candidates are provided in the Appendices~\ref{fig:LC1} and \ref{fig:LC2}.

\subsection{Foreground Contamination}
To accurately identify LPV stars in the dwarf satellites of Andromeda, it is essential to remove foreground stars from the Milky Way. We cross-matched our catalog with Gaia DR3 \citep{GaiaDR3} to minimize such contamination. Milky Way foreground stars were identified using Gaia proper motions and parallaxes. The total proper motion was calculated as
\[
\mu_{\rm tot} = \sqrt{(\mu_{\rm RA})^2 + (\mu_{\rm DEC})^2},
\]
where $\mu_{\rm RA}$ and $\mu_{\rm DEC}$ are the proper motion components in right ascension and declination. Stars with $\mu_{\rm tot} > 0.28$ mas yr$^{-1}$, roughly at 2$\sigma$ above the measurement uncertainty, were classified as foreground. Additionally, stars with parallax significance $\rm Pa / \sigma_{Pa} \ge 2$ were also flagged as Milky Way stars. A star satisfying either of these two criteria was excluded from the analysis. The threshold value of 0.28 mas yr$^{-1}$ follows the approach adopted by \citet{2019ApJ...872...24V} for identifying Milky Way foreground stars in Local Group dwarf galaxy studies.

\subsection{Amplitude of Variability for Candidate Stars}

The amplitude of variability was estimated by assuming a sinusoidal light-curve. To account for photometric uncertainties, random offsets that scale with the measurement errors were included by using the observed standard deviation of the magnitudes.
The amplitude can be recovered using the equation:
\begin{equation}
A=2\sigma/0.701.
\label{eq:Amp_i}
\end{equation}

Where $\sigma$ is the observed standard deviation. Since our observations span a maximum of 2–3 pulsation cycles, the impact of semi-regular variations and changes in mean magnitude is reduced, enhancing the reliability of the derived amplitude estimates.

In this study, we prioritize stars with amplitudes exceeding 0.2 mag due to our uncertainty regarding the nature of stars with lower amplitudes.

\section{Results}\label{sec:parameter}
In this study, the photometric analysis of the target galaxies has enabled the identification of long-period variable stars and the estimation of the TRGB. For this work, all LPVs are treated as a single class, as the available data do not cover enough cycles to distinguish between Miras and semi-regular variables reliably. Determining the TRGB allows us to estimate the distance modulus, which is crucial for understanding the spatial positions of these galaxies. Additionally, the analysis provides estimates of the half-light radius, offering further insight into their structural properties. The following section discusses the implications of these results in more detail.

\subsection{Cross-Matching with DUSTiNGS}

To enhance the robustness of our findings, we cross-matched our photometric catalog results for M31's satellite galaxies with the DUST in Nearby Galaxies with \textit{Spitzer} (DUSTiNGS) catalog by \citet{Boyer15a}. Integrating our photometric results with this infrared catalog allows a better comparison of variability characteristics.

The DUSTiNGS project is an infrared observational study targeting 50 dwarf galaxies in and near the LG to identify evolved, dust-generating stars \citep{Boyer15a, Boyer15b}. Utilizing the InfraRed Array Camera (IRAC) on the \textit{Spitzer} Space Telescope (SST) in its post-cryogenic phase, the survey examined 37 dwarf spheroidal galaxies, eight dwarf irregular galaxies, and five galaxies exhibiting characteristics of both dIrr and dSph types. To improve the detection of AGB stars, known for their variability at 3.6 and 4.5 $\mu m$ wavelengths, each galaxy was observed twice, approximately six months apart. 

\begin{figure}
\includegraphics [width=0.48\textwidth]{ 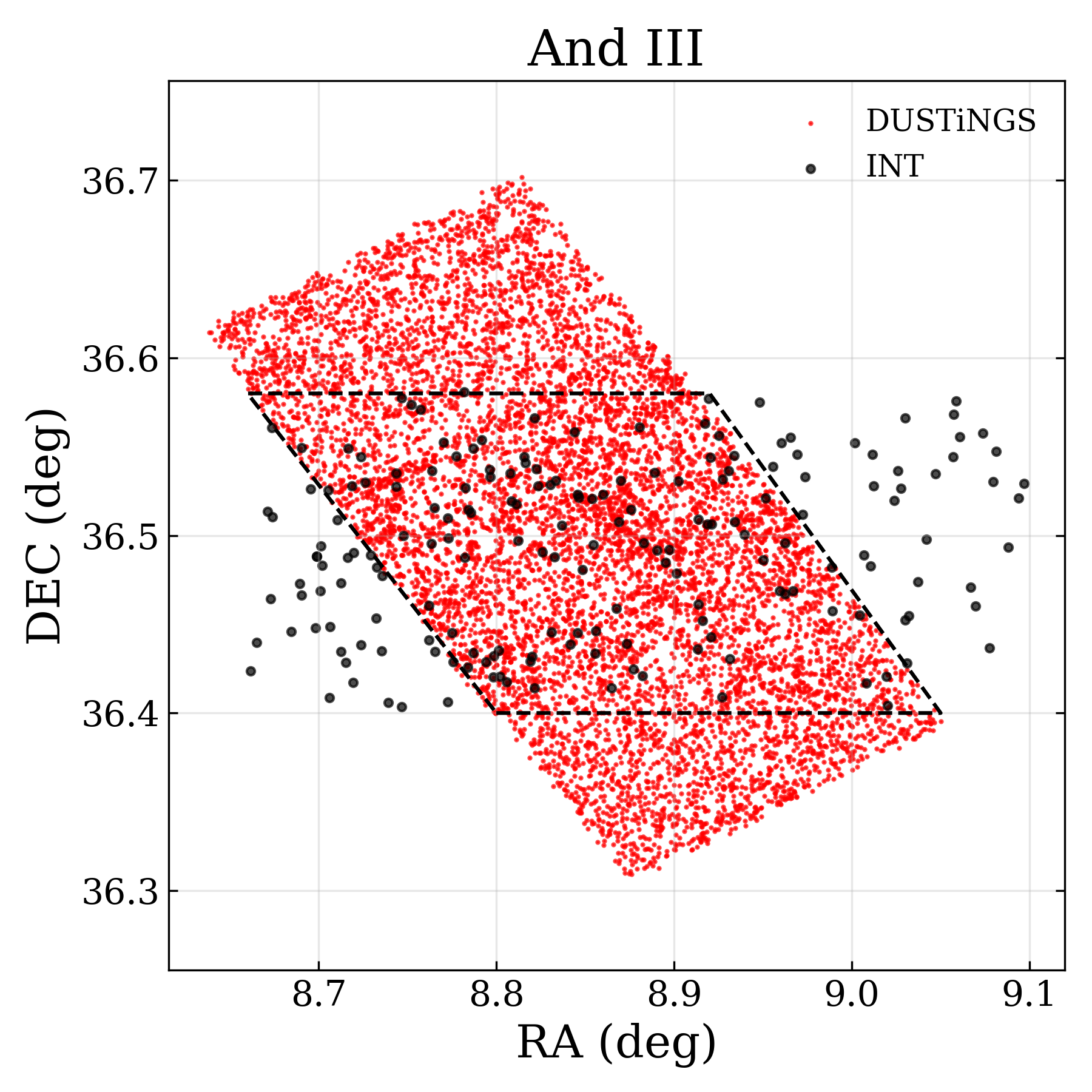}
\caption{The observed regions of the And\,III Galaxy, with black dots representing data from the Isaac Newton Telescope and red dots indicating observations from the \textit{Spitzer} Space Telescope.}
\label{fig:positions}
\end{figure}

\begin{figure}
\includegraphics [width=0.48\textwidth]{ 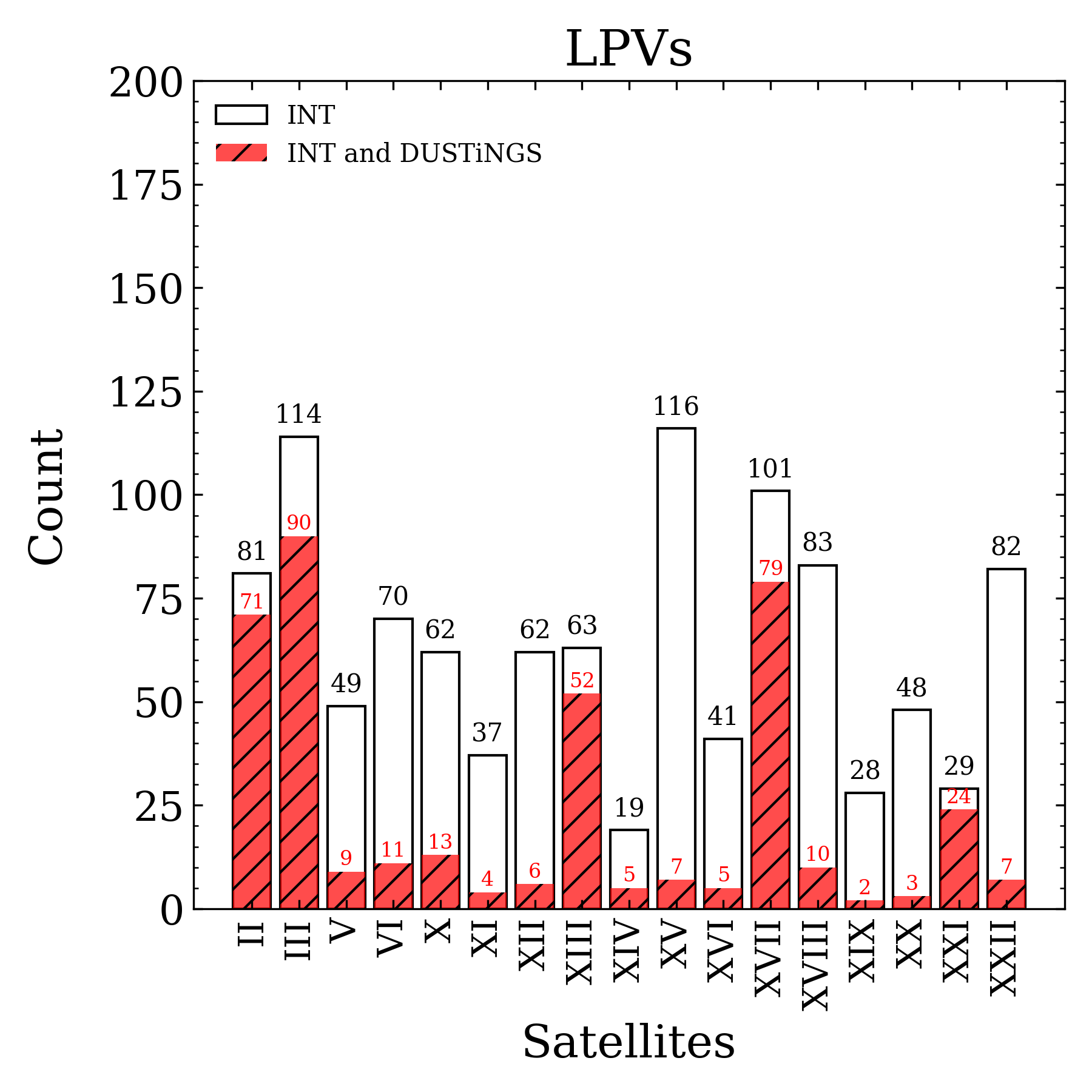}
\caption{The distribution of LPVs observed in both Isaac Newton and \textit{Spitzer} datasets. Black histograms show the total number of LPVs within the area overlapping with the \textit{Spitzer} observation, while red histograms indicate the subset of LPVs common to both observations.}
\label{fig:Dustings}
\end{figure}

To demonstrate how many LPV stars photometrically measured in this study overlap with those in the DUSTiNGS catalog, it should first be noted that the surveyed areas in the two observations differ. As illustrated in Fig.~\ref{fig:positions}, using galaxy And\,III as an example, the region enclosed by dashed-lines represents the area observed by both catalogs. For each galaxy, the number of LPVs within this overlapping region was identified, followed by determining how many of these LPVs are also included in the DUSTiNGS catalog. The results and details of these calculations are presented in Fig.~\ref{fig:Dustings}.

As seen in Fig.~\ref{fig:Dustings}, the ratio of LPVs detected in the INT and \textit{Spitzer} observations varies among galaxies. This variation primarily arises from differences in the sky coverage of the two surveys. A larger overlap area results in a ratio closer to unity, whereas a smaller overlap area or a limited number of detected LPVs leads to lower ratios.

It should be noted that the \textit{Spitzer} Space Telescope works within the infrared spectrum, which can penetrate dust clouds in a far better way. This, in turn, can select stars that are obscured in the optical range observed by the INT survey. As represented in Fig.~\ref{fig:positions}, \textit{Spitzer} detects more sources in a smaller field of view due to its infrared sensitivity, higher limits for faint objects, and effective survey strategies.

\subsection{Half-Light Radius}
\label{Section half-light}
To determine the half-light radius of each dwarf galaxy, we employed a method that involves plotting the number density and surface brightness of the stellar population as functions of the distance from the galaxy's center. Generally, as the number of stars in a region decreases, the surface brightness also diminishes. However, this trend is not always observed, as the distribution of faint and luminous stars can vary across different areas, leading to deviations from direct proportionality between star count and brightness.

To calculate the half-light radius, stars were first sorted by distance from the galaxy center. They were then grouped into radial bins, each containing an equal number of stars, so that each bin's width was adjusted to the local stellar density. For each bin, we computed the surface brightness and number density.
 
Following the approach of \citet{2006MNRAS.365.1263M}, the Exponential (Equation~\ref {equ: exp}) and Plummer (Equation~\ref{equ: Plummer}) profiles were used to estimate the half-light radius \citep{Faber83, Plummer11}. The findings from the fitting of the King profile \citep{king66} exhibited disparities when compared to those derived from the Exponential and Plummer profiles. Subsequently, we elected to utilize the S\'ersic profile (Equation~\ref{equ: Sersic}) to attain results that are more consistent and precise \citep{Sersic68}.

\begin{equation}
\label{equ: exp}
 I(x) = a\times e^{(-x/b)} + c
\end{equation}

\begin{equation}
\label{equ: Plummer}
I(x) = a\times(b^2/(b^2+x^2)^2) + c
\end{equation}

\begin{equation}
\label{equ: Sersic}
I(x) = a\times e^{(-b\times((x/c)^{(1/n)}-1))}
\end{equation}

\begin{figure}[H]
    \centering
    \includegraphics[width=0.4\textwidth]{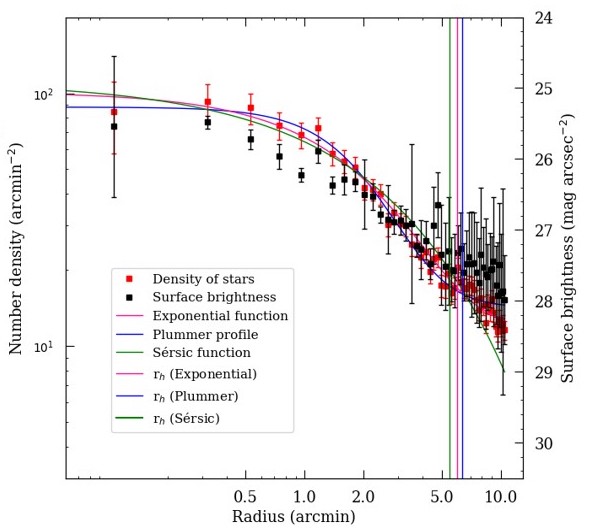}
   \caption{Stellar number density (red squares) and surface brightness (black squares) profiles of And\,II dwarf galaxies as a function of galactocentric distance, shown together with the best-fitting exponential (pink), Plummer (blue), and S\'ersic (green) models. Dashed-lines mark the half-light radii derived from the corresponding exponential, Plummer, and S\'ersic fits. Vertical error bars represent Poisson uncertainties in the stellar counts.}
    \label{fig:hl}
\end{figure}

Table~\ref{table:objects} contains the estimated half-light radii obtained from Exponential, Plummer, and S\'ersic fits. For more comparison, the half-light radii from \cite{2016ApJ...833..167M} and \cite{2012AJ....144....4M} are referenced in Table~\ref{table:objects}. As an example, Fig.~\ref{fig:hl} presents the fitted profiles employed in the half-light radius calculation for the And\,II dwarf galaxy.

% %%%%%%%%%%%%%%%%%%%%%%%%%%%%%%%%%%%%%%%%%%%%%%%%%%%%%%%%%%%%%%%%%%%% 

\subsection{The Tip of the Red Giant Branch as a Distance Indicator}
\label{sec:TRGB}

The TRGB serves as a crucial distance indicator for calibrating the cosmic distance ladder, thus playing an essential role in determining cosmological parameters such as the Hubble constant (H$_{0}$).
In this study, we employ the Sobel filter to estimate the TRGB of the stellar population within two half-light radii \citep{Lee93, Sakai96}. The Sobel filter, a well-established edge detection technique in image processing, is used here to highlight sharp discontinuities in the luminosity function of red giant branch (RGB) stars that correspond to the TRGB.

The filter computes the gradient magnitude by convolving the luminosity function with one-dimensional kernels, effectively identifying rapid changes in stellar counts with magnitude. The maximum gradient marks the TRGB location, where the luminosity function changes most abruptly. Depending on the noise level and crowding in the data, either the [-2, -1, 0, 1, 2] kernel, which provides smoother results in noisy or crowded fields, or the simpler [-1, 0, 1] kernel may be adopted \citep{2023AJ....166....2M}.

The estimated TRGB magnitudes and corresponding distance moduli for each of Andromeda’s dwarf satellites are listed in Table~\ref{table:TRGB}. The fourth column reports the TRGB values derived by \citet{2012ApJ...758...11C} using the same $i$-band filter, while the fifth and sixth columns show the minimum and maximum TRGB estimates from other studies. The two rightmost columns indicate the Galactic extinction and its references. 

For each galaxy, the absolute TRGB magnitude in the Sloan $i$-band was determined individually using PARSEC stellar evolution models in the SDSS photometric system, following the methodology of \citet{2012ApJ...758...11C} and accounting for the metallicity of the RGB population. Galactic extinction was corrected using $A_i = 1.698 \times E(B-V)$, where the SDSS $i$-band extinction coefficient was adopted from \citet{2011ApJ...737..103S} and the reddening values E(B−V) were taken from \citet{1998ApJ...500..525S}. 

Discrepancies between our results and those in the literature can arise from several factors, including photometric depth, instrumental characteristics, and data reduction methods. In particular, the level of stellar crowding influences the accuracy of the TRGB determination \citep{Dolphin02, Makarov06, Rizzi07}. 

Excluding the most crowded central regions could, in principle, yield a cleaner TRGB detection by minimizing blending effects, though this comes at the cost of reduced RGB statistics and increased field contamination. Future analyses could explore this trade-off in more detail using artificial-star tests.

The TRGB magnitudes obtained here correspond to red magnitudes observed in the $i$-band, which show only a weak dependence on metallicity and age \citep{Lee93}. This relative insensitivity makes the TRGB a robust extragalactic distance indicator. Compared to RR~Lyrae and Cepheid variables, the TRGB method benefits from being intrinsically brighter than RR~Lyrae stars and less affected by extinction than Cepheids, although contamination from AGB stars near the tip can occasionally complicate edge detection. Figs.~\ref{fig:TRGB1}--\ref{fig:TRGB3} present Sobel filter response for the TRGB determination.

\begin{figure}
\includegraphics [width=0.48\textwidth]{ 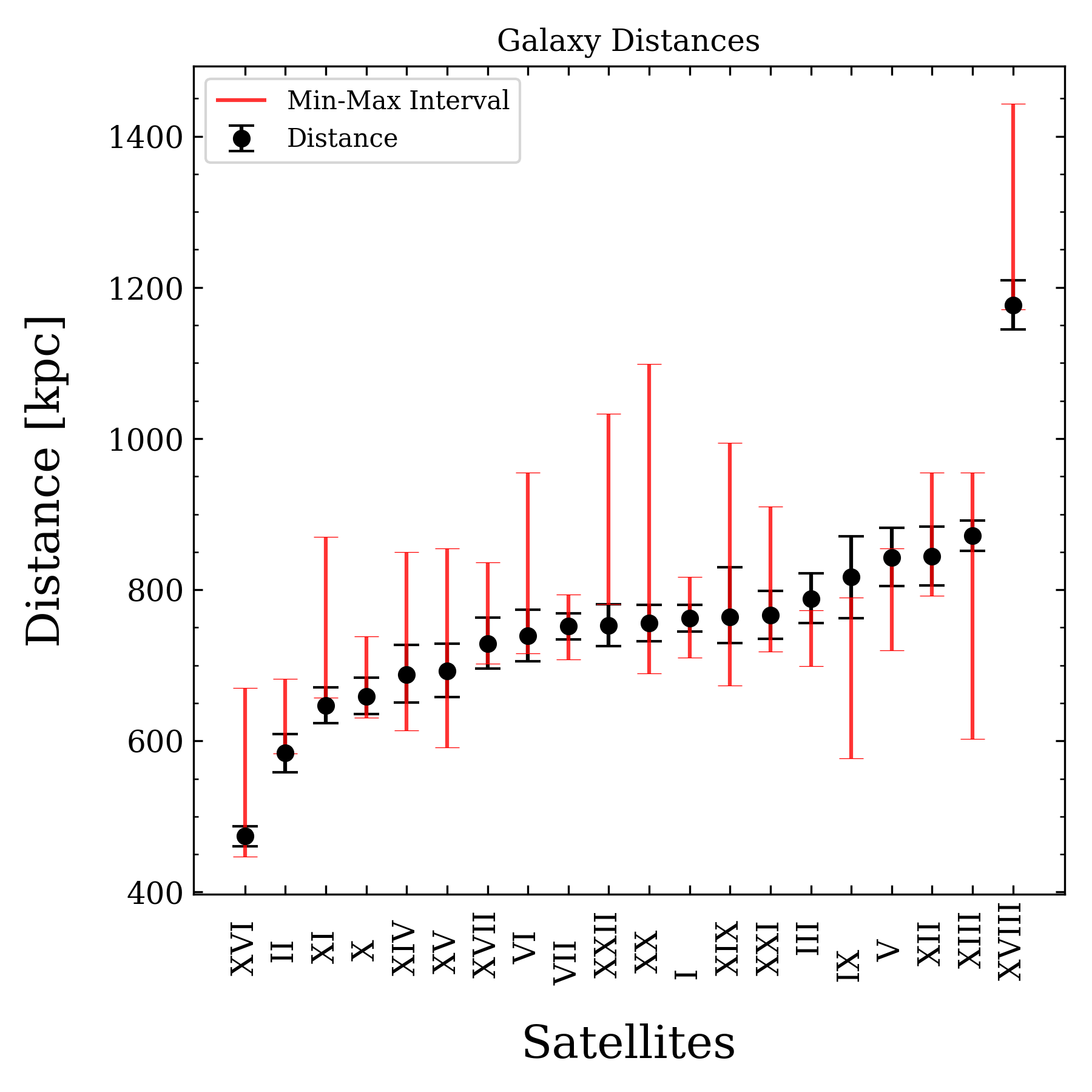}
\caption{Distance results for Andromeda's satellite galaxies.
The range depicted in this plot is derived from \citet{2012ApJ...758...11C}.}
\label{fig:distance}
\end{figure}

Fig.~\ref{fig:distance} presents the distances derived for each galaxy using the TRGB method. Our estimates, shown as black circles with uncertainties, are compared with the range of literature values from \citet{2012ApJ...758...11C} (red error bars). Overall, our results are consistent with previous determinations within uncertainties. A more detailed comparison and discussion of individual systems is provided in Section~\ref{sec:Discussion}.

\begin{table*}
\caption{Physical characteristics of the targets.}
\small % Adjust the font size to make the table smaller
\setlength{\tabcolsep}{3.3pt}
\label{table:TRGB}
\centering
\begin{tabular}{llllllll}
\hline\hline
 \noalign{\smallskip}
{Galaxy} &
TRGB$_{\,this\,work}$ (mag) &
$\mu_{\,this\,work}$ (mag) &
TRGB$^a$ (mag)&
$\mu_{Min}$ (mag) &
$\mu_{Max}$ (mag)&
A$_V$$^s$ (mag) 
&
A$_i$$^s$ (mag)\\
&  &Distance (kpc) &  & Distance (kpc) & Distance (kpc) &&\\

 \noalign{\smallskip}
\hline
 \noalign{\smallskip}
And\,I$^\star$      & $21.00\pm0.05$ & $24.41\pm0.05$ & $20.98\pm0.05$ &$24.31\pm0.05$ $^a$ & $24.56\pm0.09$ $^b$ & 0.145 & 0.090\\
            & & $762.08\pm17.75$& &$727.28\pm16.95$ & $816.58\pm 34.56$& &\\
            
And\,II     & $20.40\pm0.10$ & $23.83\pm0.10$ & $20.69\pm0.05$ &$23.83\pm0.42$ $^c$ & $24.17\pm0.12$ $^b$ & 0.167 & 0.104 \\
            & & $584.45^{+27.54}_{-26.31}$& &$583.96\pm54.69$ & $682.34\pm 38.77$& &\\

And\,III   & $21.14\pm0.09$ & $24.48\pm0.09$ & $20.98\pm0.07$ & $24.18\pm0.11$ $^d$ & $24.48\pm0.19$ $^b$ & 0.152 & 0.094\\
           & & $788.21^{+33.36}_{-32.00}$ & &$685.44\pm30.98$ & $787.17\pm68.41$ & & \\

And\,V     & $21.27\pm0.10$ & $24.63\pm0.10$ & $21.17\pm0.07$&$24.35\pm0.07$ $^a$ & $24.66\pm0.13$ $^e$ & 0.342 & 0.212\\
           & & $842.46^{+39.70}_{-37.92}$& & $741.31\pm24.29$ & $855.07\pm 52.75$ & &\\

And\,VI    & $21.01\pm0.10$ & $24.34\pm0.10$ & - & $24.37\pm0.09$ $^d$ & $24.94\pm0.09$ $^f$ & 0.173 & 0.107 \\
           & &$738.93^{+34.83}_{-33.26}$& & $748.17\pm31.66$ & $973.99\pm40.42$ & &\\

And\,VII$^\star$   & $21.30\pm0.05$ & $24.38\pm0.05$ & - &$24.25\pm0.25$ $^g$ & $24.50\pm0.10$ $^h$ & 0.532& 0.329\\
           & & $751.62\pm17.15$& &$707.95\pm86.38$ & $794.33\pm 37.43$ & & \\

And\,IX$^\star$    & $21.20\pm0.15$ & $24.56\pm0.15$& $20.61\pm0.31$ &$23.89\pm0.31$ $^a$ & $24.48\pm0.20$ $^i$ & 0.206 & 0.128\\
           & &$816.58\pm54.50$& &$599.79\pm92.04$ & $787.05\pm 75.93$ & &\\

And\,X     &$20.77\pm0.08$ & $24.09\pm0.08$ & $20.95\pm0.28$& $24.01\pm0.07$ $^j$ & $24.23\pm0.10$ $^k$ & 0.354 & 0.219\\
           & & $659.19^{+24.74}_{-23.84}$& &$633.87\pm20.77$ & $701.46\pm33.05$ & &\\

And\,XI    & $20.75\pm0.08$& $24.05\pm0.08$ & $21.14\pm0.31$& $24.17\pm0.09$ $^d$ & $24.70\pm0.20$ $^l$ & 0.219 & 0.136\\
           & & $646.89^{+24.28}_{-23.40}$& & $682.34\pm28.87$ & $870.96\pm 84.03$ & & \\

And\,XII   & $21.38\pm0.10$ & $24.63\pm0.10$ & $21.63\pm0.34$&$24.44\pm0.12$ $^f$ & $24.84\pm0.34$ $^a$ & 0.302 & 0.187\\
           & & $843.9^{+39.77}_{-37.99}$& & $773.76\pm47.26$ & 928.97$\pm 157.46$ & & \\

And\,XIII  & $21.40\pm0.05$ & $24.70\pm0.05$ & $21.30\pm0.07$ &$24.40\pm0.49$ $^a$ & $24.80\pm0.35$ $^m$ & 0.226 & 0.140\\
           & & $871.27^{+20.29}_{-19.83}$ & &$758.58\pm192.02$ & $912.01\pm 159.51$ & & \\

And\,XIV   & $20.85\pm0.12$ & $24.19\pm0.12$ & $21.18\pm0.56$ &$24.33\pm0.33$ $^n$ & $24.50\pm0.56$ $^a$ & 0.164
& 0.101\\
           & & $688.06^{+39.10}_{-37.00}$ & &$734.51\pm120.56$ & $794.33\pm 223.69$ &&\\
           
And\,XV    & $20.84\pm0.11$ & $24.21\pm0.11$ & $21.02\pm0.05$ &$23.98\pm0.26$ $^a$ & 24.66 $^o$ & 0.128 & 0.079\\
           & & $692.43^{+35.98}_{-34.20}$& &$625.17\pm 79.52$ & 855.07 &&\\

And\,XVI   & $20.05\pm0.06$ & $23.38\pm0.06$ & $20.27\pm0.08$ &$23.30\pm0.19$ $^a$ & $24.13\pm0.04$ $^p$ & 0.182 & 0.113\\
           & & $473.79^{+13.27}_{-12.91}$& &$467.74\pm 42.76$ & $669.88\pm12.46$ &&\\

And\,XVII  & $21.00\pm0.10$ & $24.31\pm0.10$ & $21.12\pm0.07$ &$24.31\pm0.11$ $^a$ & $24.50\pm0.10$ $^q$ & 0.204 & 0.126\\
           & & $728.67^{+34.34}_{-32.80}$ & &$727.78\pm37.82$ & $794.33\pm37.43$ &&\\
           
And\,XVIII & $22.09\pm0.06$ & $25.35\pm0.06$ & $22.20\pm0.08$ &$25.42\pm0.08$ $^a$ & $25.66\pm0.13$ $^r$ & 0.290 & 0.180\\
           & & $1176.74^{+32.97}_{-32.07}$& & $1213.39\pm45.54$ & $1355.19\pm 83.61$ &&\\
           
And\,XIX   & $21.08\pm0.10$ & $24.41\pm0.10$ & $21.24\pm0.09$ &$24.52\pm0.23$ $^s$ & $24.85\pm0.13$ $^r$ & 0.170 & 0.105\\
           & & $763.74^{+35.99}_{-34.38}$ & &$802.35\pm109.65$ & $933.25\pm 57.58$ &&\\
           
And\,XX    & $21.05\pm0.07$ & $24.39\pm0.07$ & $21.03\pm0.08$ &$24.35\pm0.16$ $^a$ & $24.52\pm0.74$ $^r$ & 0.161 & 0.100\\
           & & $755.62^{+24.75}_{-23.97}$ & & $741.31\pm56.68$ & $801.68\pm325.52$ &&\\
           
And\,XXI   & $21.14\pm0.09$ & $24.42\pm0.09$ & $21.19\pm0.07$&$24.40\pm0.17$ $^s$& $24.67\pm0.13$ $^t$ & 0.256 & 0.158\\
           & & $766.33^{+32.43}_{-31.11}$& & $827.94\pm27.13$ & $859.01\pm53.00$ &&\\
           
And\,XXII  & $21.07\pm0.08$ & $24.38\pm0.08$ & $21.11\pm0.07$ &$24.65\pm0.07$ $^u$ & $24.87\pm0.09$ $^u$  & 0.216 & 0.133\\
           & & $752.54^{+28.24}_{-27.24}$& &$853.45\pm32.98$ & $940.12\pm41.70$  &&\\
\hline
\end{tabular}
\tablecomments{
\footnotesize{The physical parameters in this Table are based on references; 
a~ \citep{2012ApJ...758...11C}, 
b \citep{2017ApJ...850..137M} , 
c \citep{1993AJ....106.1819K}, 
d \citep{2013MNRAS.435.3206D}, 
e \citep{2000ApJ...529..745F}, 
f \citep{2014ApJ...789..147W}, 
g \citep{1999AstL...25..332T}, 
h \citep{2009AJ....138..332J}, 
i \citep{2004ApJ...612L.121Z}, 
j \citep{2011ApJ...729...23B}, 
k \citep{2007ApJ...659L..21Z}, 
l \citep{2006MNRAS.371.1983M}, 
m \citep{2010MNRAS.407.2411C}, 
n \citep{2007ApJ...670L...9M}, 
o \citep{2017ApJ...837..102S}, 
p \citep{2016ApJ...819..147M}, 
q \citep{2008ApJ...676L..17I}, 
r \citep{2008ApJ...688.1009M}, 
s \citep{2013ApJ...779....7C}, 
t \citep{2009ApJ...705..758M}, 
and u \citep{2013MNRAS.430...37C}}\\
}
\footnotesize{Galaxies with star signs have been previously studied by the INT group; And\,I \citep{2021ApJ...923..164S}, And\,VII \citep{2021ApJ...910..127N}, and And\,IX \citep{2023ApJ...948...63A}. }
\end{table*}

%@@@@@@@@@@@@@@

%%%%%%%%%%%%%%%%%%%%%%%%%%%%%%%%%%%%%%%%%%%%%%%%%%%%%%%%%%%%%%%%%%%%
\subsection{Color-Magnitude Diagram} \label{sec:CMD}

\begin{figure}
\includegraphics [width=0.48\textwidth]{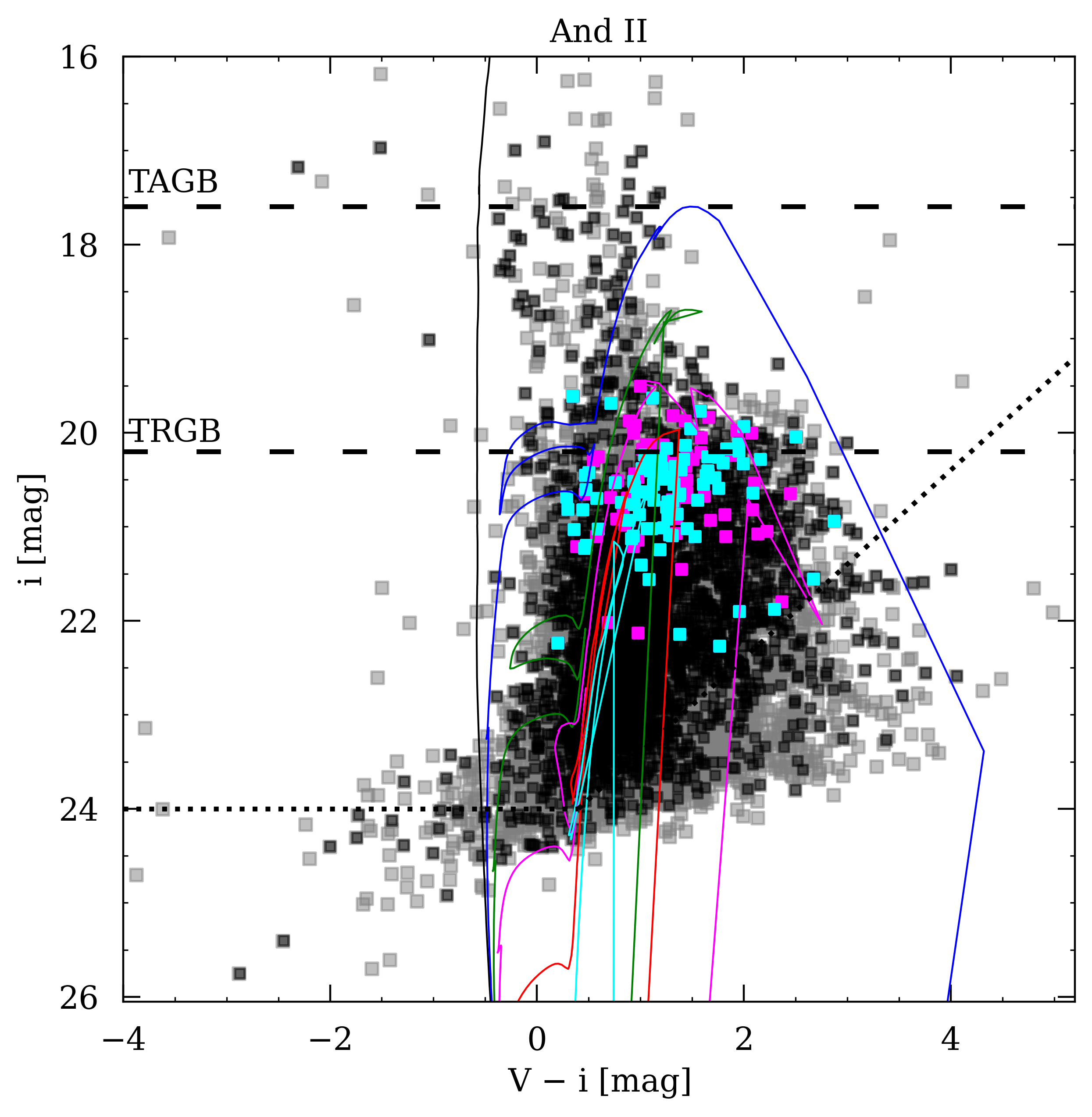}
\caption{Color–magnitude diagram (CMD) of LPVs in the And\,II galaxy (additional CMDs are provided in Appendices~\ref{fig:CMDs1}--\ref{fig:CMDs3}). Gray markers represent all stars photometrically identified in the final images (Appendices~\ref{fig:Location1}--\ref{fig:Location3}). Black points denote stars located within two half-light radii of the galaxy. Magenta points indicate all LPV candidates across the observed field, while cyan points highlight LPV candidates within twice the half-light radius. The isochrones are shown as colored curves, ranging from log(t/yr) = 6.6 (black), 8 (blue), 8.6 (green), 9 (purple), 9.4 (red), and 10 (cyan). The dashed-lines mark the positions of the TAGB and TRGB, while the dotted-line indicates the photometric completeness limit.}
\label{fig:CMD_And_II}
\end{figure}

Fig.~\ref{fig:CMD_And_II} shows the color–magnitude diagrams (CMDs) of And\,II as an example of the stellar population distribution in the Hertzsprung–Russell diagram. CMDs for the other galaxies are provided in Appendices~\ref{fig:CMDs1}--\ref{fig:CMDs3}. In these diagrams, gray markers represent all stars photometrically identified in the final images (Appendices~\ref{fig:Location1}--\ref{fig:Location3}), black points denote stars within two half-light radii of the galaxy, and magenta points indicate all LPV candidates across the observed field. Figs. \ref{fig:Location1}--\ref{fig:Location3} show the distribution of LPVs in the photometry field. The half-light radius of each galaxy is indicated by a solid black line, while the dashed-line illustrates twice the half-light radius.

The PARSEC-COLIBRI isochrones \citep{Marigo17}, along with relevant metallicity data (Table~\ref{table:objects}) and our estimated distance modulus discussed in Section~\ref{sec:TRGB}, are overlaid on these plots. The black dashed-lines indicate the AGB and TRGB tips. 

We used the classic core-luminosity relation to estimate the theoretical AGB tip, determining that the absolute bolometric magnitude for a Chandrasekhar core mass is approximately $-$7.1 mag, which we then converted to an apparent magnitude based on our distance modulus calculation.

PARSEC group isochrones were chosen for several reasons. They provide a sufficient range of birth masses ($0.8 < M/\msun < 30$) by integrating models for intermediate-mass stars ($M < 7 \msun$) with those for more massive stars ($M > 7 \msun$; \citep{Bertelli94}). These isochrones cover the thermal pulsing AGB phase through the post-AGB stage consistently with preceding evolutionary models. Importantly, they account for mechanisms such as the third dredge-up mixing in stellar envelopes during helium-burning pulses and the increased luminosity of massive AGB stars due to hot bottom burning \citep{Iben83}. Additionally, these isochrones include molecular opacities, which are vital for describing the cool atmospheres of red giants. 
%This feature allows the differentiation between oxygen-rich M-type AGB stars and carbon stars within a birth mass range of approximately 1.5 to 4 $\msun$ \citep{Girardi07}. 
They also incorporate predictions for dust production in the winds of LPVs, associated reddening effects, forecasts for radial pulsations, and adjustments for various optical and infrared photometric systems, all of which are easily accessible through a user-friendly online platform.

\section{Discussion} 
\label{sec:Discussion}

This section discusses the results of our time-series analysis for the satellite galaxies of Andromeda observed as part of the INT Survey. Previous analyses of other satellites in the INT Survey, including NGC 147, NGC 185, and IC 10, have revealed large populations of LPV stars and demonstrated their potential as tracers of intermediate-age stellar populations and dust production in Local Group dwarfs \citep{2023ApJ...948...63A, 2023ApJ...942...33P, 2021ApJ...923..164S, 2021ApJ...910..127N}. 
The present work extends this effort to the wider M31 satellite system, allowing a comparative assessment of stellar content, TRGB distances, and LPV properties across diverse galactic environments.

\subsection{And\,II}

And\,II was first identified by Sidney Van den Bergh from photographic plates obtained with the 48-inch Schmidt telescope at Palomar Observatory during 1970–1971 \citep{1972ApJ...171L..31V}. 
Distance estimates for And\,II have ranged from $23.83\pm0.42$ mag ($583.96\pm54.69$ kpc) based on the CMD method \citep{1993AJ....106.1819K} to $24.17\pm0.12$ mag ($682.34\pm38.77$ kpc) using the TRGB method and BaSTI evolutionary tracks \citep{2017ApJ...850..137M}.

As observed in the subplot concerning the And\,II galaxy in Fig.~\ref{fig:Location1}, the CCD1, CCD3, and CCD4 of WFC had to be used for photometry due to the substantial size of this galaxy. Consequently, the number of stars (11230) and detected LPVs (155) is more significant in comparison to other target galaxies. The results obtained for TRGB and distance modulus are found to be consistent with those of the study conducted by \citet{McConnachie04}.

\subsection{And\,III}

The region corresponding to twice the half-light radius of And\,III was encompassed within CCD4. Consequently, only CCD4 underwent photometric analysis. Among the 5878 photometric stars examined, 185 LPV candidates were identified.

The least distance modulus calculated for this dwarf galaxy is $24.18\pm0.11$ mag ($685.44\pm30.98$ kpc) \citep{2013MNRAS.435.3206D} while the farthest distance is approximately 24.48 mag ($787.17\pm62.68$ kpc) \citep{2017ApJ...850..137M} using the RR~Lyrae stars. To estimate the least distance for RR~Lyrae variables, statistical parallax is utilized. The method involves using proper motions from the US Naval Observatory CCD Astrograph Catalogue (UCAC) and apparent magnitudes from the Wide-Field Infrared Survey Explorer (WISE) to determine the parameters of the velocity distribution of the Galactic RR~Lyrae population \citep{2013MNRAS.435.3206D}. 
The calculations in \cite{2017ApJ...850..137M} rely on the PL relation specific to RR~Lyrae stars. This method is based on the direct relationship between the pulsation period of RR~Lyrae stars and their absolute brightness, which is then compared to their apparent brightness.

\subsection{And\,V}
And\,V was first observed using a digital filtering technique applied to images from the second Palomar Sky Survey (POSS-II). Covering an area of 1550 square degrees around the Andromeda galaxy, this method was employed as part of a survey to identify low surface brightness galaxies near the Andromeda galaxy \citep{1998AJ....116.2287A}.

The distance modulus for And\,V was found to vary in the range of $24.35\pm0.07$ mag ($741.31\pm24.29$ kpc) \citep{2012ApJ...758...11C} to $24.66\pm0.855$ mag ($855.07\pm52.75$ kpc) \citep{2000ApJ...529..745F} using the TRGB method.
The main distinction between the two estimations arises from statistical perspectives. The \cite{2000ApJ...529..745F} utilizes Cepheid variable stars in 25 galaxies to calibrate secondary distance indicators such as TRGB, while the \cite{2012ApJ...758...11C} paper applies a Bayesian statistical framework to estimate TRGB.
In this galaxy, 112 LPV candidates were identified out of 7916 stars that underwent photometry. The calculations conducted in this study exhibit a closer alignment with the minimum distance previously reported for this galaxy.

\subsection{And\,VI}
The distances to And\,VI span from $24.37\pm0.09$ mag ($748.17\pm31.66$ kpc) utilizing the RR~Lyrae method \citep{2013MNRAS.435.3206D} to 24.94 mag ($973.99\pm40.42$ kpc) using the CMD method \citep{2014ApJ...789..147W}. Using the TRGB as a distance indicator, we obtained a distance modulus of $24.32\pm0.10$ mag ($731.14\pm34.46$ kpc).

The variations among reported distance moduli can be attributed to several factors, including differences in methodology, calibration, and the treatment of interstellar dust. For instance, \cite{1998ApJ...500..525S} utilized infrared data to trace dust emission, while \cite{2011ApJ...737..103S} used higher-resolution optical data from the Sloan Digital Sky Survey (SDSS) to recalibrate those dust maps. The latter study revised the original dust extinction values upward by about 14\%, which directly affects the correction for attenuation of starlight and hence the inferred brightness of stars used in distance determinations. These differences in extinction correction, therefore, lead to corresponding variations in the derived distance modulus. 

This galaxy contains 6342 stars, including 117 LPV candidates.

\subsection{And\,X}
The discovery of And\,X was reported by \cite{2007ApJ...659L..21Z}. The discovery was based on stellar photometry from the SDSS. Follow-up imaging data allowed the researchers to estimate the distance modulus ranging from $24.01\pm0.07$ mag ($633.87\pm20.77$ kpc) \citep{2011ApJ...729...23B} with the Horizontal Branch method to $24.23\pm0.10$ mag ($701.46\pm33.05$ kpc) \citep{2007ApJ...659L..21Z} with the TRGB method. 
\cite{2011ApJ...729...23B} employed deep wide-field photometry from the Large Binocular Camera to scrutinize the stellar and structural properties of the galaxies. 

In this galaxy, among the 3502 stars observed in CCD4, 123 LPV candidates were detected. The distance modulus and TRGB stars also show good concordance with existing literature.

\subsection{And\,XI}
\cite{2006MNRAS.371.1983M} highlighted the discovery of And\,XI, And\,XII, and And\,XIII dwarf galaxies using data collected from the Canada-France-Hawaii Telescope (CFHT). These discoveries were obtained from a MegaCam survey that encompassed the southern quadrant of the Andromeda galaxy, covering a projected distance of approximately 50 to 150 kpc. Based on the overdensity of stars aligned along the RGB tracks in their CMDs, the detections were determined \citep{2006MNRAS.371.1983M}. The least distance estimation is $24.17\pm0.09$ mag ($682.34\pm28.87$ kpc) with RR~Lyrae methods \citep{2013MNRAS.435.3206D}, while the furthest distance is about $24.70\pm0.20$ mag ($870.96\pm84.03$ kpc) through the TRGB method \citep{2006MNRAS.371.1983M}.

In this small and faint galaxy, 84 LPV candidates have been identified using the index method from a pool of 5219 photometric stars. Furthermore, factoring in the margin of error in calculations, the results obtained for the distance modulus align well with the documented observations.

\subsection{And\,XII}

Based on our calculations, And\,XII, which contains 116 LPVs and 3552 stellar sources, is located at a distance of 915 kpc as derived from the TRGB method. Previous distance estimates range from $24.44\pm0.12$ mag ($774\pm47$ kpc) using the CMD method \citep{2014ApJ...789..147W} to $24.84\pm0.34$ mag ($929\pm157$ kpc) using the TRGB method \citep{2012ApJ...758...11C}.

\cite{2007ApJ...662L..79C} obtained spectroscopic observations of red giant branch stars in And\,XII to measure their radial velocities rather than their distances directly. These velocities were then combined with assumed distances from photometric estimates to model the galaxy’s orbit within the gravitational potential of M31. Their analysis indicated that And\,XII has an unusually high radial velocity (approximately $-556~\mathrm{km\,s^{-1}}$), approaching the local escape speed of the Andromeda system, suggesting it may be on its first infall into the Local Group.

Our TRGB-derived distance of 915~kpc places And\,XII slightly beyond the canonical distance to M31 ($\sim780~\mathrm{kpc}$). This position is consistent with the scenario proposed by \cite{2007ApJ...662L..79C}, where And\,XII is located behind M31 and moving rapidly toward it. Thus, our independent distance determination supports the interpretation of And\,XII as a dynamically extreme satellite, possibly on its first approach to the Andromeda galaxy.

\subsection{And\,XIII}

In \citet{2010MNRAS.407.2411C}, And\,XIII is described as having a "very poorly populated luminosity function." As shown in the subplot for this galaxy in Fig.~\ref{fig:TRGB2}, the limited number of observed AGB stars increases the uncertainty in locating the TRGB. Nevertheless, our photometry (4052 stars, including 111 LPV candidates) yields a TRGB-based distance of $859\pm20$ kpc.

\citet{Yang12} advocate the use of RR~Lyrae stars for And\,XIII and report a RRab-based true distance modulus of $24.62\pm0.05$, which corresponds to $840\pm18$ kpc . The two independent estimates differ by about 20 kpc, which is comparable to the quoted uncertainties and therefore consistent within the errors. Owing to the sparse AGB population in And,XIII, the RR~Lyrae-based distance offers a reliable and complementary constraint on its distance.

\subsection{And\,XIV}

The discovery of this dwarf galaxy is notable because And\,XIV shows a heliocentric radial velocity of $-481$ km s$^{-1}$, corresponding to $-206$ km s$^{-1}$ relative to the Andromeda galaxy \citep{2007ApJ...670L...9M}. This indicates that And\,XIV is approaching M31 at a velocity close to the galaxy’s escape speed. If And\,XIV is gravitationally bound to Andromeda, this would imply a higher total mass for M31. Alternatively, if it is not bound, And\,XIV may be entering the Local Group for the first time, representing a dwarf galaxy that evolved in isolation before its current interaction with M31.

Out of the 7823 stars subjected to photometry in CCD4, 48 LPV candidates were recognized. Our calculations indicate a distance of $680\pm38.60$ kpc for this galaxy, a finding that aligns with the conclusions drawn by \citet{2007ApJ...670L...9M}.

\subsection{And\,XV}

A deep photometric survey of the Andromeda galaxy, carried out with the wide-field cameras of the Canada-France-Hawaii Telescope and the Isaac Newton Telescope, led to the discovery of And\,XV \citep{2007ApJ...671.1591I}. The survey extended to nearly 150 kpc from the center of M31 and revealed numerous faint stellar substructures, including stellar streams and an extended smooth halo component, offering valuable insights into the structure and formation history of the Andromeda system.

The distances to these galaxies varied between $23.98\pm0.26$ mag ($625.17\pm79.52$ kpc) \citep{2012ApJ...758...11C} and 24.66 mag (855.07 kpc) \citep{2017ApJ...837..102S} through the TRGB method. The other estimations fall within these ranges. Following the identification of 214 LPV stars among 7228 stellar populations, the distance modulus computed for this galaxy stands at $24.18\pm0.11$ mag ($685.49\pm35.62$ kpc).

\subsection{And\,XVI}

In this galaxy, we discovered 78 LPV candidates out of 2979 stars that underwent photometry. The distance modulus we computed for this galaxy is determined to be $23.35\pm0.06$ mag ($467.74\pm13.10$ kpc).
The minimum distance for And\,XVI is $23.35\pm0.19$ mag ($467.74\pm42.76$ kpc) \citep{2012ApJ...758...11C} up to now, and the maximum inferred distance, which stands at $24.13\pm0.04$ mag ($669.88\pm12.46$ kpc) \citep{2016ApJ...819..147M} utilizing the TRGB method. The other estimation varied between these ranges.

\subsection{And\,XVII}

And\,XVII was discovered using the Wide Field Camera on the Isaac Newton Telescope \citep{2008ApJ...676L..17I}. \citet{2008ApJ...676L..17I} measured its line-of-sight distance using the TRGB method, obtaining $794\pm40$ kpc (distance modulus 24.5 mag), placing it well within the halo of M31.

In our study, 174 LPV candidates were identified among 11097 photometered stars. Using the TRGB method, we derived a distance modulus of $24.28\pm0.10$ mag ($718\pm34$ kpc). Our value is slightly lower than that reported by \citet{2008ApJ...676L..17I}, but remains broadly consistent within the uncertainties, confirming the galaxy’s location within the M31 halo.

\subsection{And\,XVIII}

And\,XVIII was observed with ACS/HST on June 20, 2014, in the F606W and F814W filters. \citet{Makarova17} analyzed these images to determine the star-formation history and derived a TRGB distance of 1.33 Mpc, placing the galaxy at roughly 579 kpc from Andromeda.

In our ground-based photometry with the Isaac Newton Telescope, we identified 123 LPV candidates among 2544 stars. Using the TRGB method, we obtained a distance of $1153\pm32$ kpc. This revised distance brings And\,XVIII somewhat closer to M31, but it remains at a large enough separation that it is not definitively bound to the Andromeda system. Therefore, while it could be associated, And\,XVIII may also be an isolated dwarf galaxy within the Local Group.

\subsection{And\,XIX}

And\,XIX is the most extended dwarf galaxy known in the Local Group, with a half-light radius of $14.2\pm3.4$ arcmin \citep{2022MNRAS.517.4382C}.
The lowest distance estimate for this particular dwarf galaxy is approximately $24.52\pm0.23$ mag ($802.35\pm109.65$ kpc), as determined by utilizing the RR~Lyrae method \citep{2013ApJ...779....7C}.

As depicted in Fig.~\ref{fig:Location1}, the expanse of this galaxy is substantial enough to encompass all four CCDs of the WFC camera. The total count of photometric stars within this galaxy amounts to 7470, yielding 179 LPV candidates. The distance modulus computed for this galaxy stands at $24.39\pm0.10$ mag ($755.09\pm35.59$ kpc).

\subsection{And\,XX}

Galaxy And\,XX, being the smallest target in this study in terms of size, with 4065 stars, hosts 120 LPV candidates observed in CCD4 along its line of sight. The distance modulus calculated for this galaxy is $24.37\pm0.07$ mag ($748.17\pm24.51$ kpc), a value supported by the findings of \citet{2008ApJ...688.1009M}, accounting for calculation errors.

\subsection{And\,XXI}

As part of the Pan-Andromeda Archaeological Survey, the galaxies And\,XXI and And\,XXII were observed. This survey is dedicated to the exploration of the galactic cluster encompassing the Andromeda and Triangulum galaxies.
Both galaxies were identified as spatial overdensities of stars aligning with the metal-poor red giant branches \citep{2009ApJ...705..758M}. And\,XXI was observed with less estimation at a distance of $24.40\pm0.17$ mag ($827.94\pm27.13$ kpc) \citep{2015ApJ...806..200C} using the RR~Lyrae method, and the furthest distance was observed at $24.67\pm0.13$ mag ($859.01\pm53.00$ kpc) \citep{2009ApJ...705..758M} using the TRGB method. Following the identification of 52 LPV stars among 4078 detected stars, the distance modulus we computed for this galaxy stands at $24.39\pm0.09$ mag ($759.09\pm31.96$ kpc).

\subsection{And\,XXII}

And\,XXII lies significantly closer to the Triangulum galaxy (M33) than to Andromeda, suggesting a potential association with M33 \citep{2009ApJ...705..758M}. Literature distance estimates using the Horizontal Branch method range from $24.65\pm0.07$ mag ($853\pm33$ kpc) \citep{2013MNRAS.430...37C} to $24.87\pm0.09$ mag ($940\pm42$ kpc) \citep{2013MNRAS.430...37C}. In our photometry, we identified 158 LPV candidates among 4651 stars in CCD4. Using the TRGB method, we derived a distance modulus of $24.35\pm0.08$ mag ($741\pm28$ kpc).

This revised distance places And\,XXII closer to M33, strengthening the case that it could indeed be a satellite of the Triangulum galaxy rather than Andromeda. However, given uncertainties in the three-dimensional position and potential line-of-sight projection effects, a definitive assignment to M33 cannot be confirmed.

\subsection{Integrated Characteristics of the Andromeda Satellite LPV Population}

The spatial distribution of LPVs within the M31 satellites provides important information on their membership and concentration. Appendices~\ref{fig:Location1}--\ref{fig:Location3} show the positions of LPVs in each dwarf galaxy. In some systems, such as And\,VI, the LPVs are clearly concentrated within the galaxy boundaries, while in others, like And\,XII, the clustering is less obvious. 

To quantify membership, the surface density of LPVs inside each galaxy’s nominal boundary was compared to the surrounding control fields. This analysis indicates that, although some LPVs may belong to the background halo of M31 or M33, a substantial fraction are consistent with genuine membership in their respective dwarfs. LPVs located outside the nominal galaxy boundaries are unlikely to be foreground contamination. To further assess this, Gaia DR3 data were examined for the Andromeda field, where epoch photometry is available for individual sources. None of the candidate LPVs are listed in Gaia DR3, likely due to their faint magnitudes. 

The LPV stars identified in this study provide an integrated perspective on late stellar evolution across diverse galactic environments. Approximately 2800 LPV candidates were detected among roughly $1.2\times10^{5}$ resolved stars in the combined photometric catalog. The composite luminosity distribution of these variables closely matches that observed in well-studied, intermediate-age systems such as NGC 147 and NGC 185, confirming that luminous asymptotic giant branch (AGB) stars are reliable tracers of evolved stellar populations throughout the M31 satellite system \citep{Lorenz11, Hamedani17, Mahani25}.

No strong trend is found between the LPV population and the three-dimensional distance from M31. However, satellites located within about 150 kpc of the host galaxy, such as And\,I to And\,VII, show slightly higher fractions of dusty or large-amplitude LPVs than more distant systems, including And\,XVIII to And\,XXII. This pattern suggests that weak environmental effects, possibly tidal interactions or extended periods of star formation, may enhance LPV production in the inner satellites \citep{Preston25}.

\begin{table*}
\caption{Master Catalog of the And\,II Dwarf Galaxy.}
\small % Adjust the font size to make the table smaller
\setlength{\tabcolsep}{3pt}
\label{table:And_II}
\centering
\begin{tabular}{llllllllllllll}
\hline\hline
 \noalign{\smallskip}

ID & RA & Dec & V  & V$_{err}$  & i  & i$_{err}$ &  A$_V$ & A$_i$ & I & J & K & L & N$_{observation}$ \\

 & $J2000$ & $J2000$ & (mag) & (mag) & (mag) & (mag) & (mag) & (mag) &  &  &  &  & \\
\hline
    
.......\\

372 & 19.0075 & 33.2343 & 22.643 & 0.131 & 21.308 & 0.066 & 0.490 & 0.348 & 1.639 & 0.675 & 0.797 & 0.674 & 10 \\        
378 & 19.1217 & 33.2343 & 20.591 & 0.031 & 19.529 & 0.023 & 0.103 & 0.134 & 6.763 & 3.442 & 0.870 & 3.754 & 10 \\      
379 & 19.0969 & 33.2344 & 22.094 & 0.061 & 20.457 & 0.033 & 0.137 & 0.133 & 2.153 & 1.268 & 0.893 & 1.419 & 10 \\  
380 & 19.0320 & 33.2344 & 21.021 & 0.069 & 19.084 & 0.025 & 0.158 & 0.120 & 6.950 & 3.565 & 0.890 & 3.978 & 10 \\              
381 & 19.0032 & 33.2344 & 20.505 & 0.122 & 19.553 & 0.044 & 0.353 & 0.259 & 5.312 & 2.219 & 0.873 & 2.428 & 10 \\
383 & 19.2112 & 33.2344 & 23.938 & 0.259 & 20.892 & 0.045 & 0.706 & 0.232 & 3.353 & 1.596 & 0.866 & 1.733 & 10 \\
384 & 19.0765 & 33.2345 & 21.364 & 0.086 & 19.930 & 0.032 & 0.222 & 0.172 & 7.474 & 3.009 & 0.819 & 3.088 & 10 \\ 
391 & 19.0673 & 33.2346 & 21.980 & 0.148 & 21.068 & 0.057 & 0.411 & 0.327 & 2.449 & 0.938 & 0.882 & 1.038 & 10 \\ 
392 & 18.9028 & 33.2346 & 24.229 & 0.434 & 23.707 & 0.272 & 0.883 & 0.805 & 0.230 & -1.387 & 0.969 & -1.011 & 6 \\  
394 & 19.2510 & 33.2345 & 22.863 & 0.263 & 20.974 & 0.065 & 0.902 & 0.310 & 4.172 & 2.372 & 0.943 & 2.804 & 10 \\    
395 & 19.2459 & 33.2345 & 21.853 & 0.166 & 21.207 & 0.055 & 0.463 & 0.448 & 4.395 & 2.076 & 0.924 & 2.403 & 10 \\             
....\\\\
\hline
\end{tabular}

\end{table*}

\begin{table*}
\caption{Catalog of LPV candidates for the And\,II dwarf galaxy.}
\small % Adjust the font size to make the table smaller
\setlength{\tabcolsep}{3.3pt}
\label{table:And_II_LPV}
\centering
\begin{tabular}{lllll}
\hline\hline
\noalign{\smallskip}
ID & Date & Filter & Magnitude & Error \\ 
\hline
91 & 2457612.61 & i & 20.926 & 0.046 \\           
91 & 2457613.67 & V & 21.761 & 0.036 \\          
91 & 2457681.59 & i & 21.157 & 0.073 \\           
91 & 2457683.54 & V & 22.036 & 0.062 \\           
91 & 2457783.48 & i & 21.060 & 0.093 \\           
91 & 2457967.63 & i & 20.947 & 0.041 \\            
91 & 2457997.68 & i & 20.872 & 0.035 \\           
91 & 2458033.01 & i & 21.124 & 0.058 \\          
91 & 2458034.66 & V & 22.228 & 0.103 \\          
181 & 2457428.36 & i & 21.050 & 0.054 \\           
181 & 2457612.62 & i & 20.834 & 0.038 \\           
181 & 2457613.67 & V & 23.019 & 0.090 \\           
181 & 2457681.59 & i & 21.110 & 0.059 \\        
181 & 2457683.54 & V & 23.420 & 0.189 \\          
181 & 2457783.48 & i & 21.103 & 0.074 \\            
181 & 2457967.63 & i & 20.862 & 0.027 \\           
181 & 2457997.68 & i & 20.856 & 0.035 \\           
181 & 2458033.01 & i & 21.087 & 0.052 \\           
181 & 2458034.66 & V & 22.982 & 0.178 \\            
....\\
\hline
\end{tabular}

\end{table*}

\section{Summary} \label{sec:Summary}

We conducted observations using the Wide Field Camera (WFC) on the Isaac Newton Telescope (INT) to monitor most of Andromeda's satellite dwarf galaxies. These observations primarily utilized the $i$-band filter, with supplementary $V$-band observations conducted over up to nine epochs. This study provides the initial findings for galaxies: And\,II, III, V, VI, X, XI, XII, XIII, XIV, XV, XVI, XVII, XVIII, XIX, XX, XXI, and XXII, illustrating our approach and the potential scientific insights this project can offer.

For each of these galaxies, we developed photometric catalogs focused mainly on the area covered by CCD4 of the WFC, which spans $11.26^{\prime\prime} \times 22.55^{\prime\prime}$ and, for most galaxies, encompasses approximately two half-light radii. For some dwarf galaxies, however, data from additional CCDs were also included to cover a larger area. These catalogs are comprehensive, providing both extensive photometric data and the identification of potential long-period variable stars, especially those exhibiting amplitude variations greater than 0.2 magnitudes. We derived the distance modulus for these galaxies by analyzing the TRGB stars in the photometric data. Additionally, we measured the half-light radii for these satellite galaxies (as detailed in Table~\ref{table:TRGB}). All photometric catalogs of LPV stars are made publicly available at the Centre de Donn\'ees astronomiques de Strasbourg (CDS). 
The format of these catalogs is illustrated in Tables~\ref{table:And_II} and \ref{table:And_II_LPV}. Table~\ref{table:And_II} provides an overview of the entire stellar population of And\,II, including columns for ID, coordinates, magnitude, and error in both the $i$ and $V$ bands. Meanwhile, Table~\ref{table:And_II_LPV} lists the LPV candidates in And\,II. The first column represents the ID, followed by the Date, which is recorded based on Julian dates, with the final two columns detailing the magnitude and its corresponding error.

In subsequent reports in this series, we will utilize these catalogs to explore the star-formation history and dust production of all identified LPV candidates across the monitored galaxies \citep{Mahani25, SEDust}. Furthermore, we investigate how color, and consequently temperature, changes during variability phases. By analyzing these changes and luminosity data, we aim to determine the variations in stellar radius and their correlation with mid-infrared excess.

\section*{acknowledgment}
We sincerely thank the referee for reading the manuscript and for the thoughtful comments and suggestions that have greatly helped improve the paper. Support for Hedieh Abdollahi was provided by the `SeismoLab' KKP-137523 \'Elvonal grant of the Hungarian Research, Development and Innovation Office (NKFIH).
.

\counterwithin{figure}{section}
\counterwithin{table}{section}
\appendix \section{Supplementary Material}

\begin{figure}[H]
    \centering
    \includegraphics[width=0.4\textwidth]{ 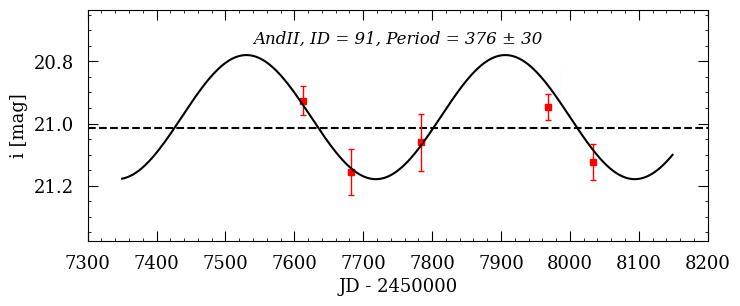}
    \includegraphics[width=0.4\textwidth]{ 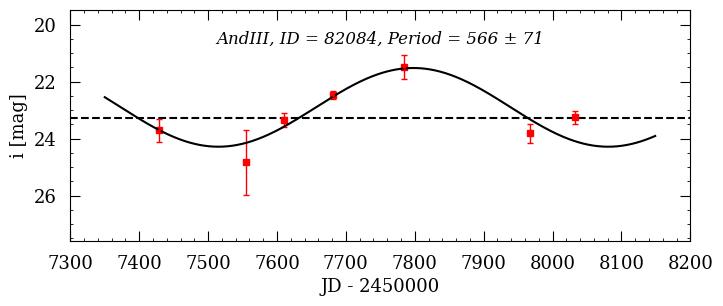}      \includegraphics[width=0.4\textwidth]{ 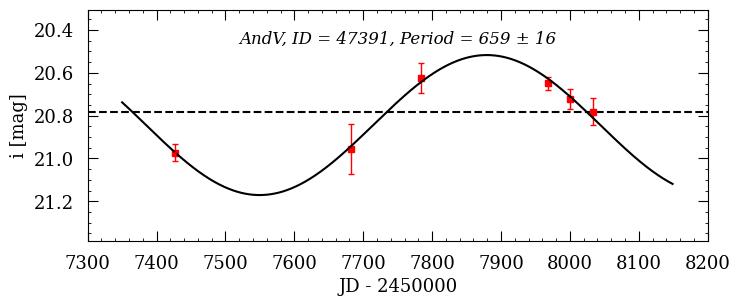}        \includegraphics[width=0.4\textwidth]{ 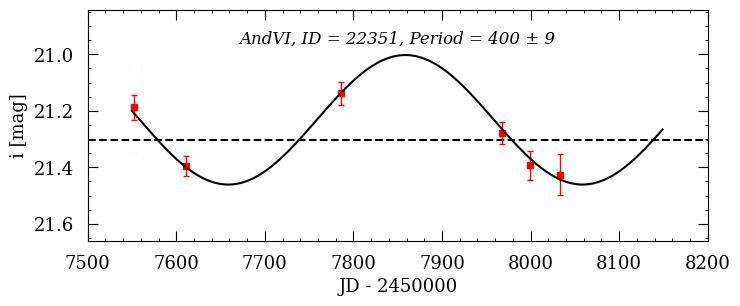}           \includegraphics[width=0.4\textwidth]{ 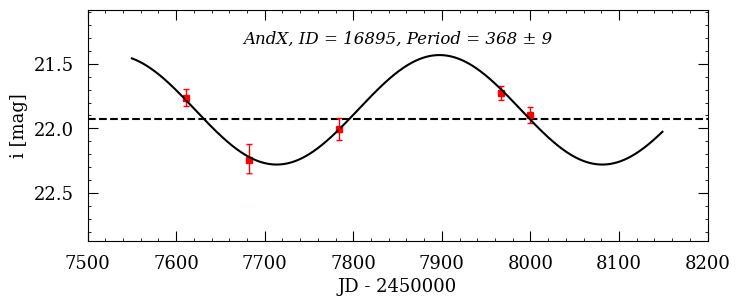}            \includegraphics[width=0.4\textwidth]{ 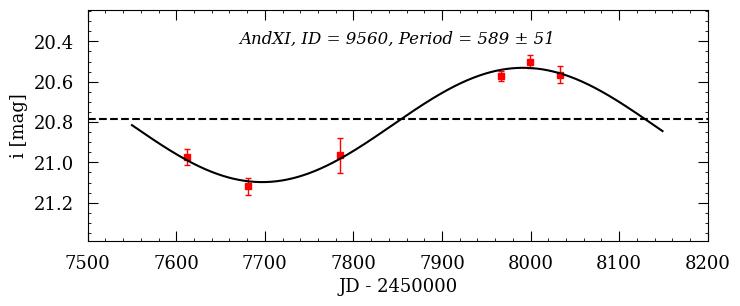}

    \includegraphics[width=0.4\textwidth]{ 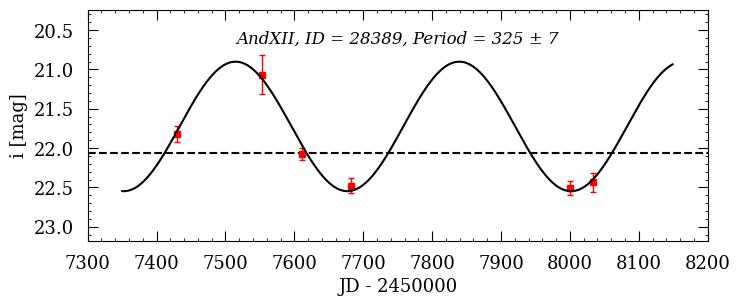}           \includegraphics[width=0.4\textwidth]{ 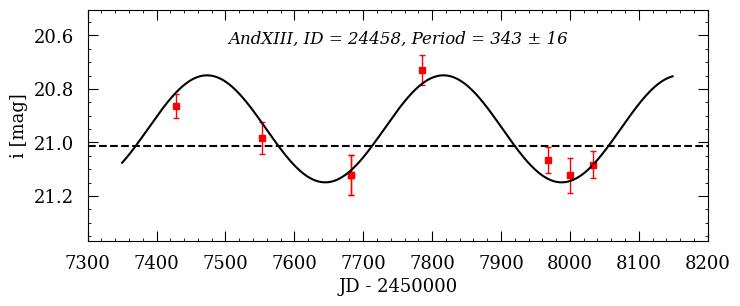}            \includegraphics[width=0.4\textwidth]{ 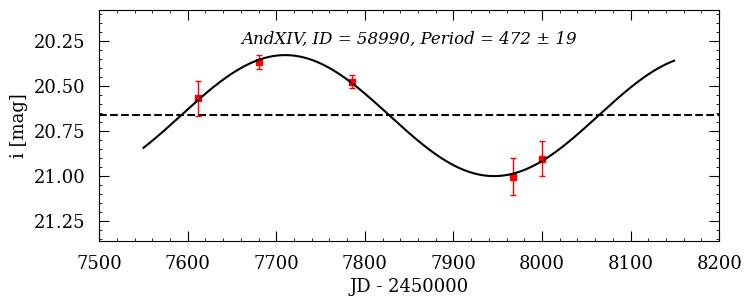}
     \includegraphics[width=0.4\textwidth]{ 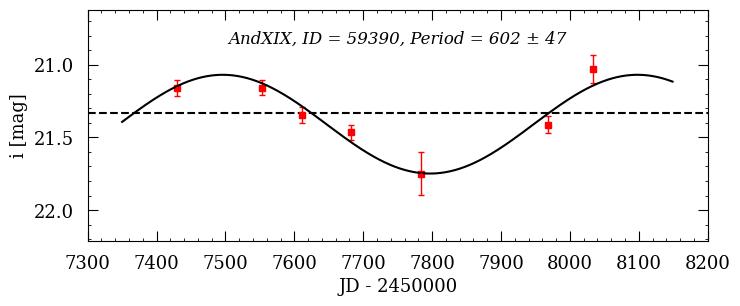}
      \includegraphics[width=0.4\textwidth]{ 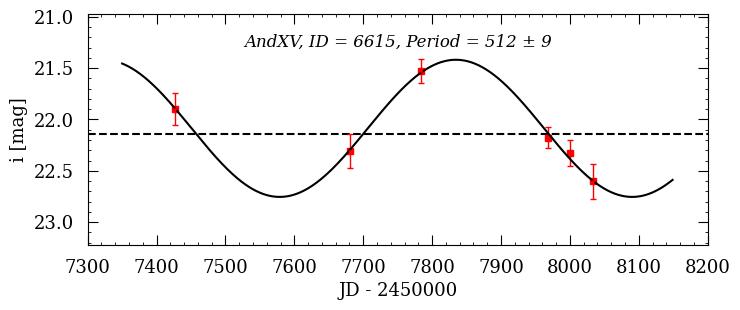}
       \includegraphics[width=0.4\textwidth]{ 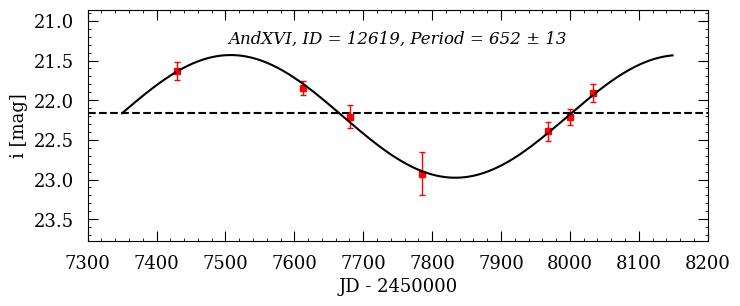}
            \caption{Light-curve example of long-period variable candidates. The black solid line represents the best fit for the observational data.}
    \label{fig:LC1}
\end{figure}

\begin{figure}[H]
    \centering
         \includegraphics[width=0.4\textwidth]{ 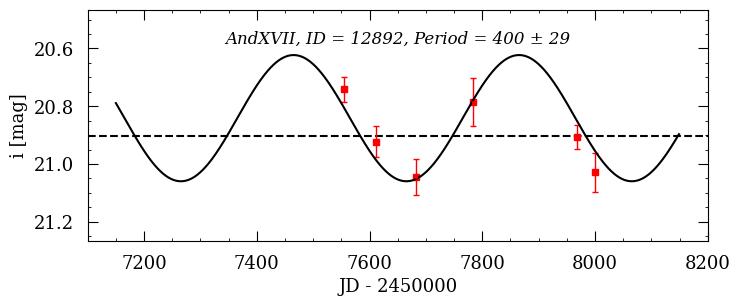}
         \includegraphics[width=0.4\textwidth]{ 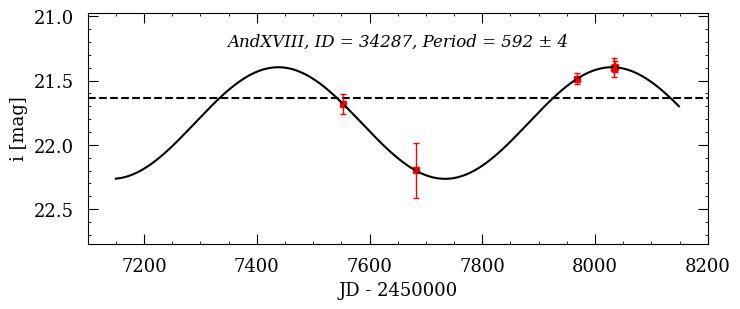}
         \includegraphics[width=0.4\textwidth]{ 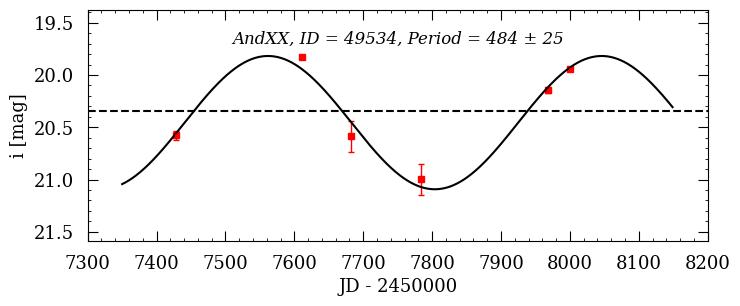}
         \includegraphics[width=0.4\textwidth]{ 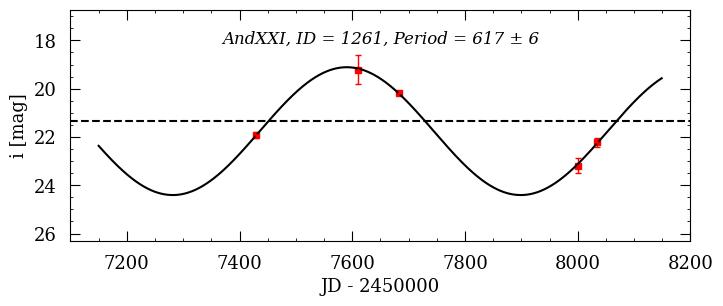}
         \includegraphics[width=0.4\textwidth]{ 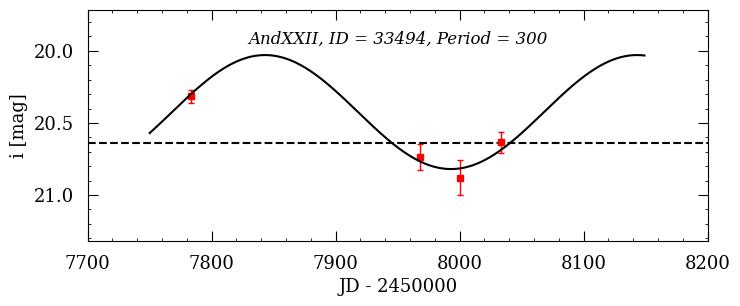}
   \caption{Light-curve example of long-period variable candidates. The black solid line represents the best fit for the observational data.}
    \label{fig:LC2}
\end{figure}

\begin{figure}[H]
    \centering
    \includegraphics[width=0.49\textwidth]{ 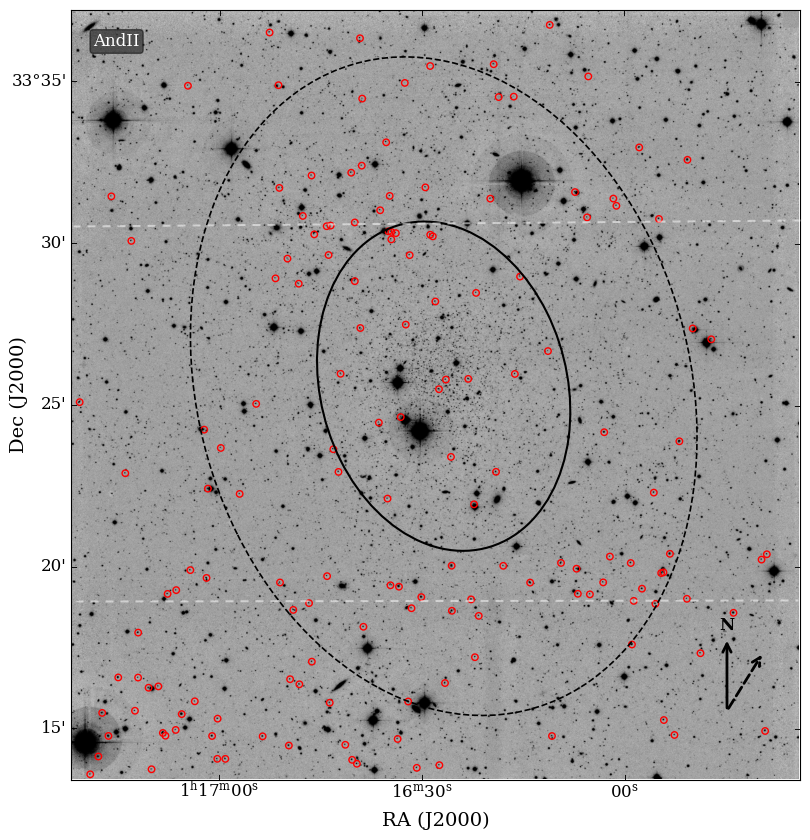}
    \includegraphics[width=0.49\textwidth]{ 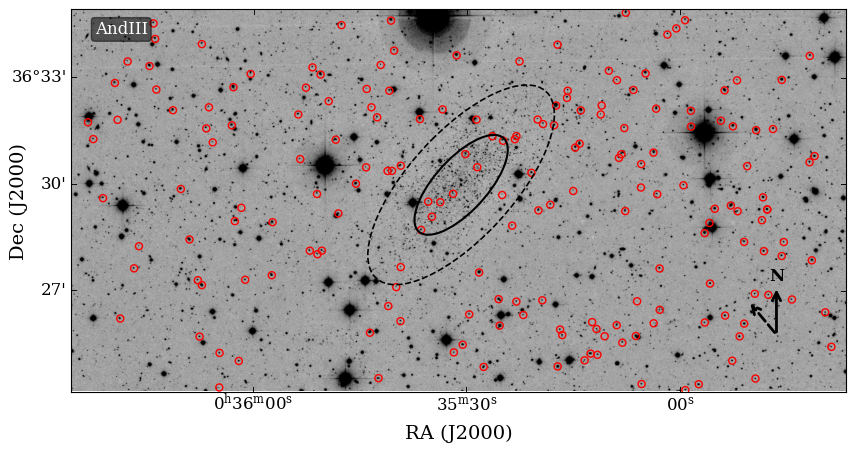}
    \includegraphics[width=0.49\textwidth]{ 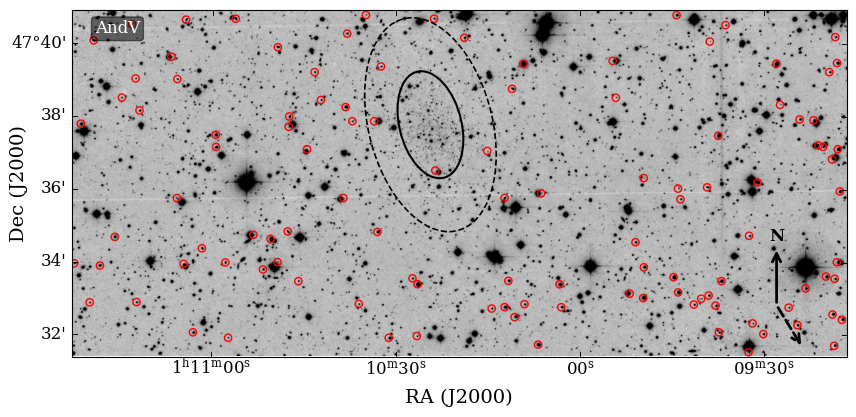}
    \includegraphics[width=0.49\textwidth]{ 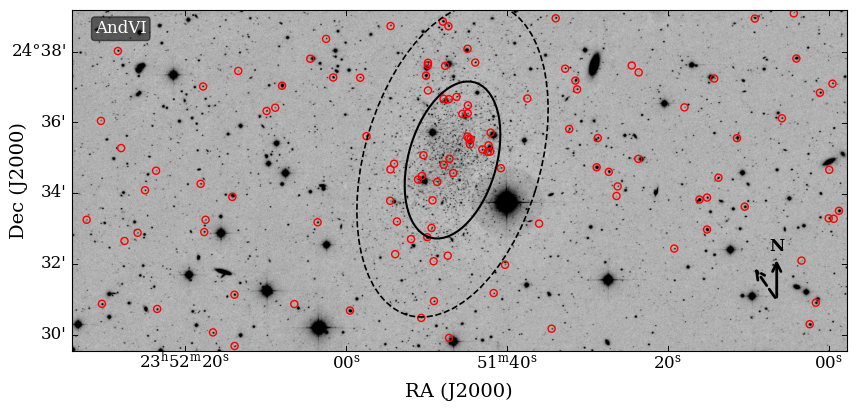}
    \includegraphics[width=0.49\textwidth]{ 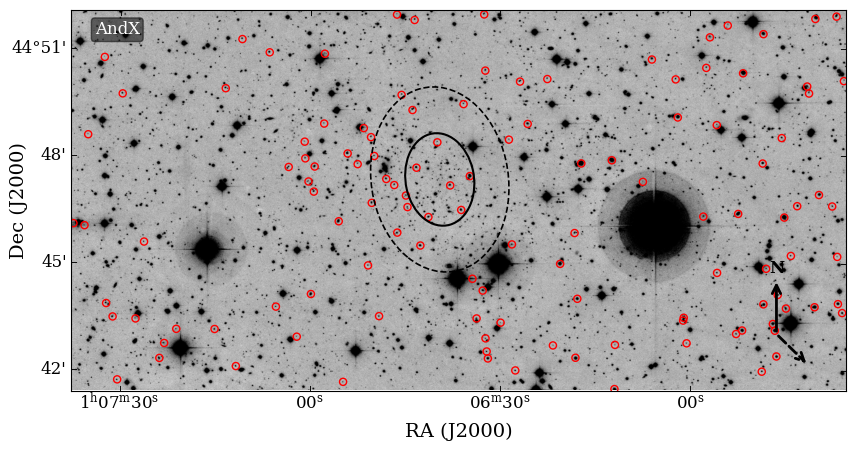}
    \includegraphics[width=0.49\textwidth]{ 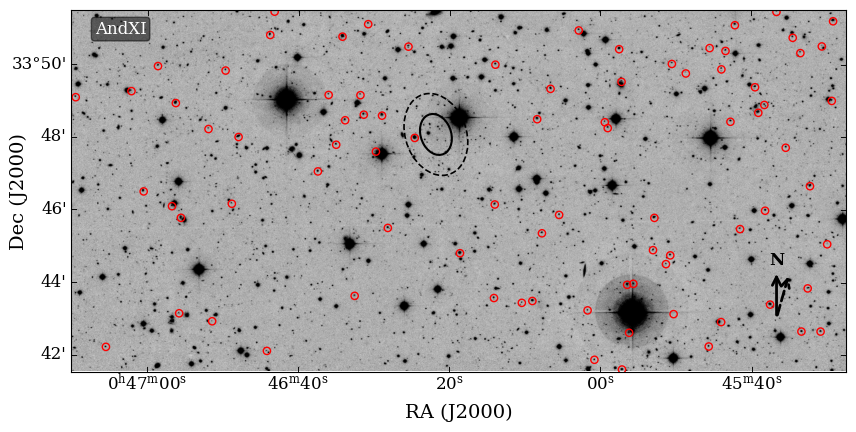}

   \caption{The distribution of LPV candidates in the studied areas is as follows: red circles denote LPVs, solid and dashed black circles represent the half-light radius and two half-light radii of a dwarf galaxy, and black arrows are directed toward the center of the Andromeda galaxy. Dwarf galaxies are And\,II, III, V, VI, X, and XI respectively.
}
    \label{fig:Location1}
\end{figure}

\begin{figure}[H]
    \centering
    \includegraphics[width=0.49\textwidth]{ 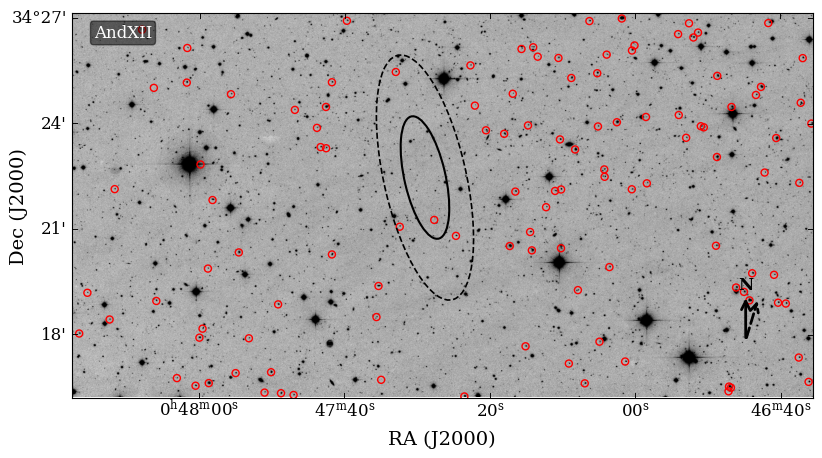}
    \includegraphics[width=0.49\textwidth]{ 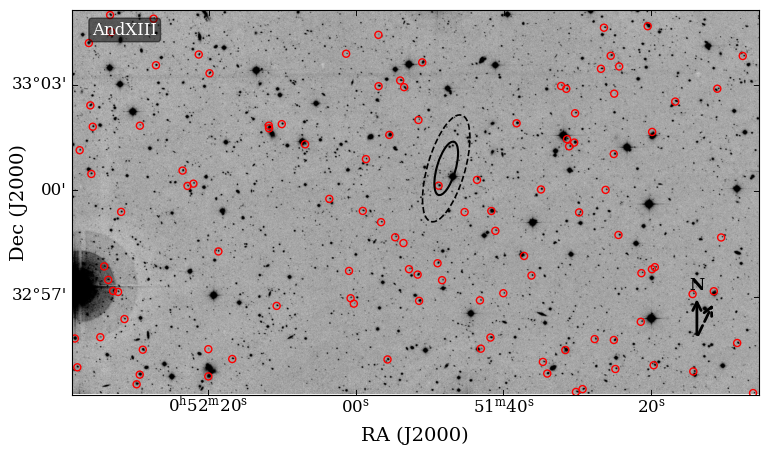}
    \includegraphics[width=0.49\textwidth]{ 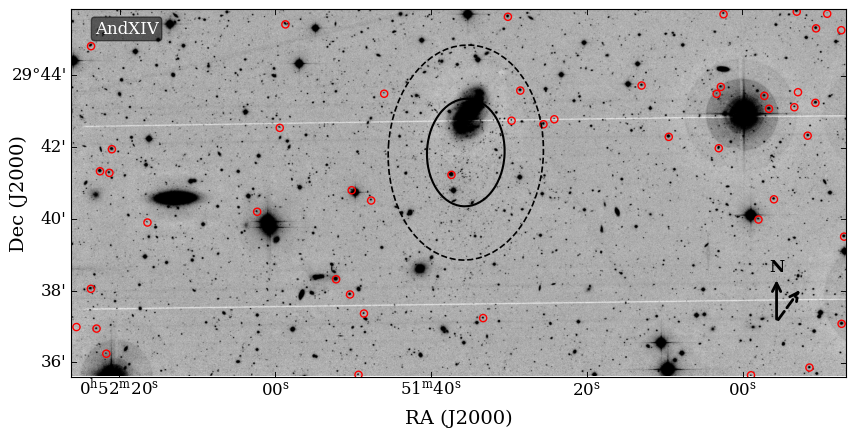}
    \includegraphics[width=0.49\textwidth]{ 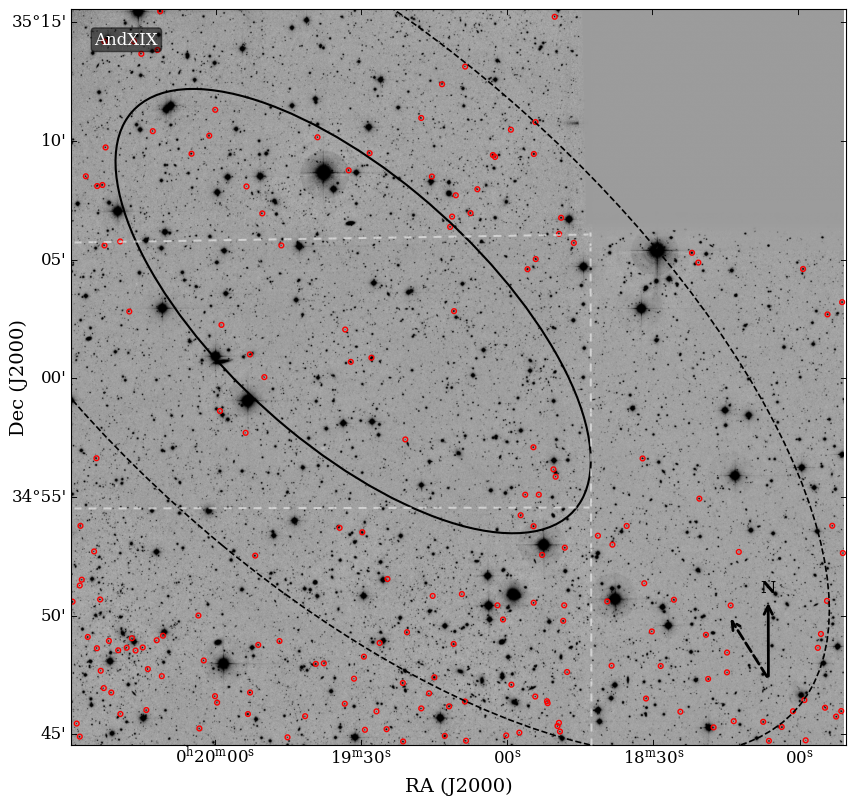}
    \includegraphics[width=0.49\textwidth]{ 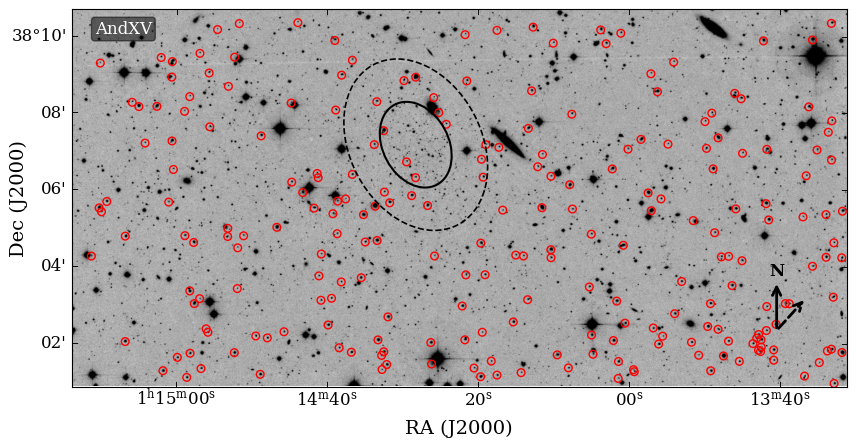} 
    \includegraphics[width=0.49\textwidth]{ 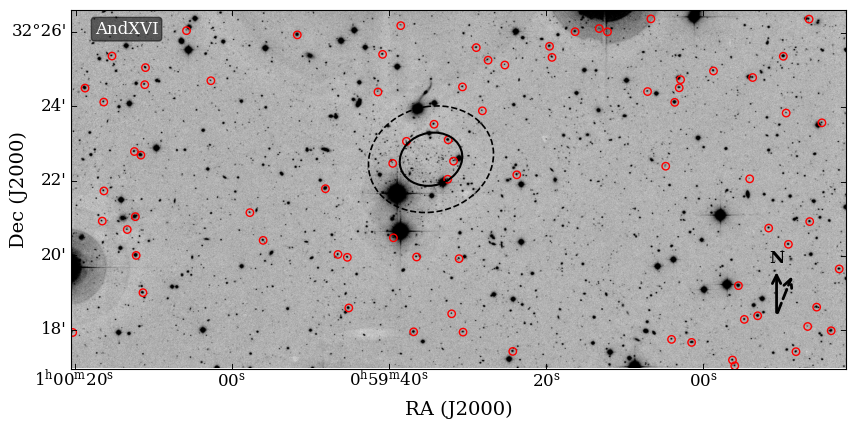}

   \caption{Dwarf galaxies are And\,XII, XIII, XIV, XIX, XV, and XVI respectively.
}
    \label{fig:Location2}
\end{figure}

\begin{figure}[H]
    \centering
  
    \includegraphics[width=0.49\textwidth]{ 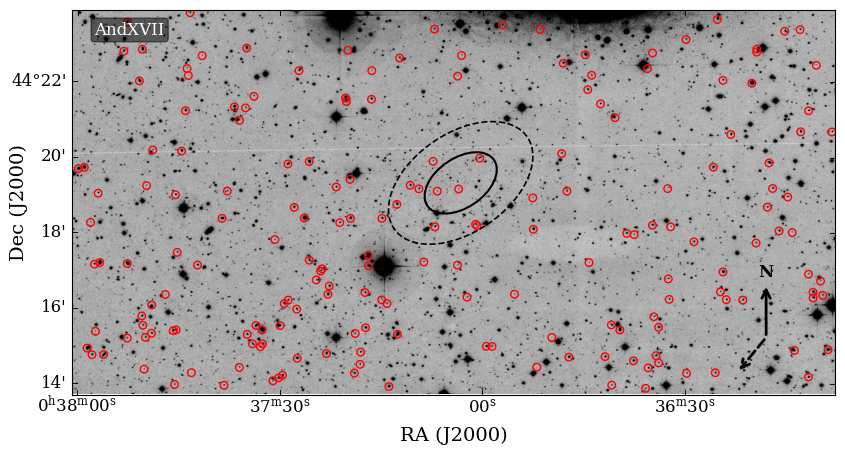}

    \includegraphics[width=0.49\textwidth]{ 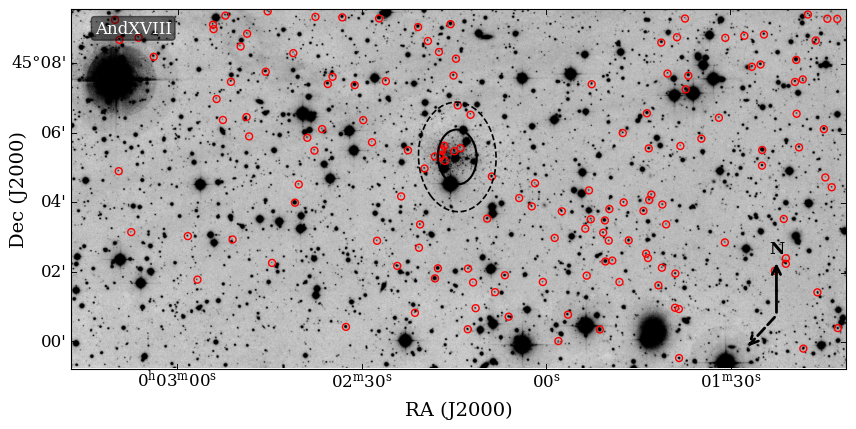}
    \includegraphics[width=0.49\textwidth]{ 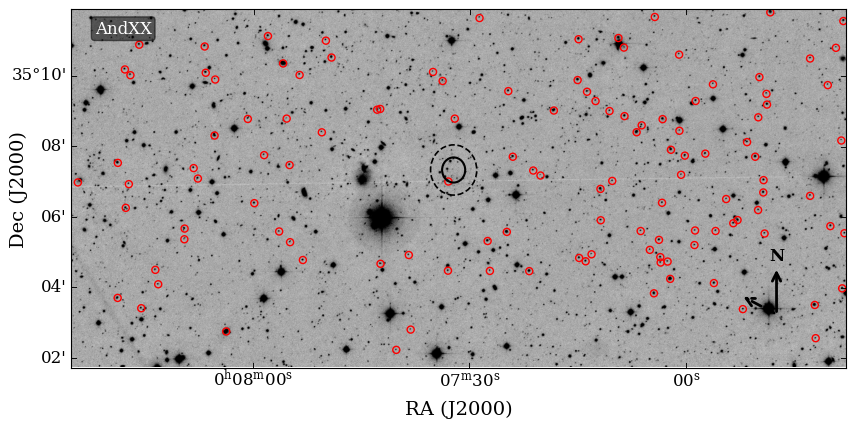}
    \includegraphics[width=0.49\textwidth]{ 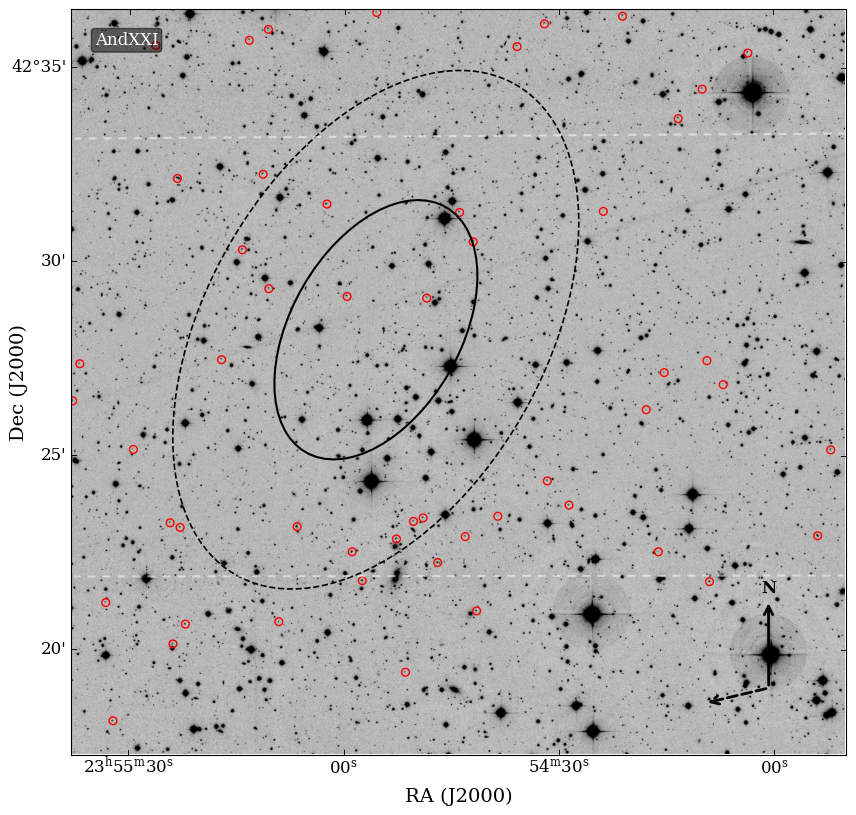}
    \includegraphics[width=0.49\textwidth]{ 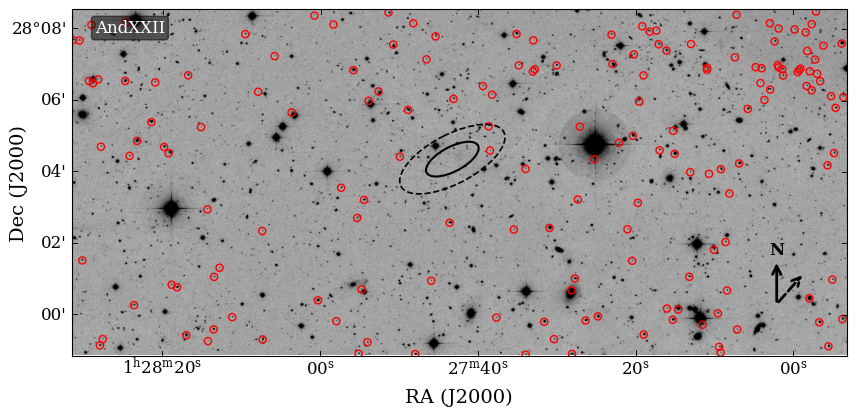}  
   \caption{Dwarf galaxies are And\,XVII, XVIII, XX, XXI, and XXII respectively.
}
    \label{fig:Location3}
\end{figure}

\begin{figure}[H]
    \centering
    \includegraphics[width=0.4\textwidth]{ 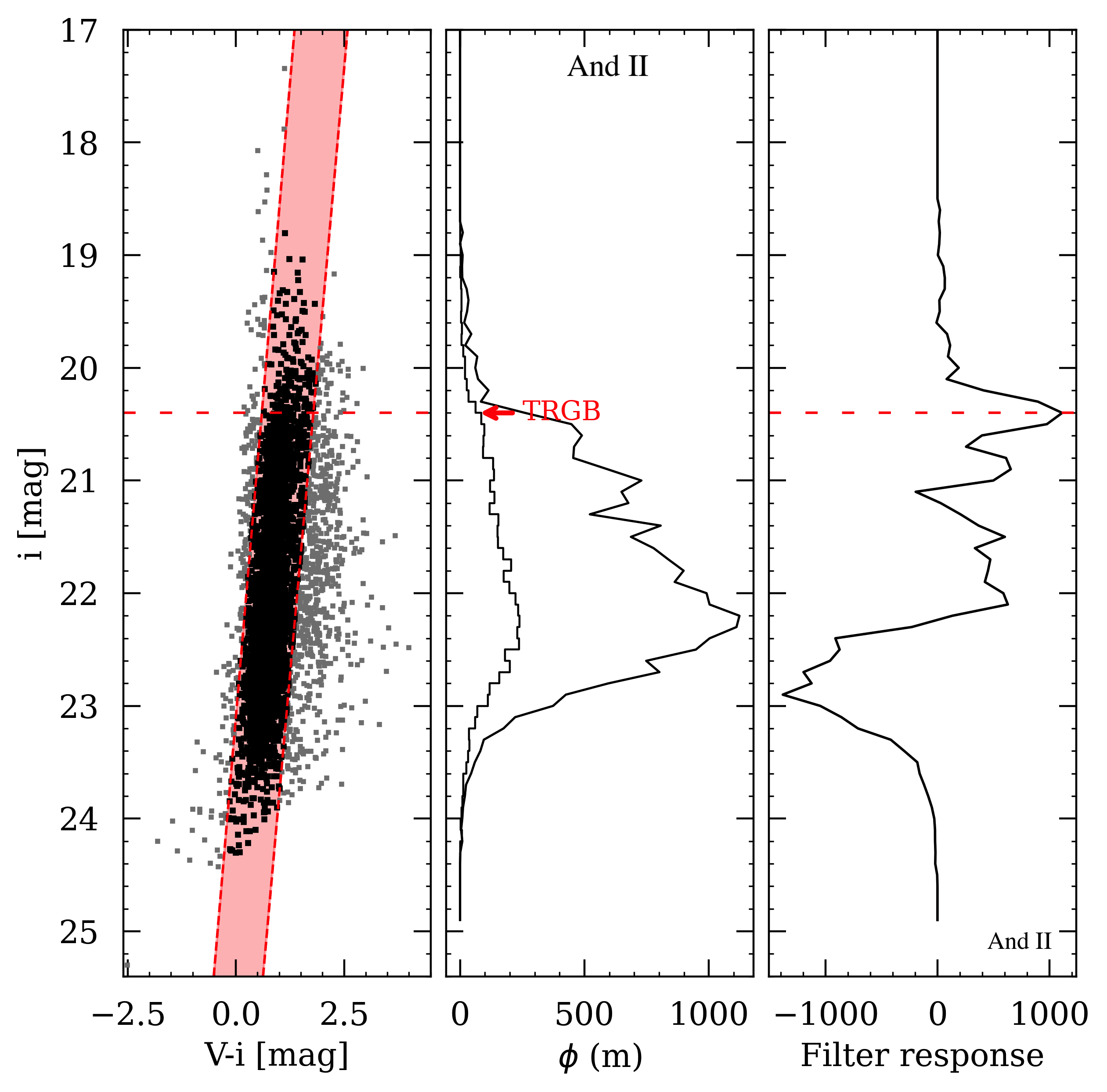}
    \includegraphics[width=0.45\textwidth]{ 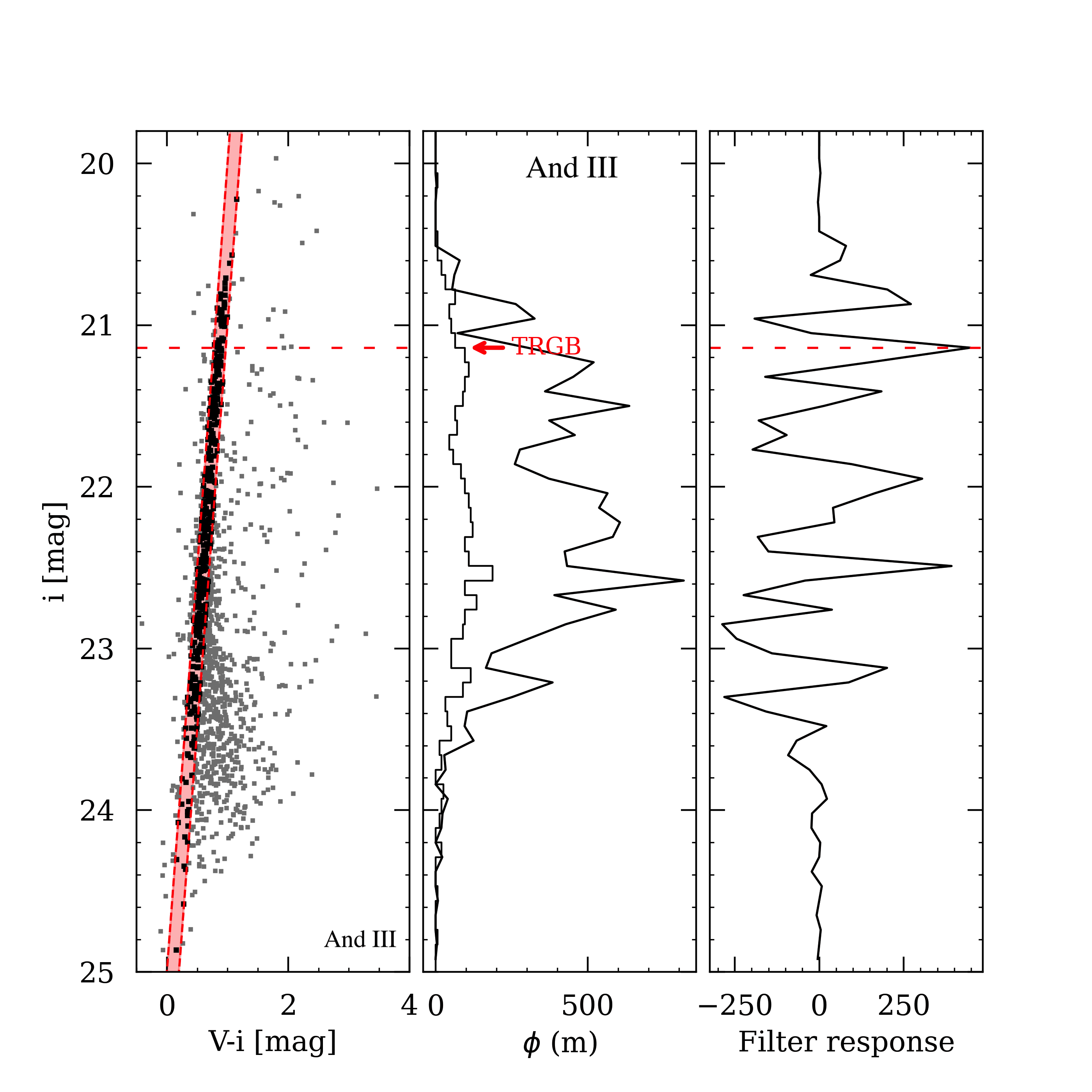}
    \includegraphics[width=0.4\textwidth]{ 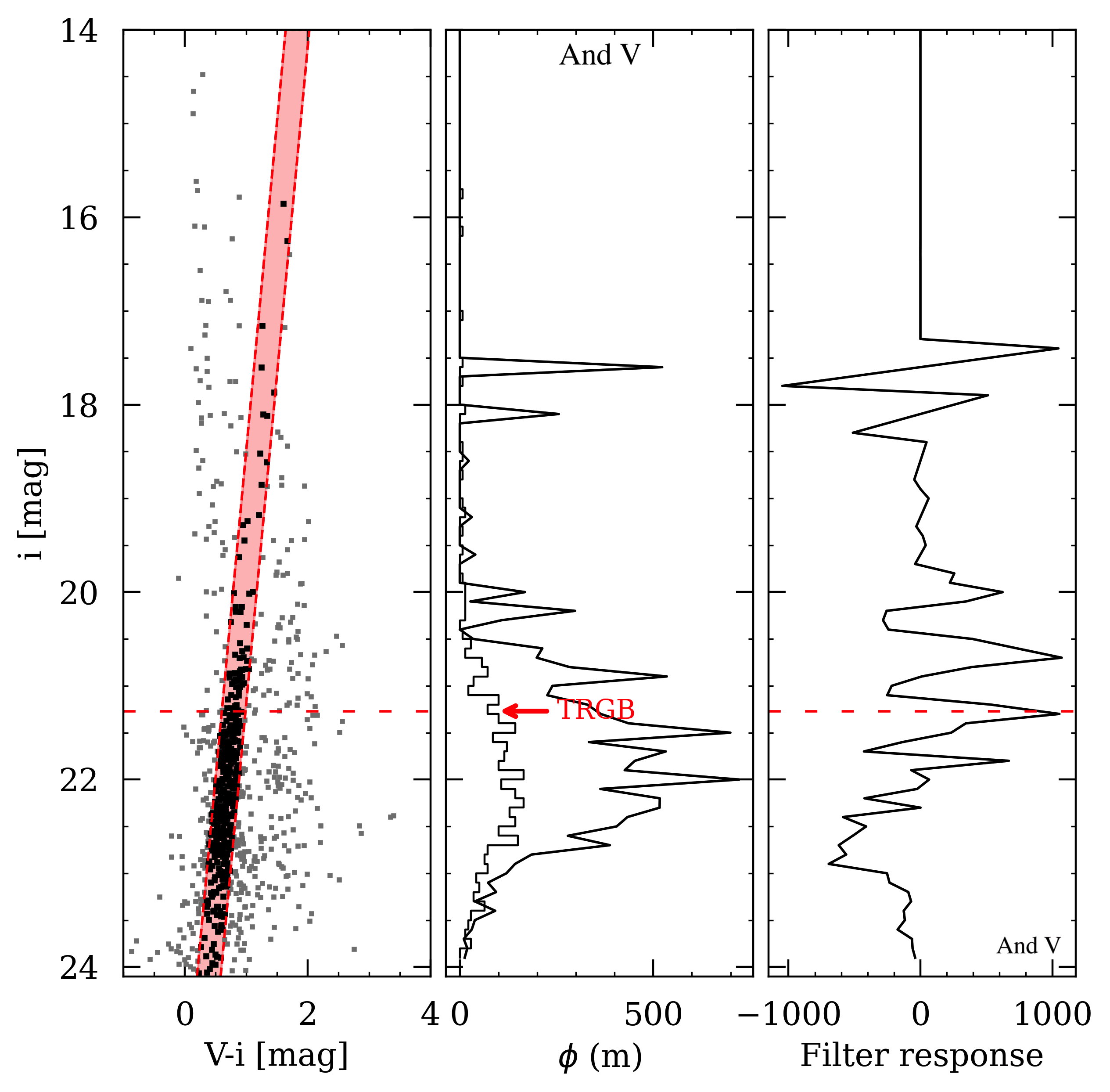}
    \includegraphics[width=0.4\textwidth]{ 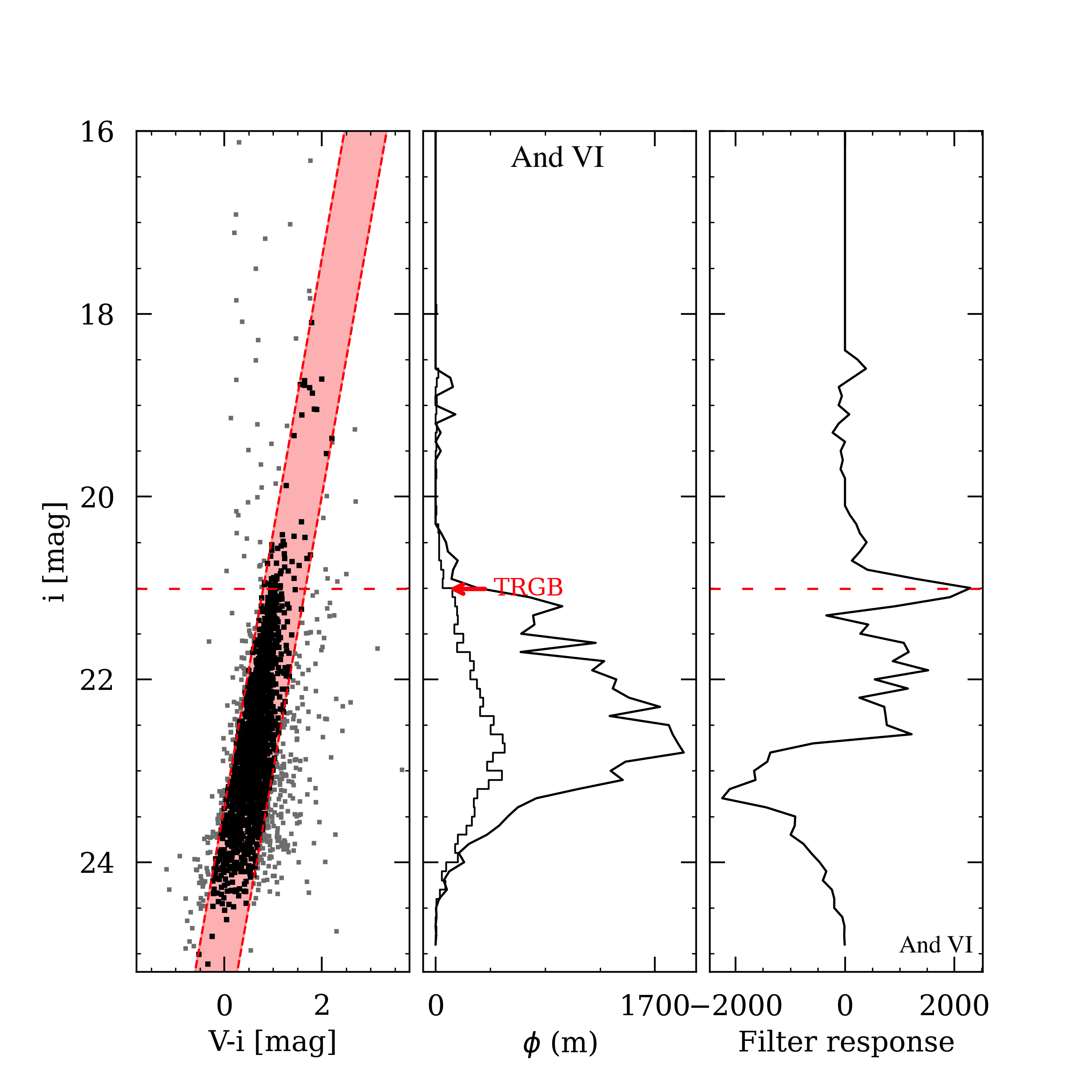}
    \includegraphics[width=0.4\textwidth]{ 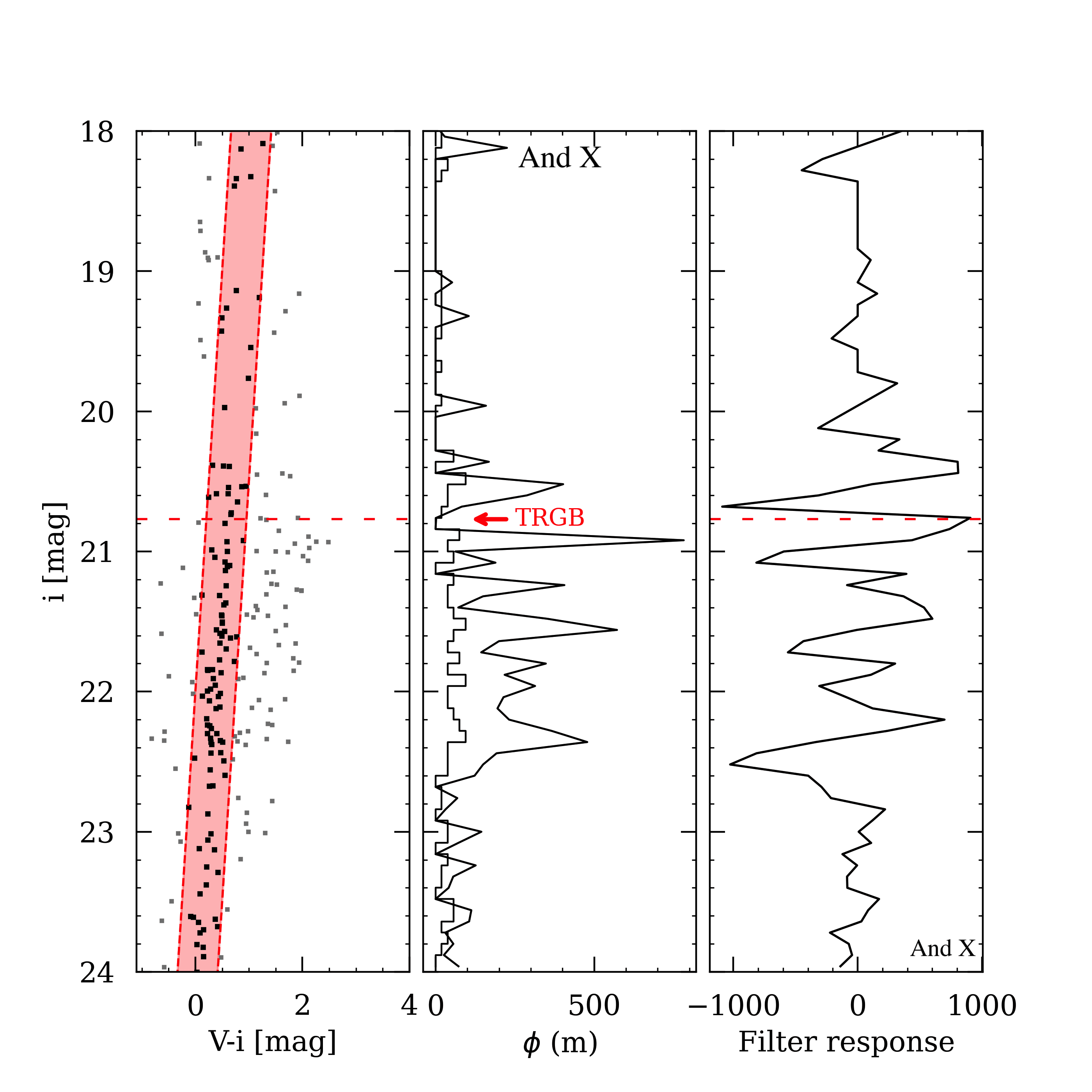}
    \includegraphics[width=0.4\textwidth]{ 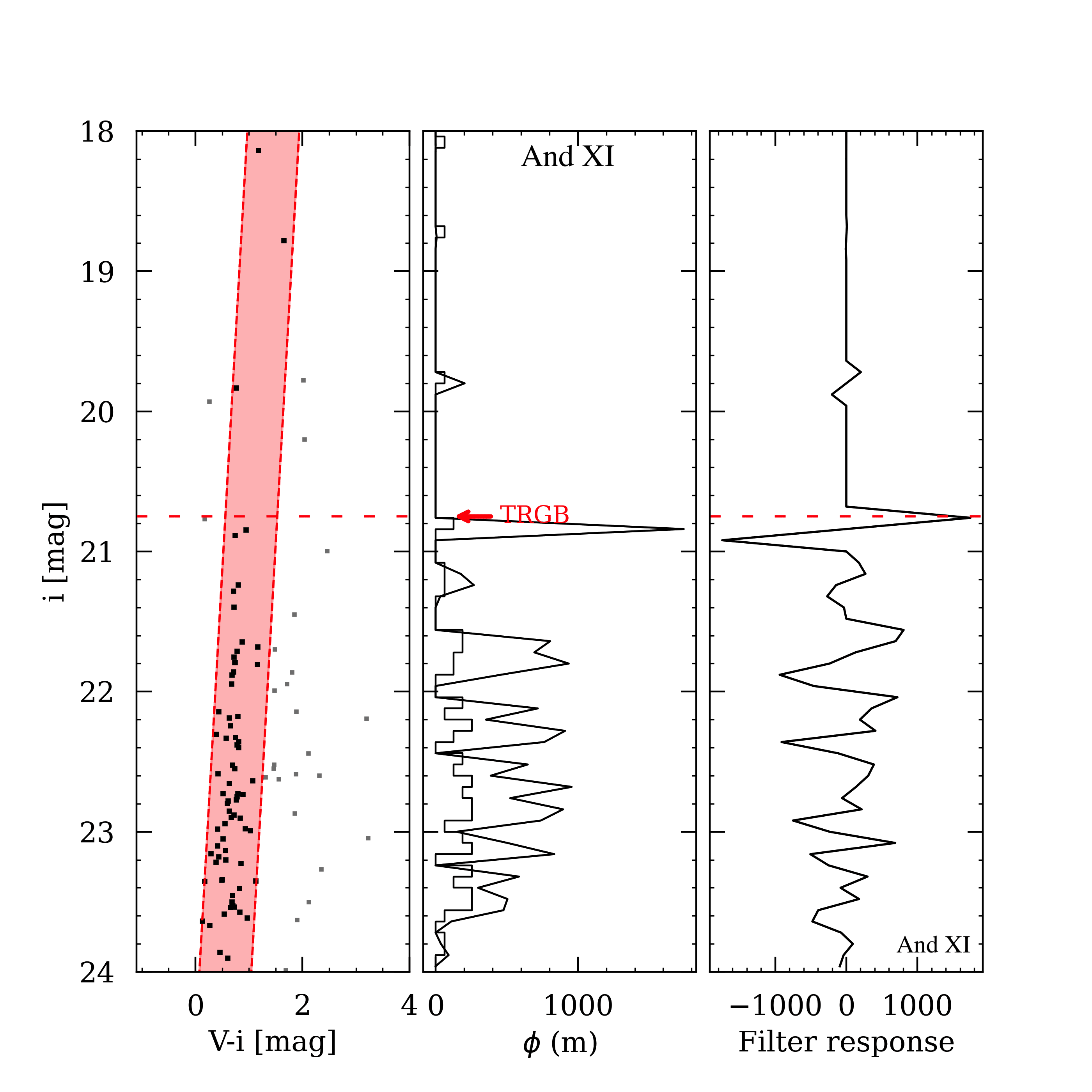}
   \caption{The left panel displays stellar sources within two half-light radii. The middle panel exhibits the histogram of the luminosity function, and the right panel demonstrates the Sobel filter response for the TRGB with edge detection. Plots are presented for And\,II, III, V, VI, X, and XI. The TRGBs are specifically denoted with red lines and arrows.}
    \label{fig:TRGB1}
\end{figure}

\begin{figure}[H]
    \centering
    \includegraphics[width=0.4\textwidth]{ 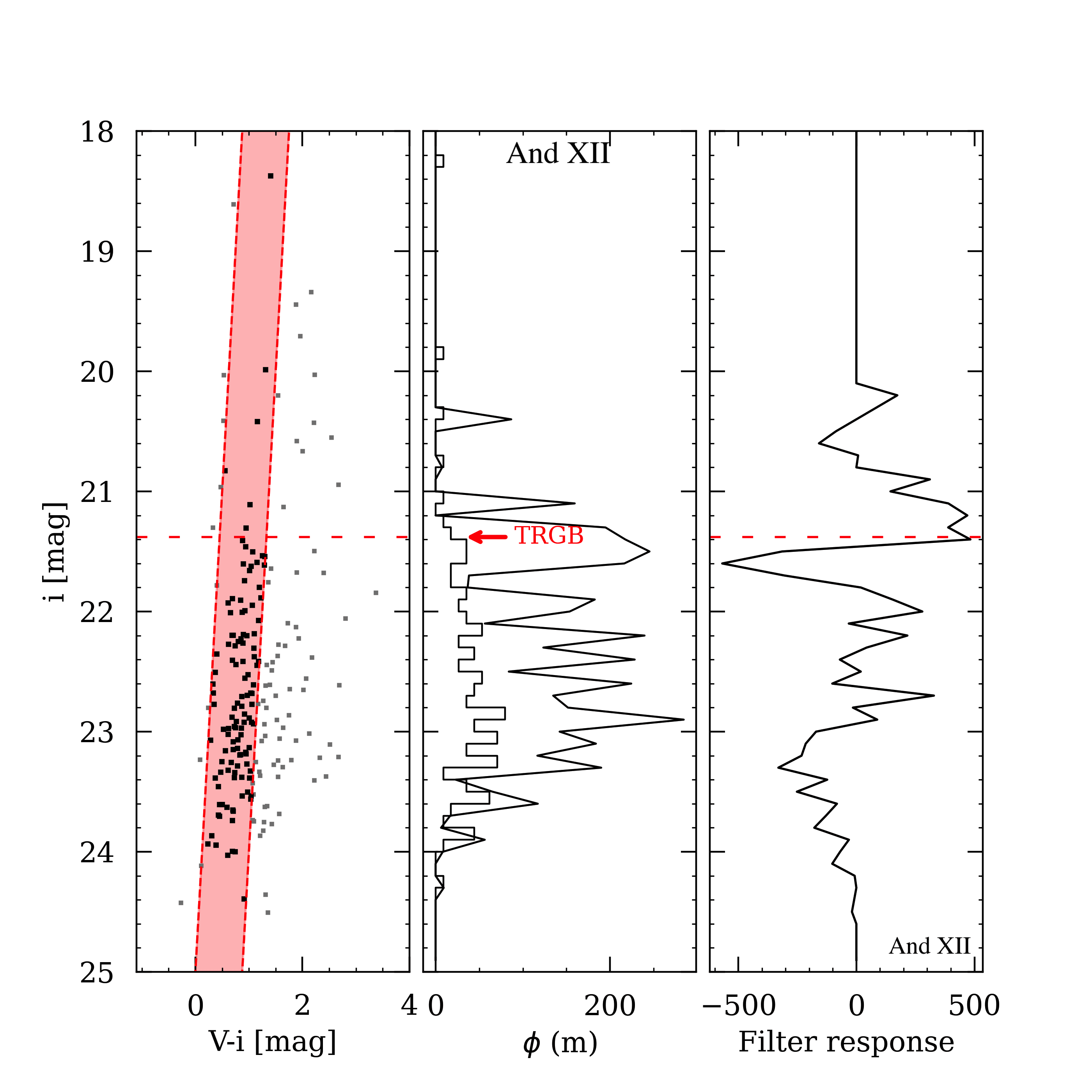}
    \includegraphics[width=0.4\textwidth]{ 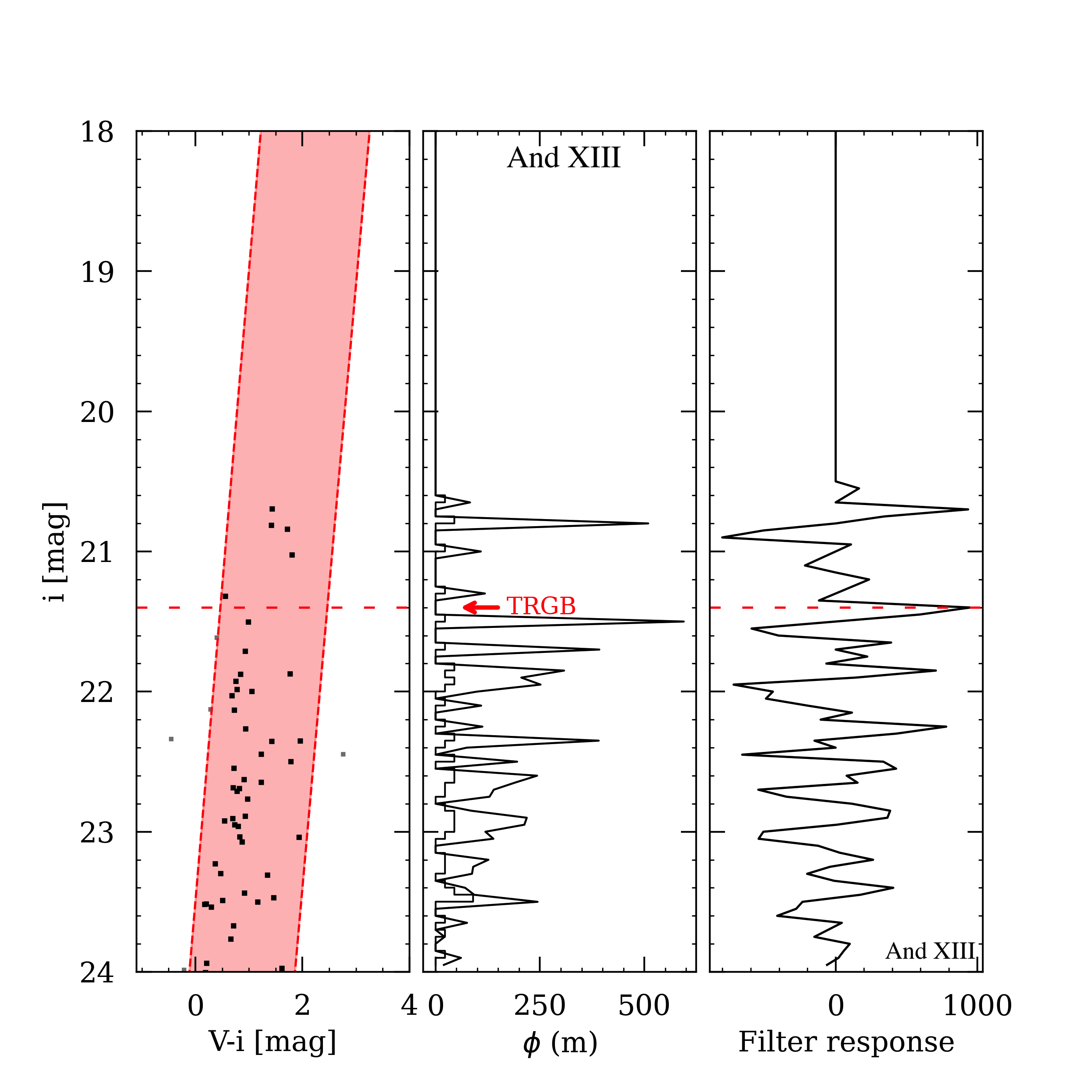}
    \includegraphics[width=0.4\textwidth]{ 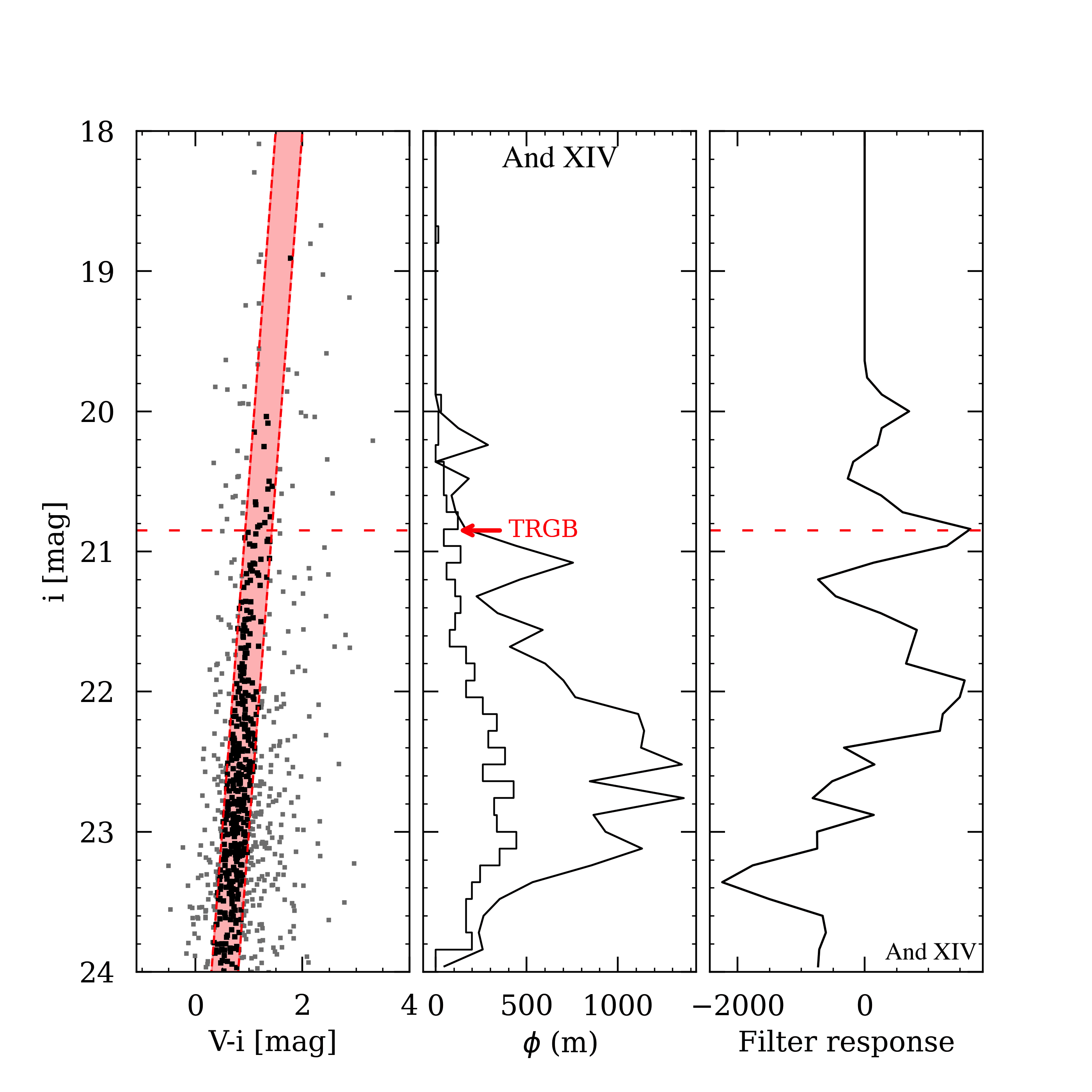}
    \includegraphics[width=0.4\textwidth]{ 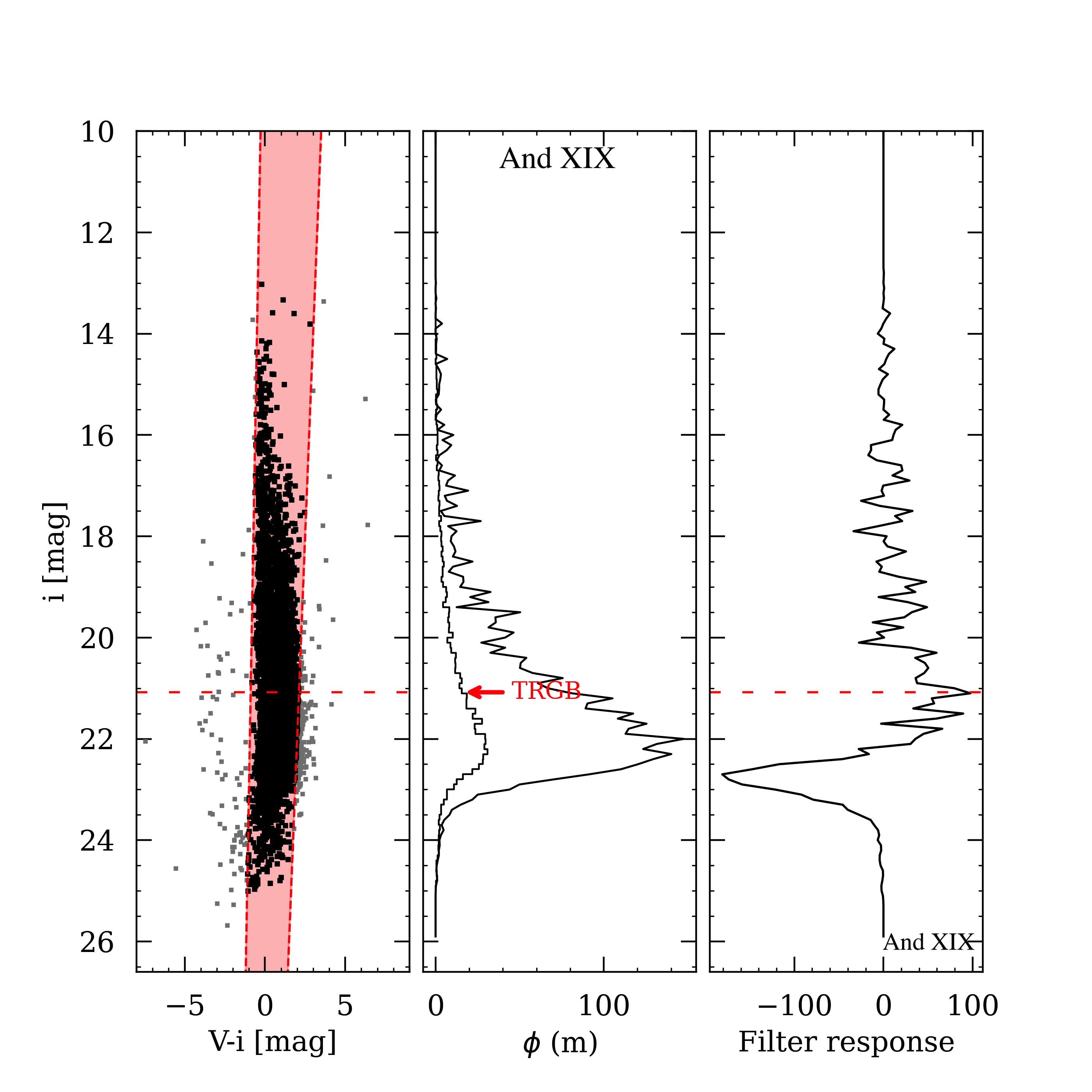}
    \includegraphics[width=0.4\textwidth]{ 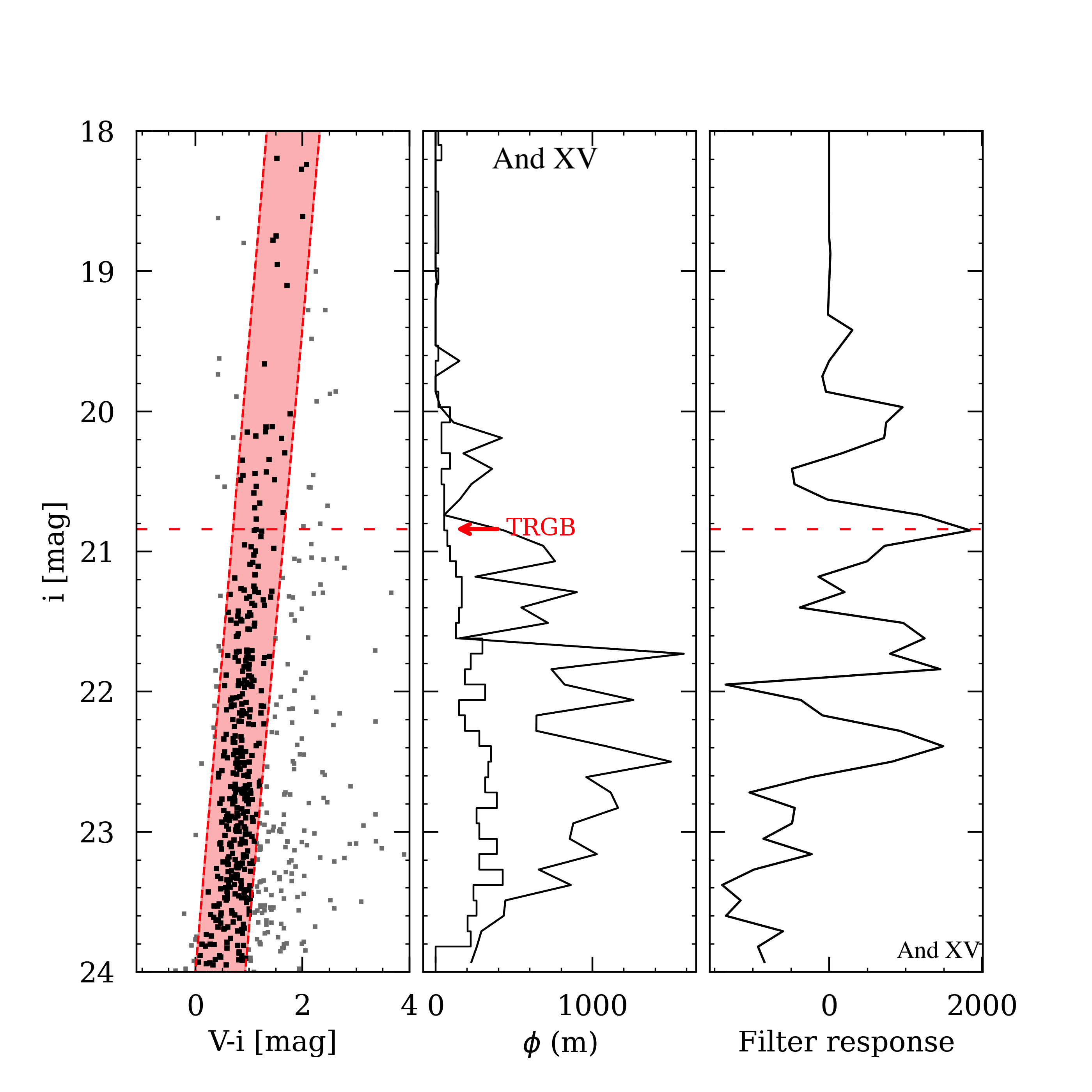}
    \includegraphics[width=0.4\textwidth]{ 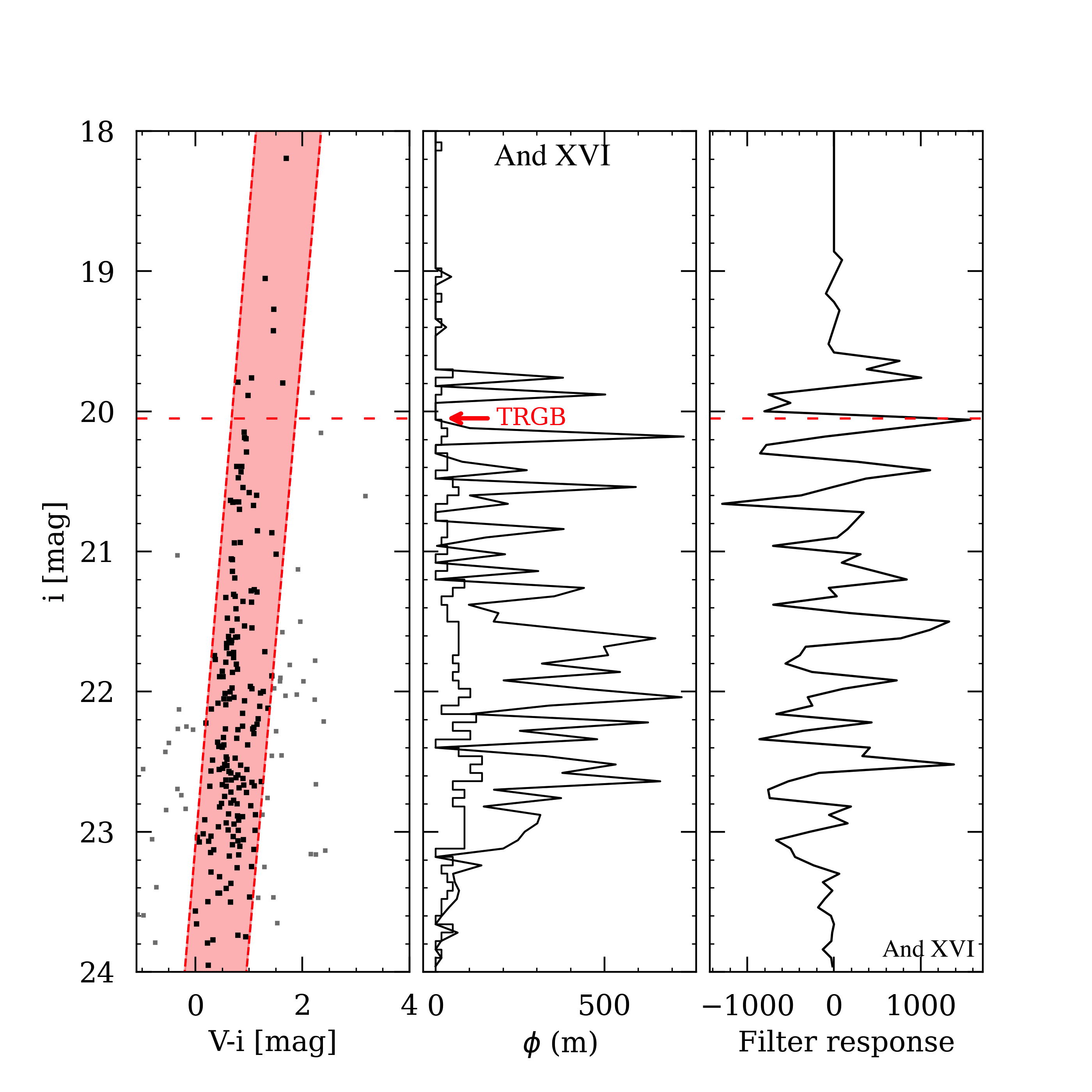}
   \caption{And\,XII, XIII, XIV, XIX, XV, and XVI, respectively.}
    \label{fig:TRGB2}
\end{figure}

\begin{figure}[H]
    \centering
    \includegraphics[width=0.4\textwidth]{ 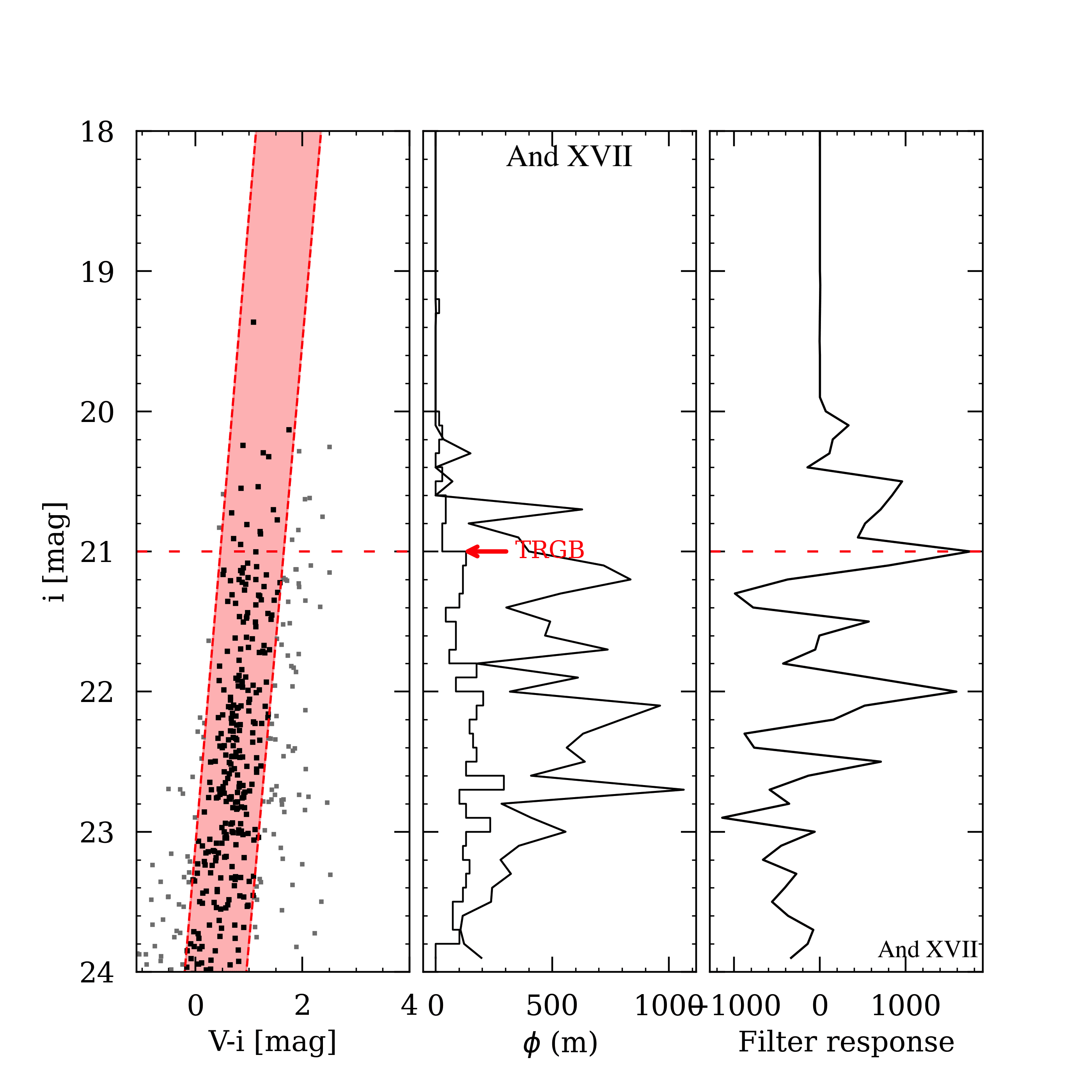}
    \includegraphics[width=0.4\textwidth]{ 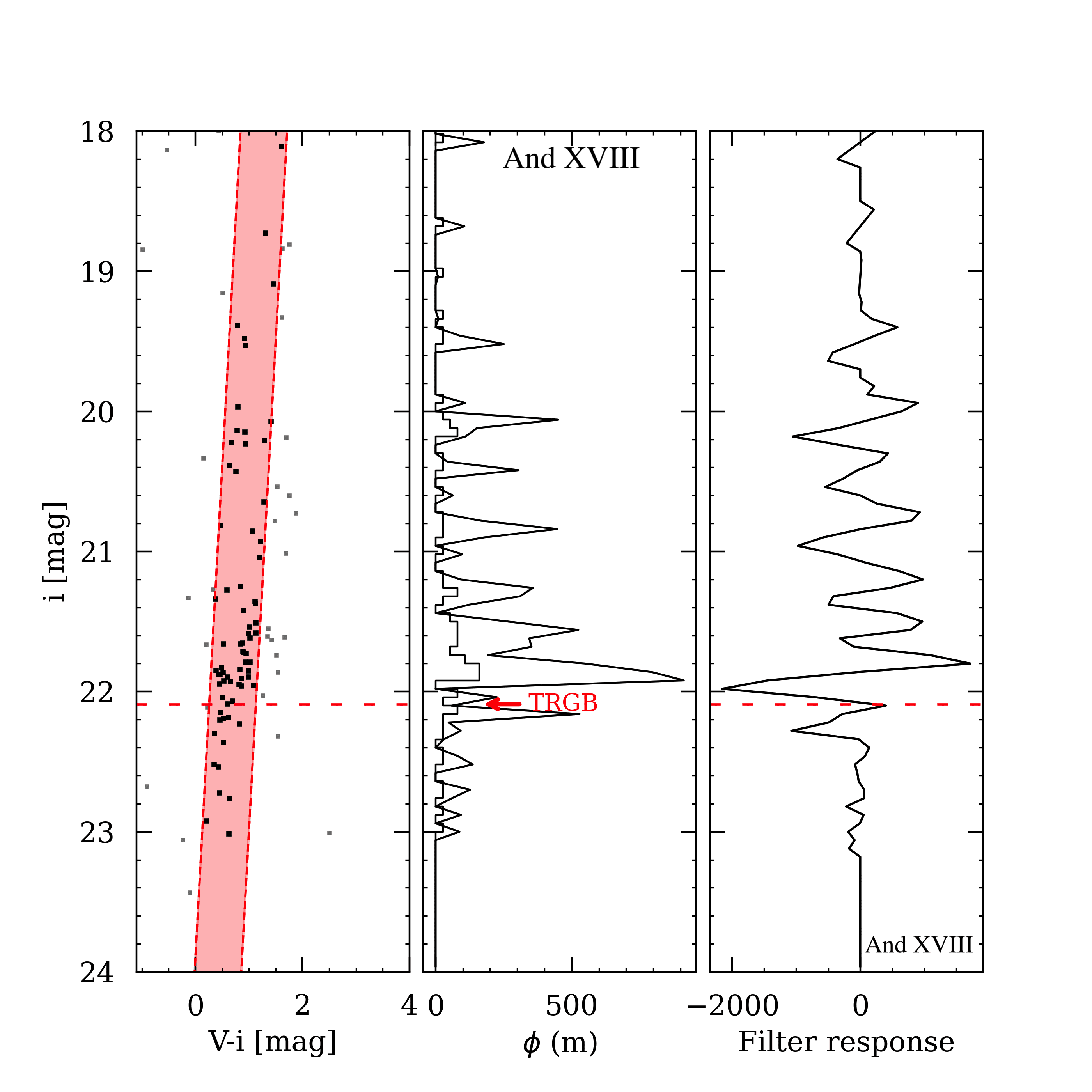}
    \includegraphics[width=0.4\textwidth]{ 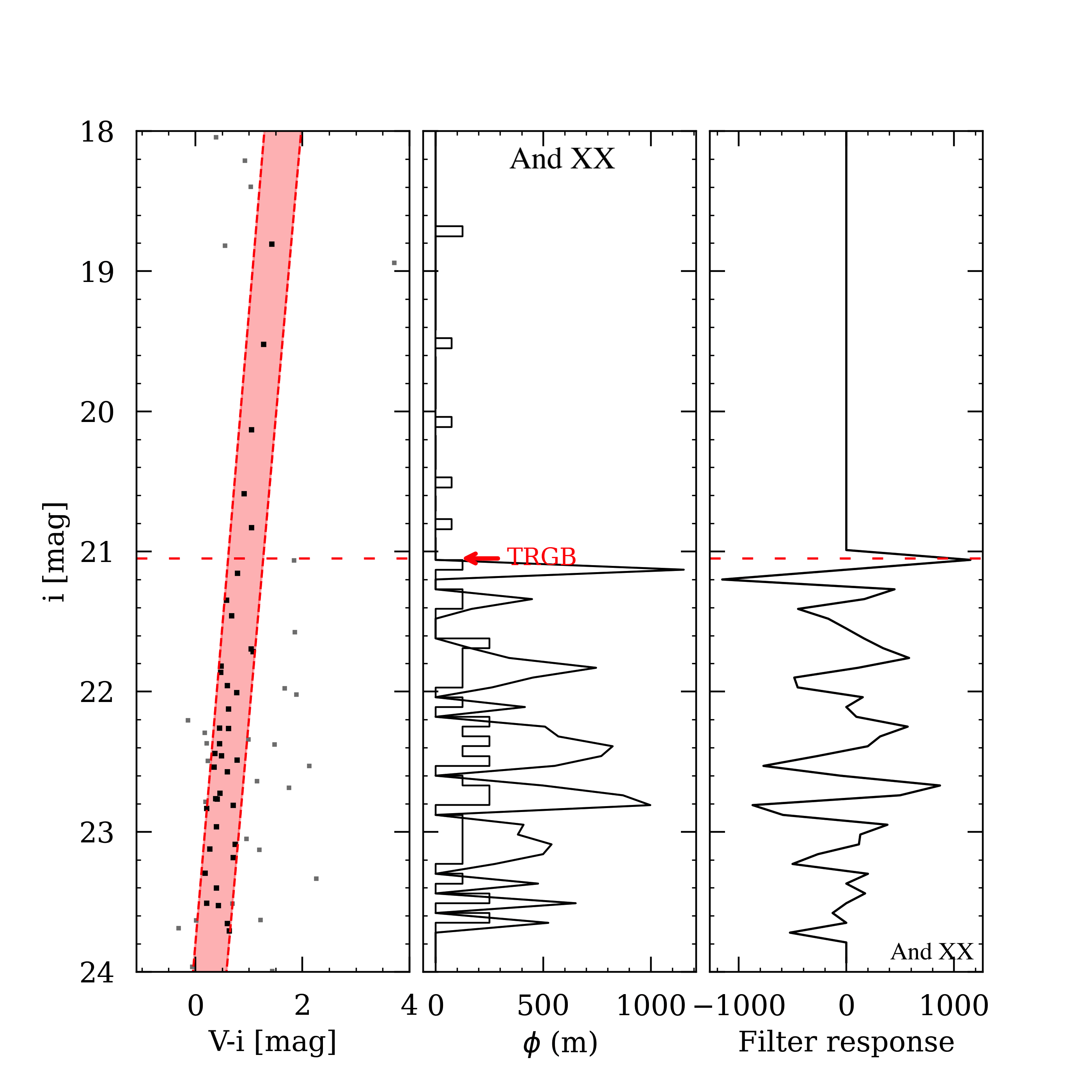}
    \includegraphics[width=0.4\textwidth]{ 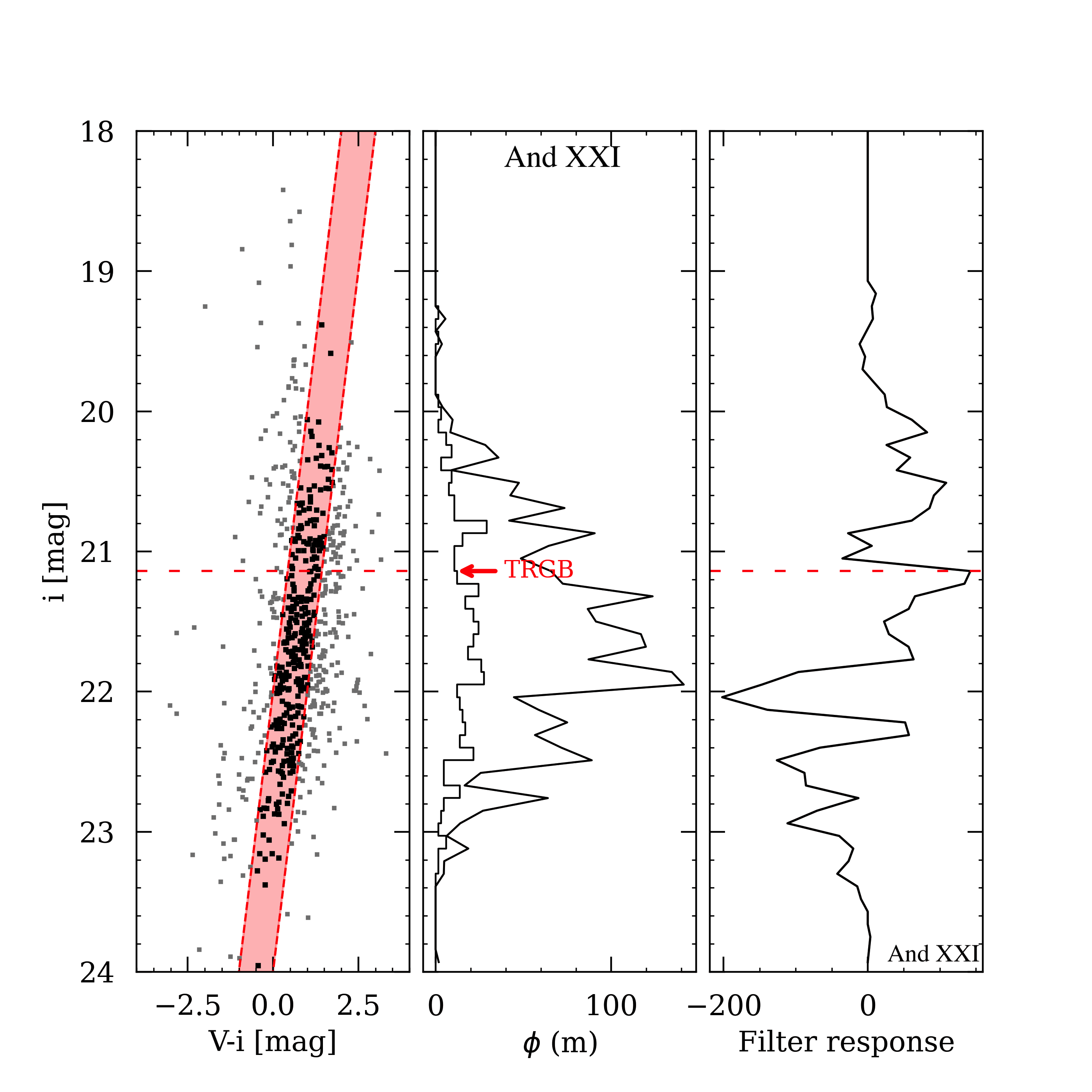}
    \includegraphics[width=0.4\textwidth]{ 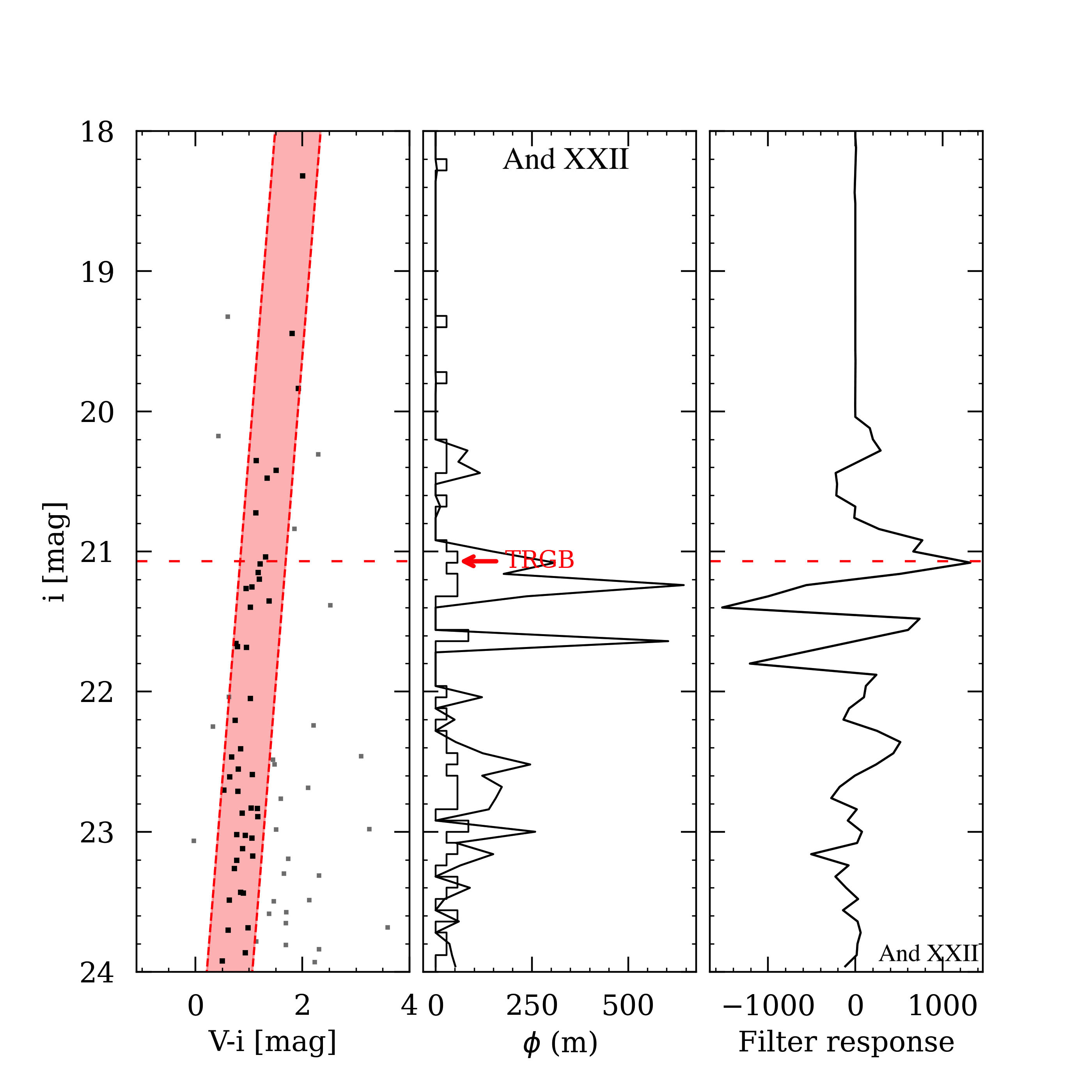}
   \caption{And\,XVII, XVIII, XX, XXI, and XXII, respectively.}
    \label{fig:TRGB3}
\end{figure}

\begin{figure}[H]
    \centering
  
    \includegraphics[width=0.4\textwidth]{ 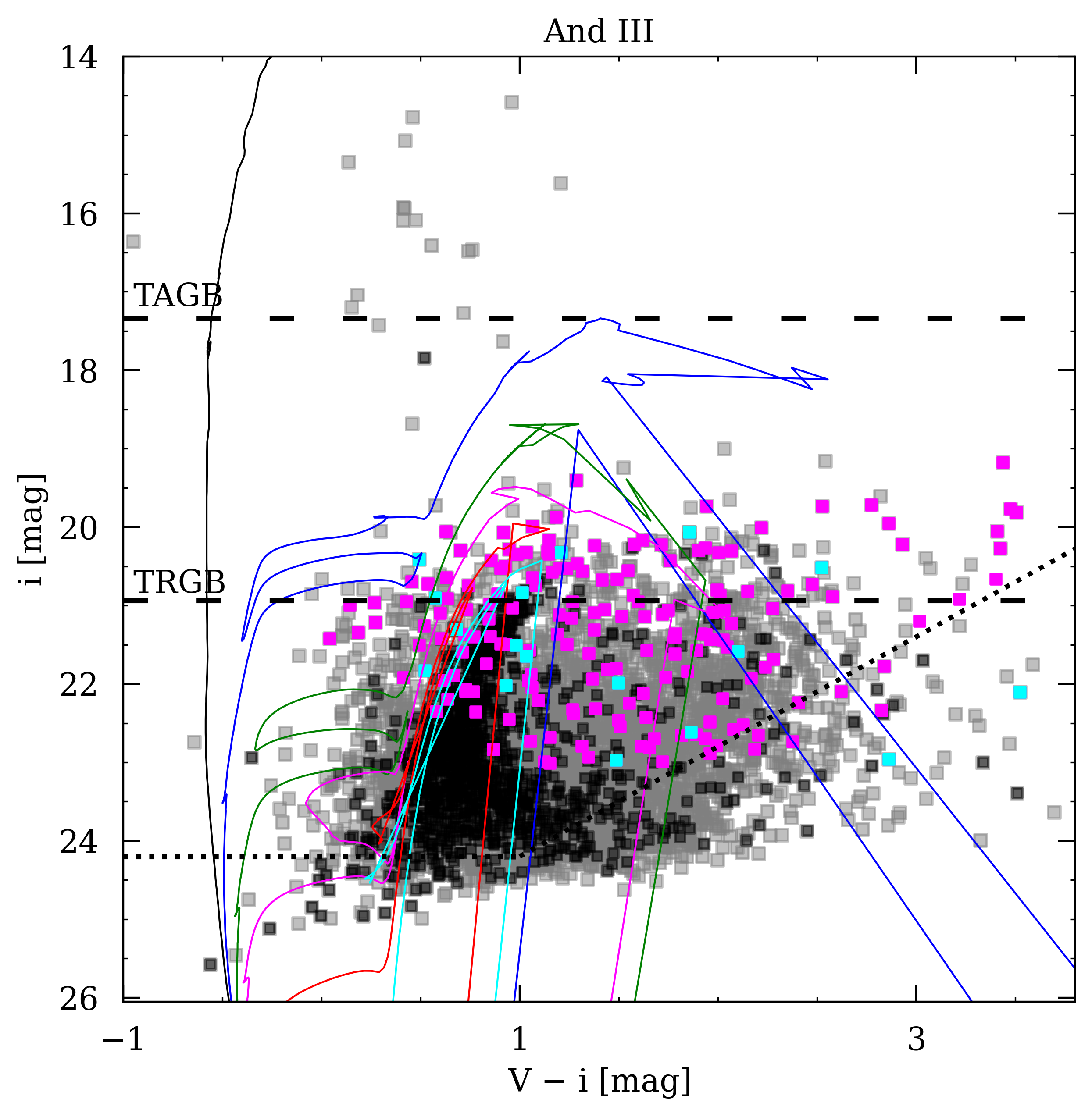}      \includegraphics[width=0.4\textwidth]{ 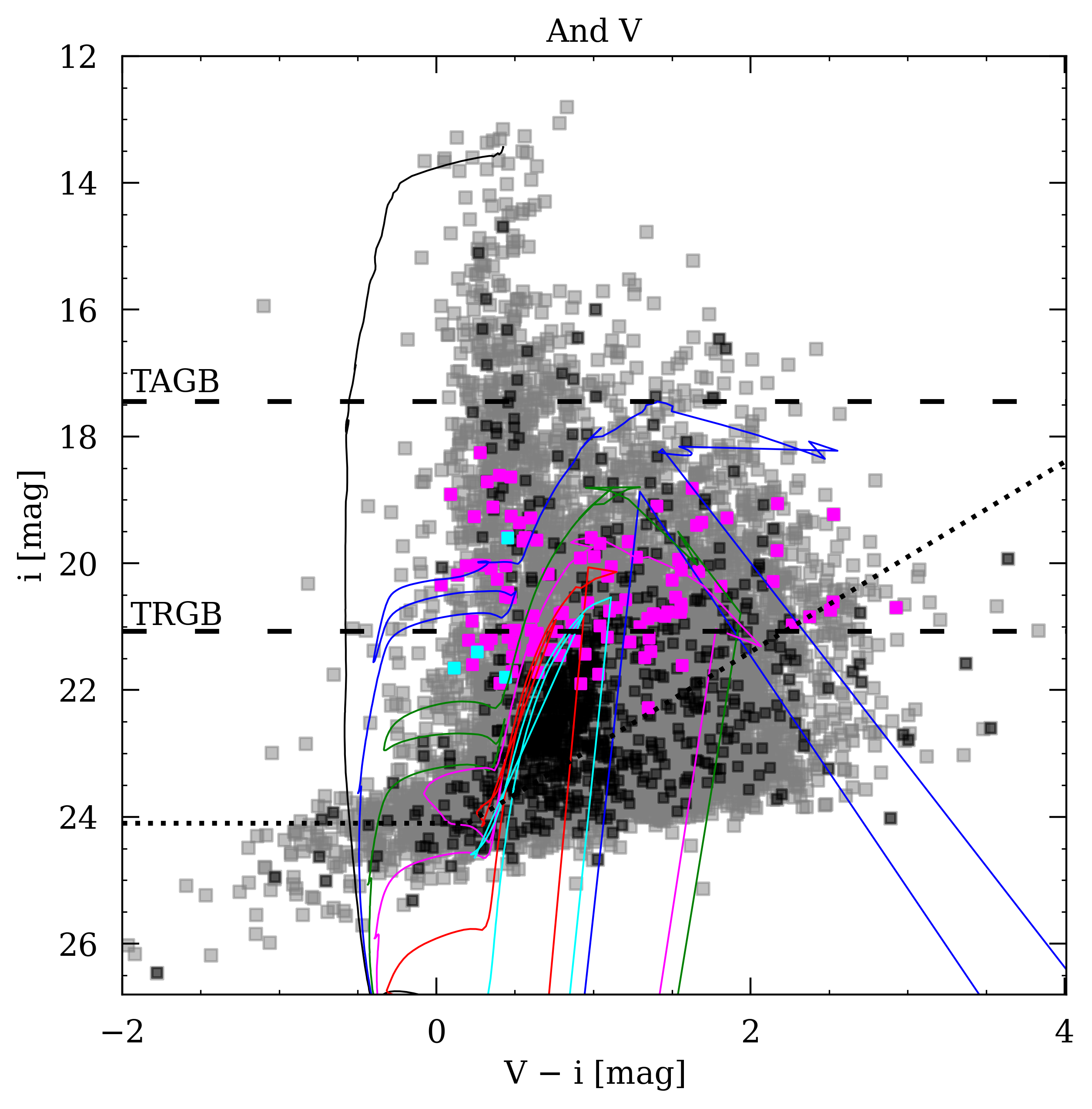}        \includegraphics[width=0.4\textwidth]{ 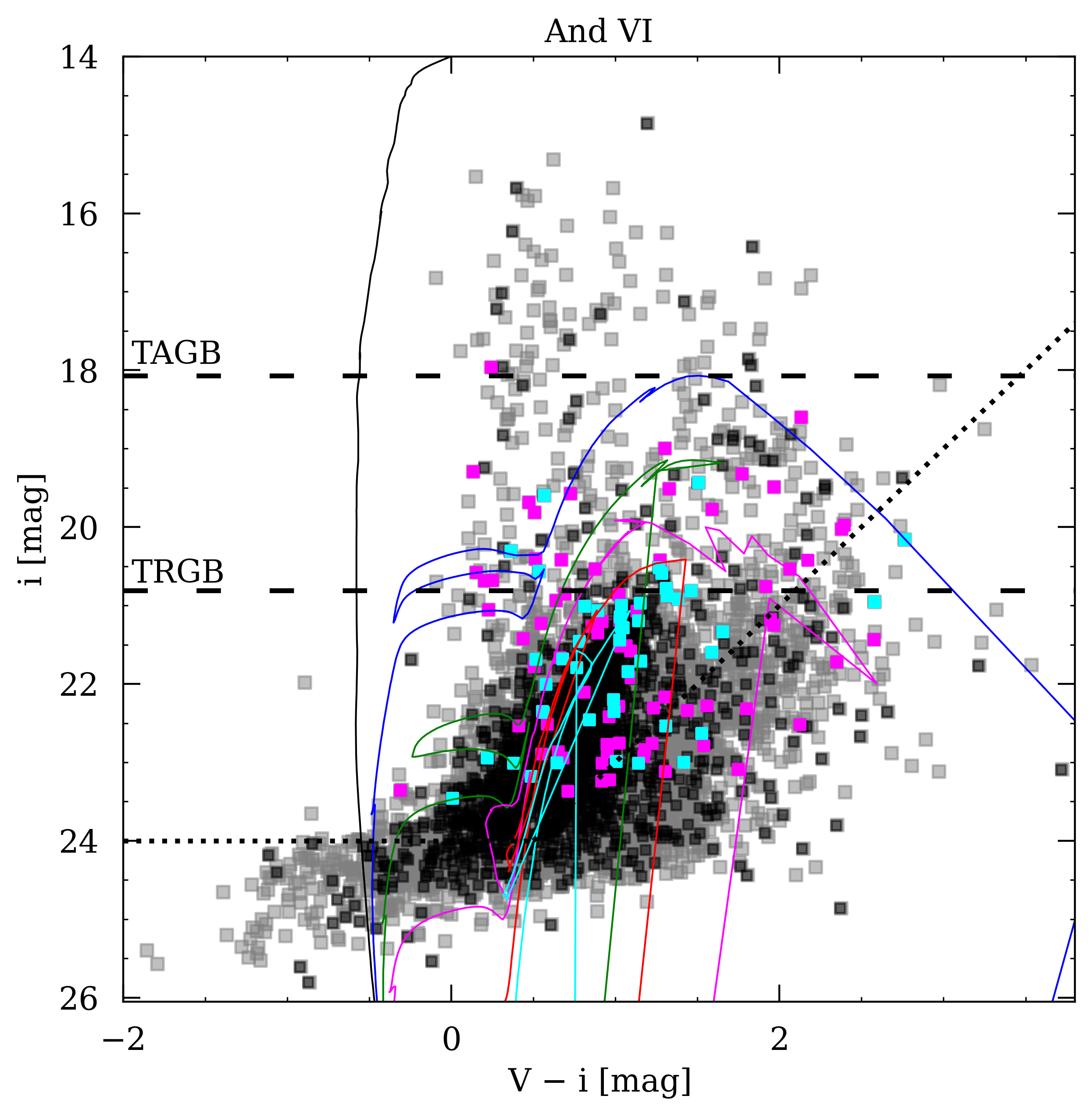}           \includegraphics[width=0.4\textwidth]{ 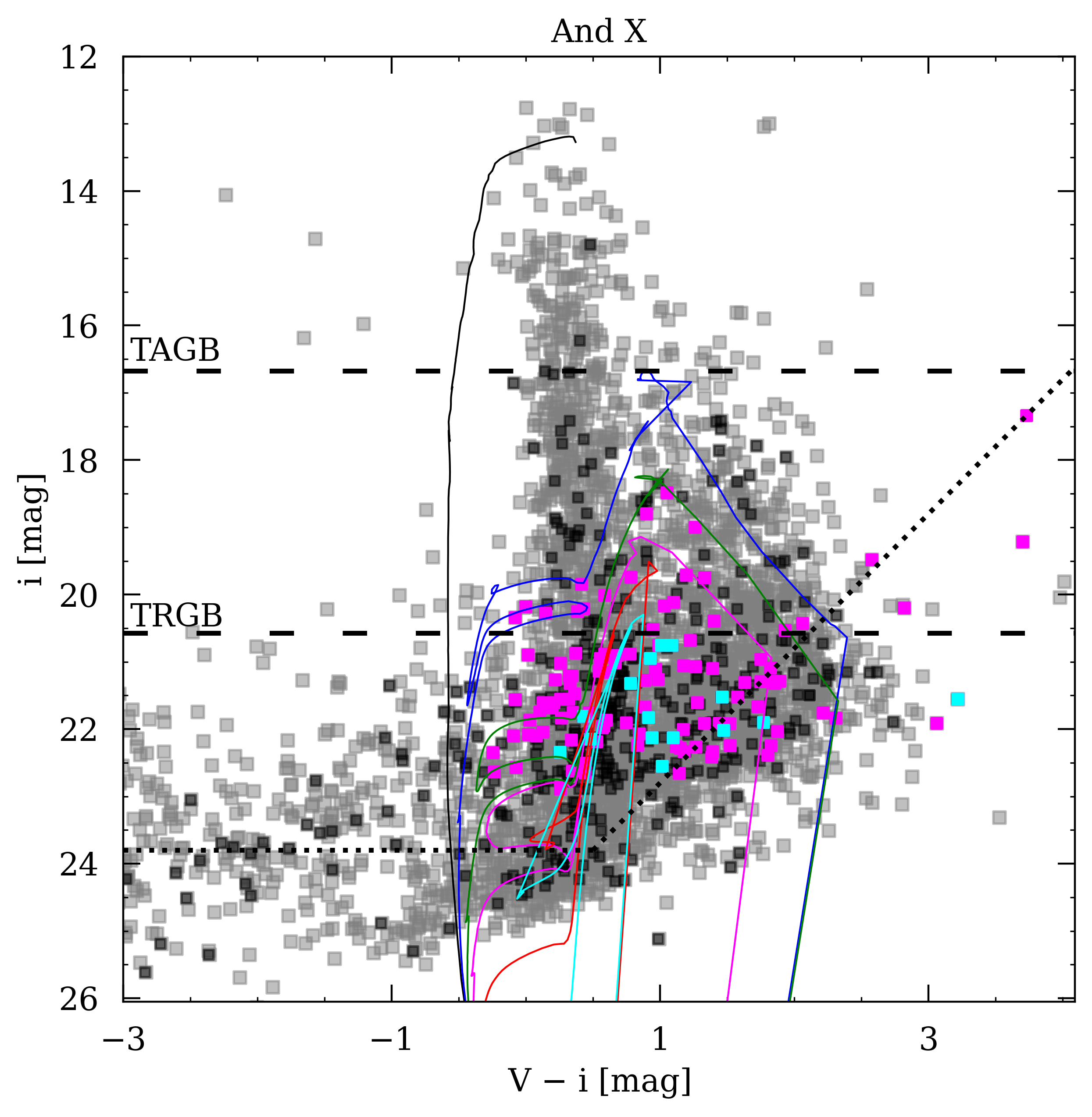}            \includegraphics[width=0.4\textwidth]{ 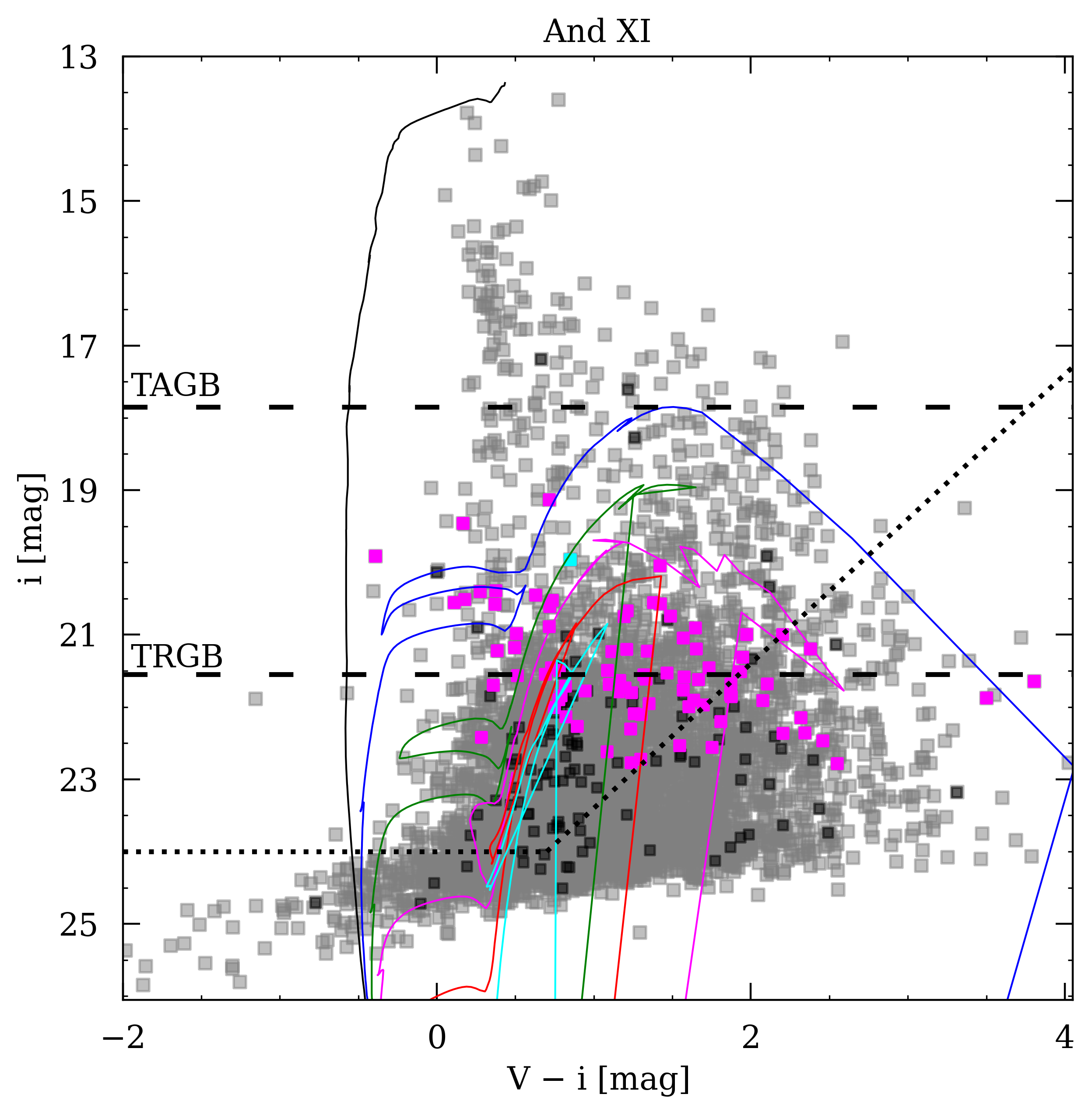}
    \includegraphics[width=0.4\textwidth]{ 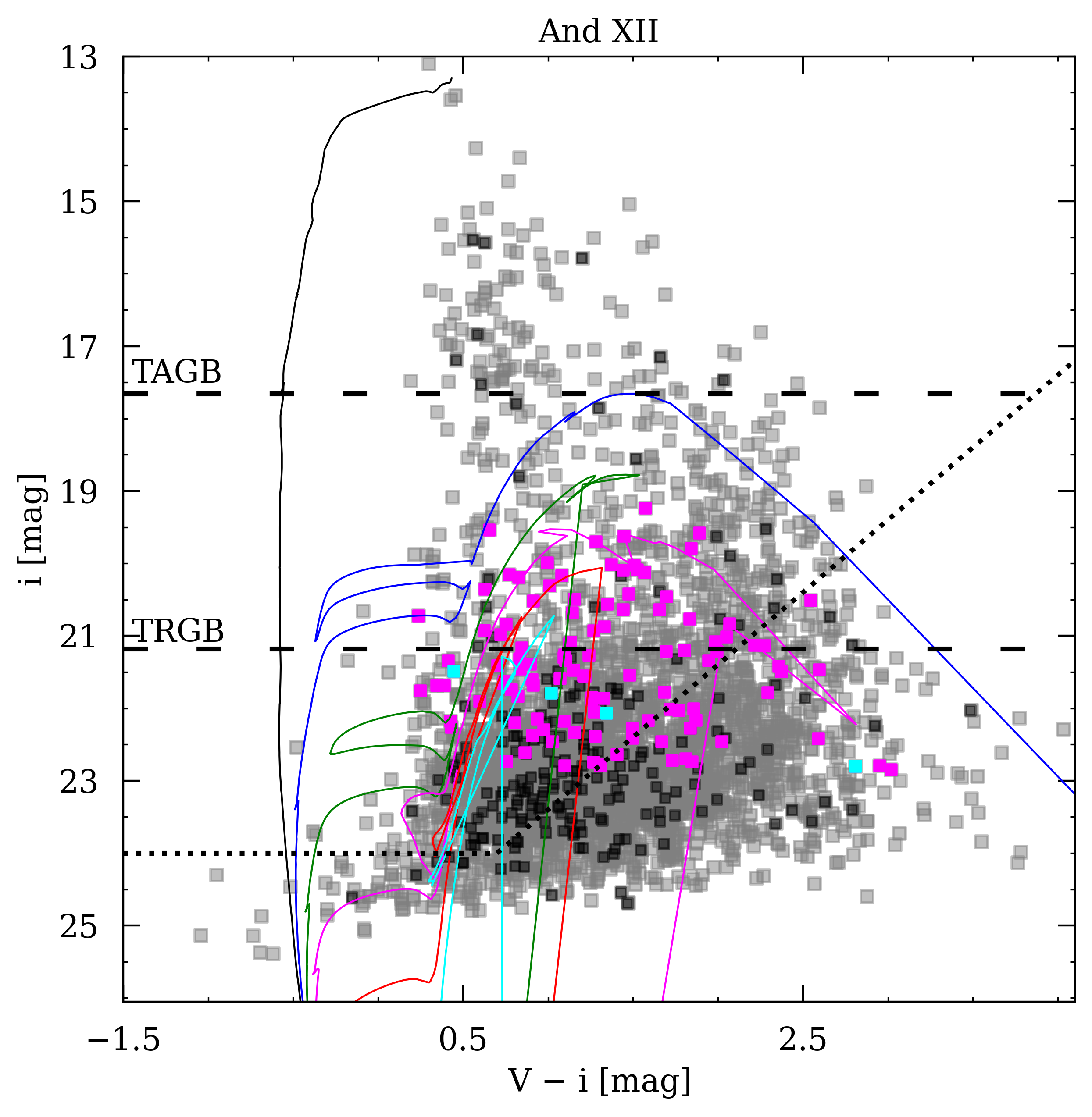}
   \caption{CMDs of dwarf satellites. Gray squares mark photometrically identified stars, and black squares show stars within twice the galaxy’s half-light radius. Magenta points mark LPV candidates in the total studied field, with cyan points highlighting LPV candidates within twice the half-light radius. The isochrones are shown as colored curves, ranging from log(t/yr) = 6.6 (black), 8 (blue), 8.6 (green), 9 (purple), 9.4 (red), and 10 (cyan). The dashed-lines mark the positions of the TAGB and TRGB, while the dotted-line indicates the photometric completeness limit.}
    \label{fig:CMDs1}
\end{figure}

\begin{figure}[H]
    \centering
    \includegraphics[width=0.4\textwidth]{ 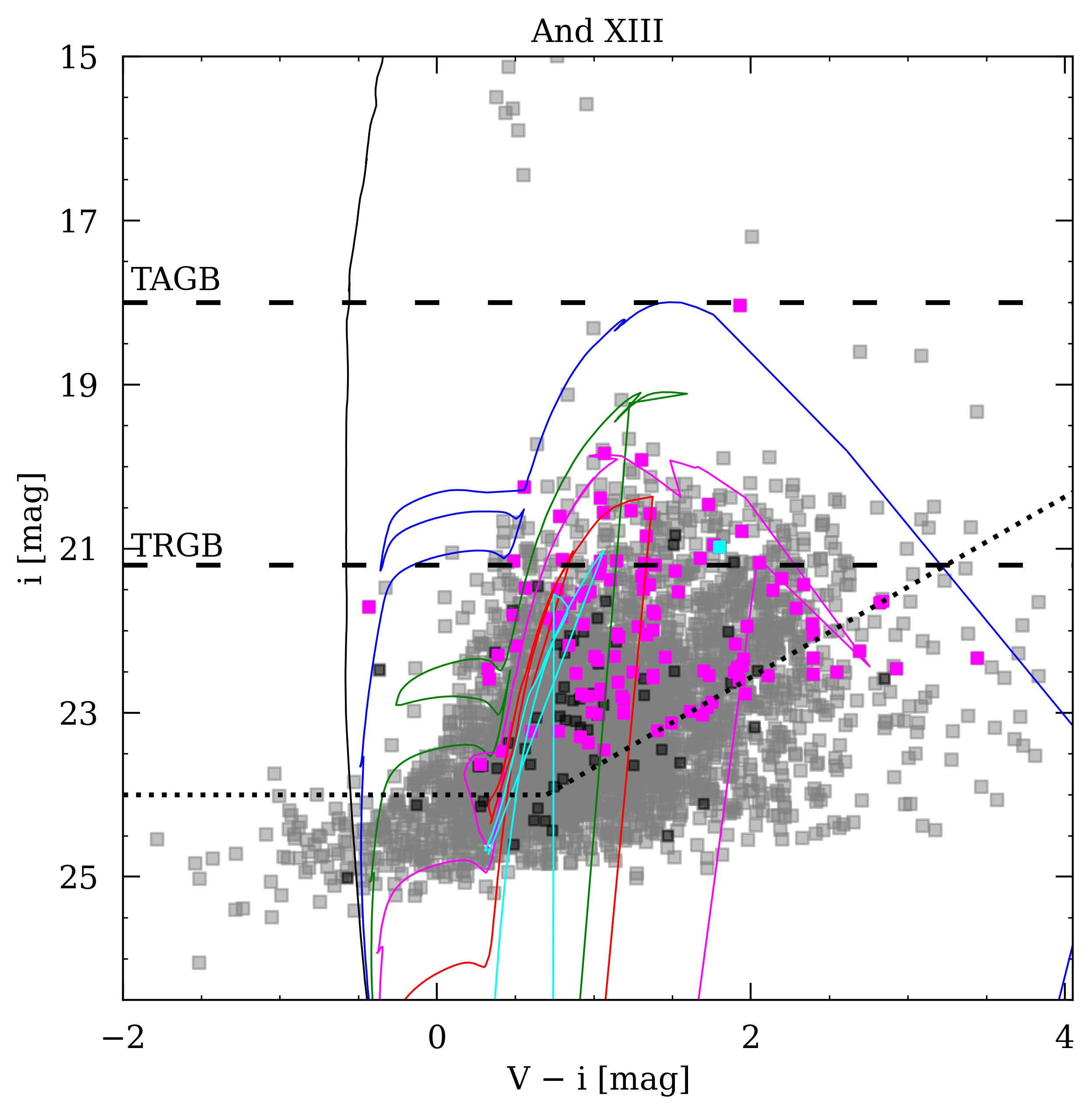}      \includegraphics[width=0.4\textwidth]{ 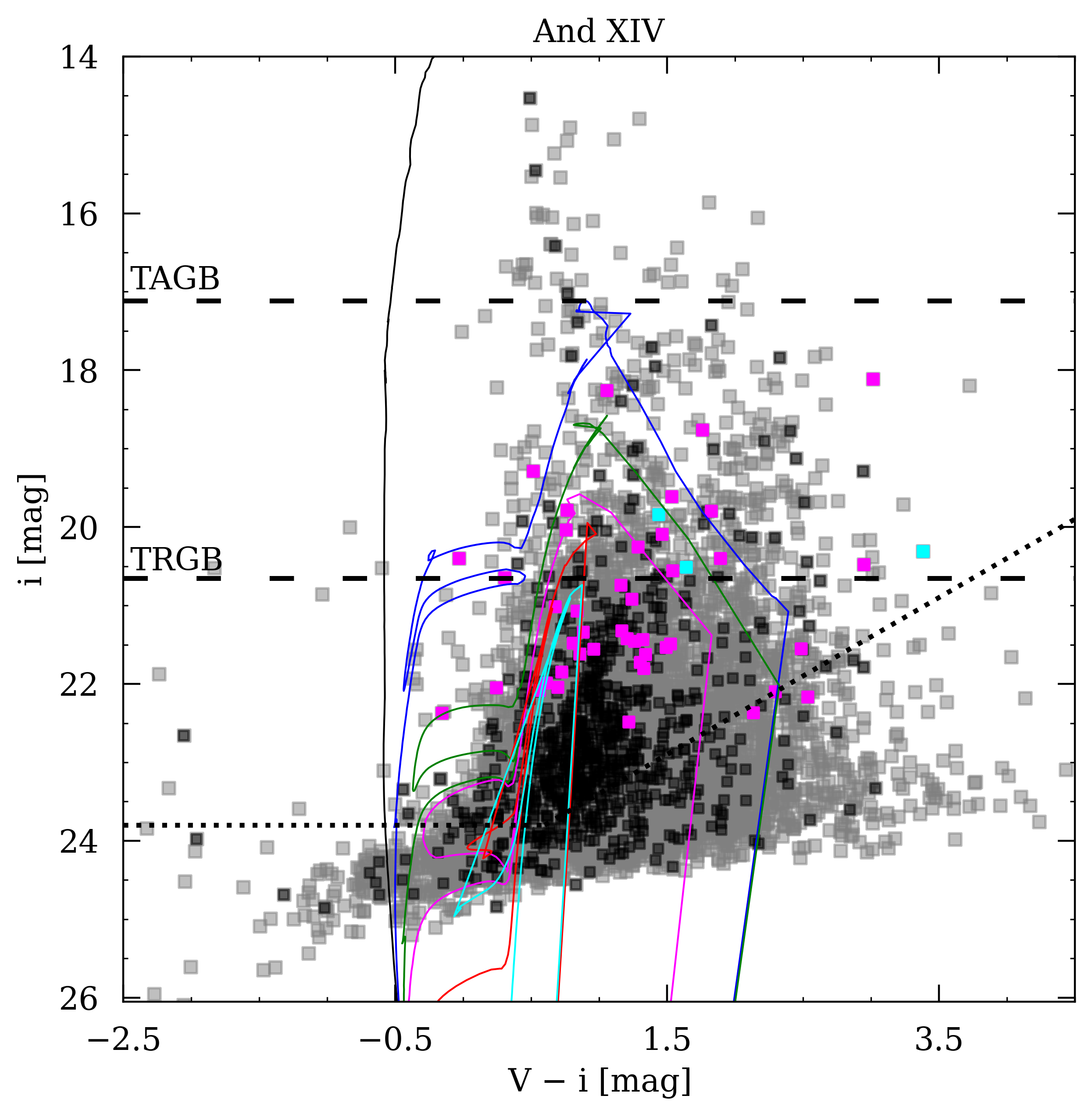}        \includegraphics[width=0.4\textwidth]{ 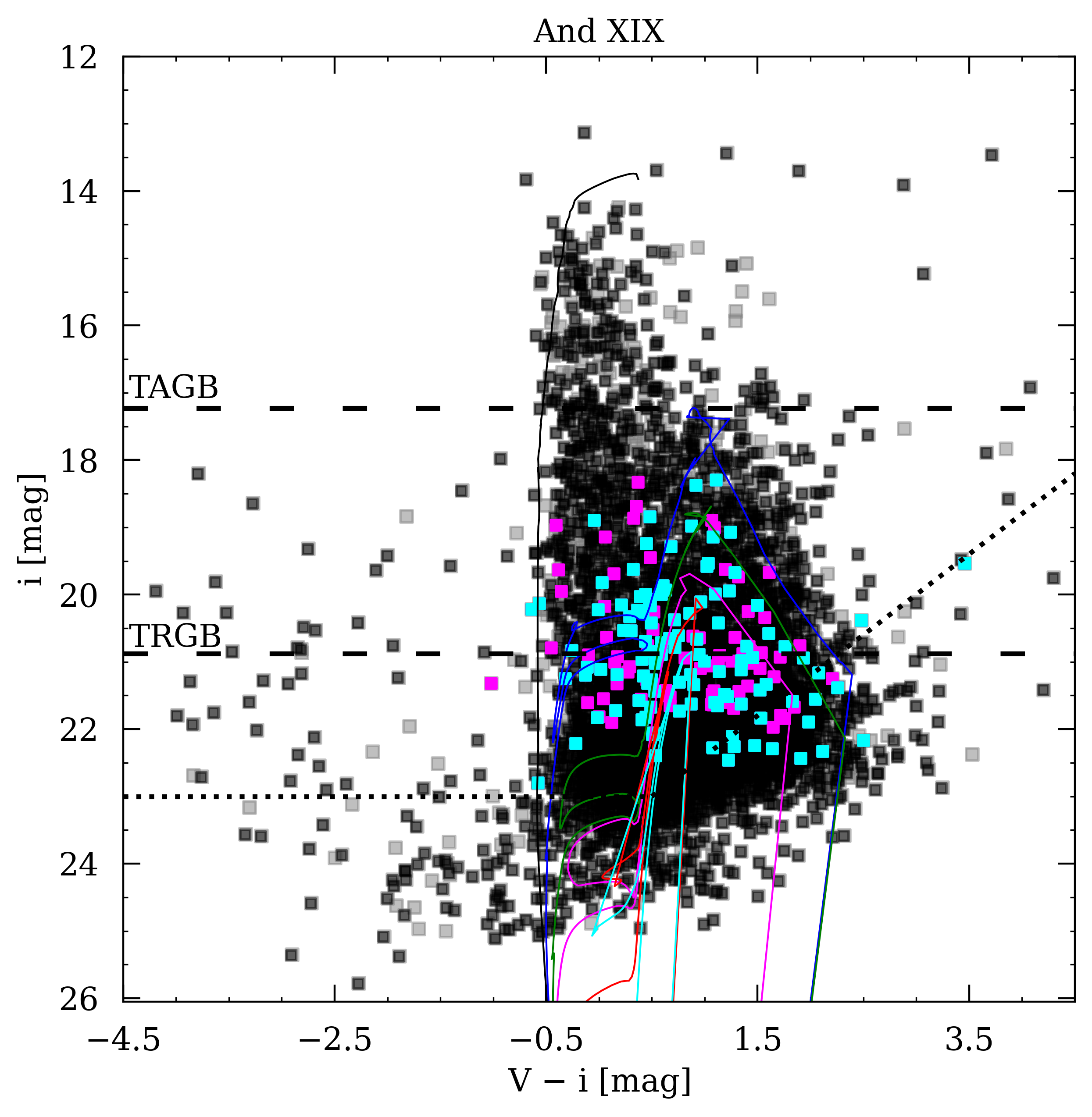}      \includegraphics[width=0.4\textwidth]{ 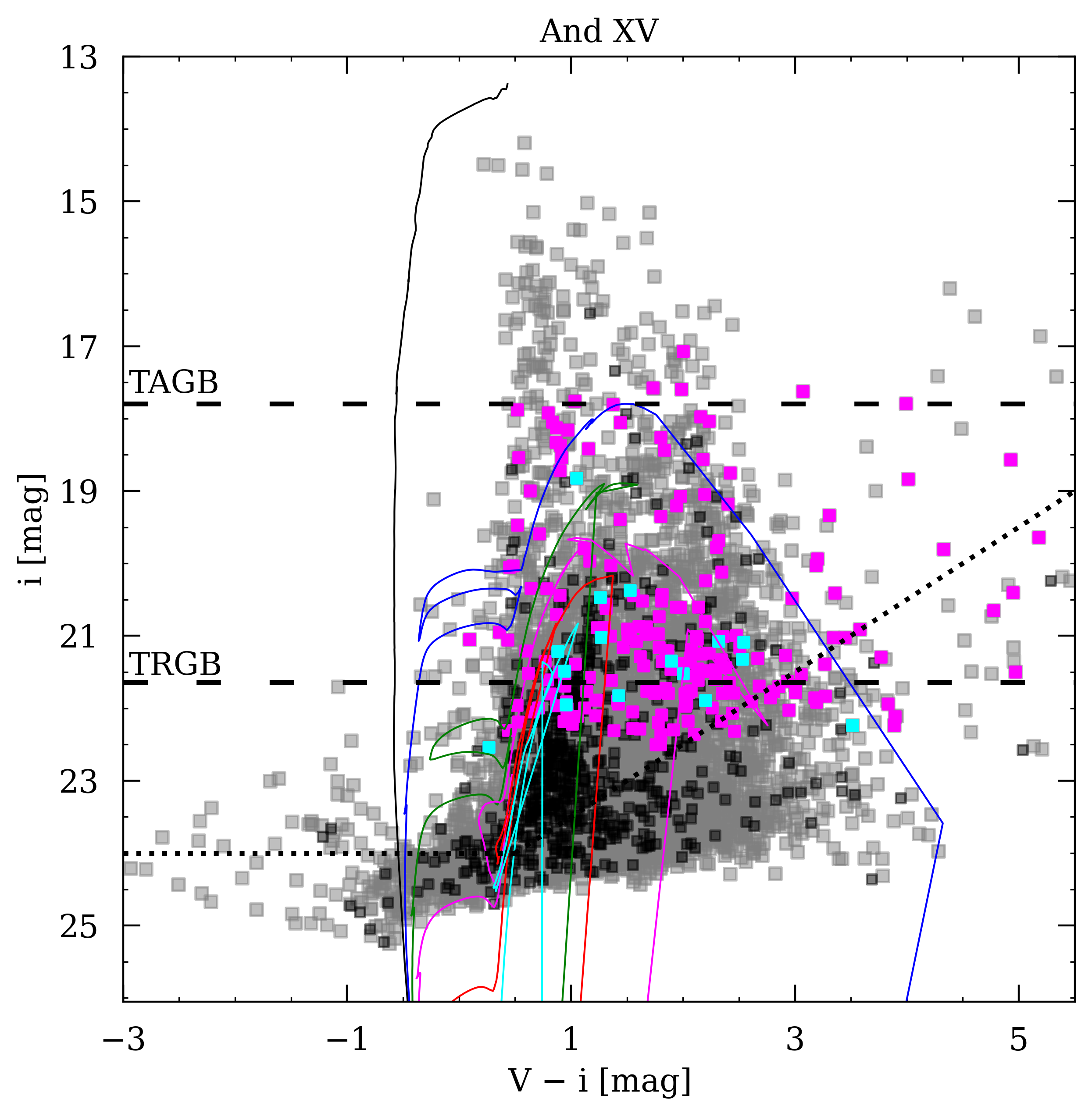}          \includegraphics[width=0.4\textwidth]{ 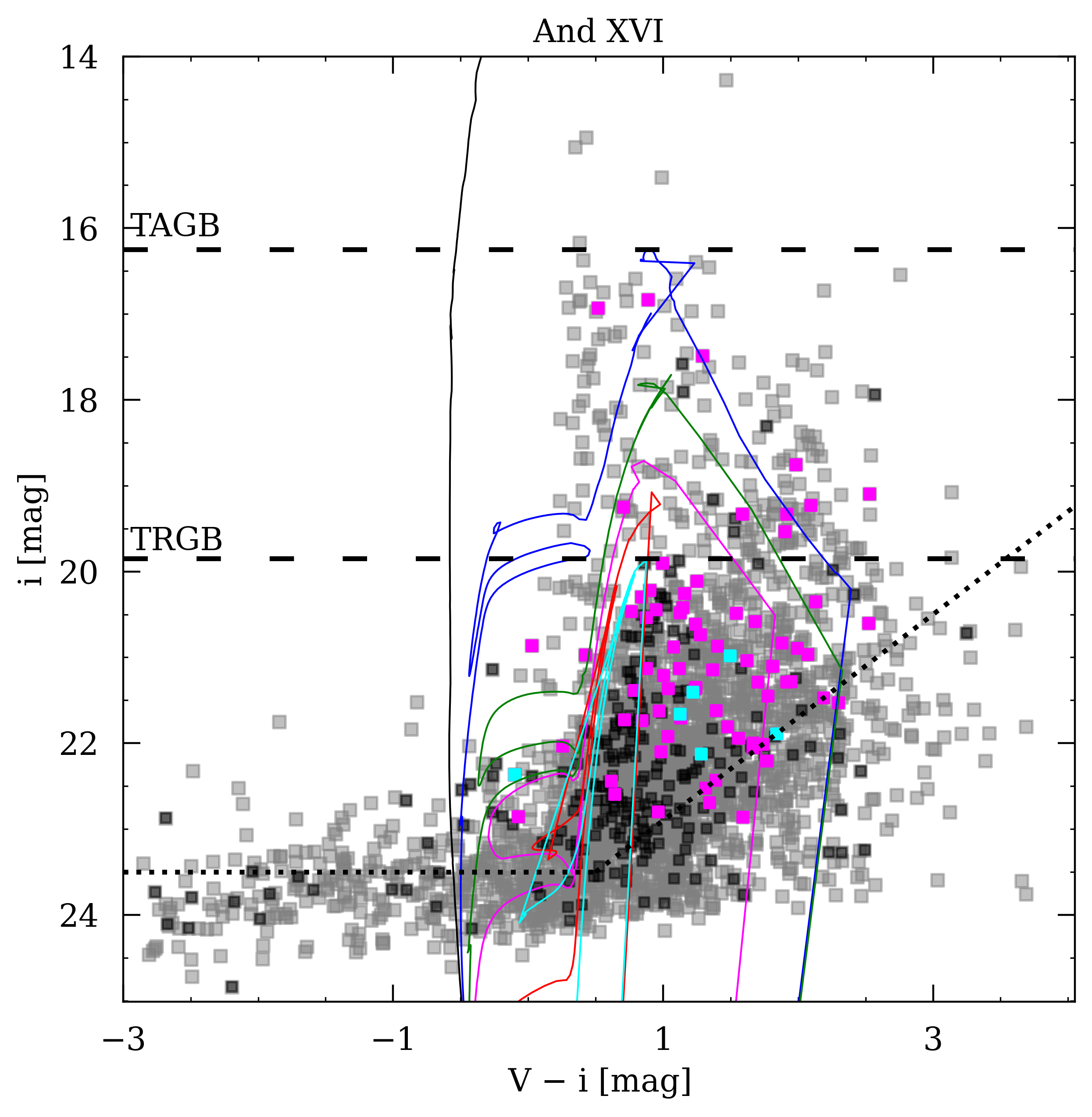}
 \includegraphics[width=0.4\textwidth]{ 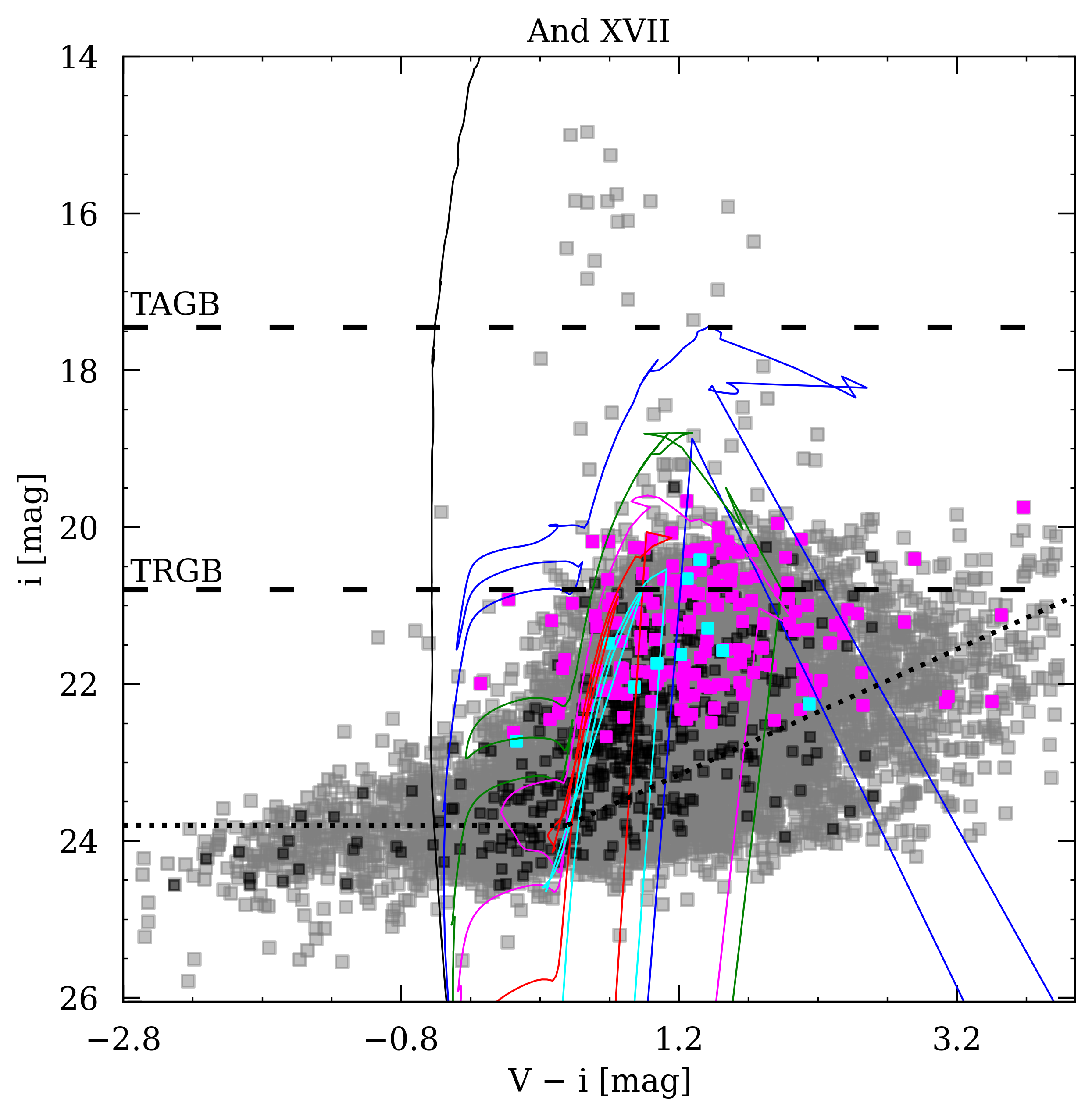}
   \caption{CMDs of dwarf satellites.}
    \label{fig:CMDs2}
\end{figure}

\begin{figure}[H]
    \centering
 
    \includegraphics[width=0.4\textwidth]{ 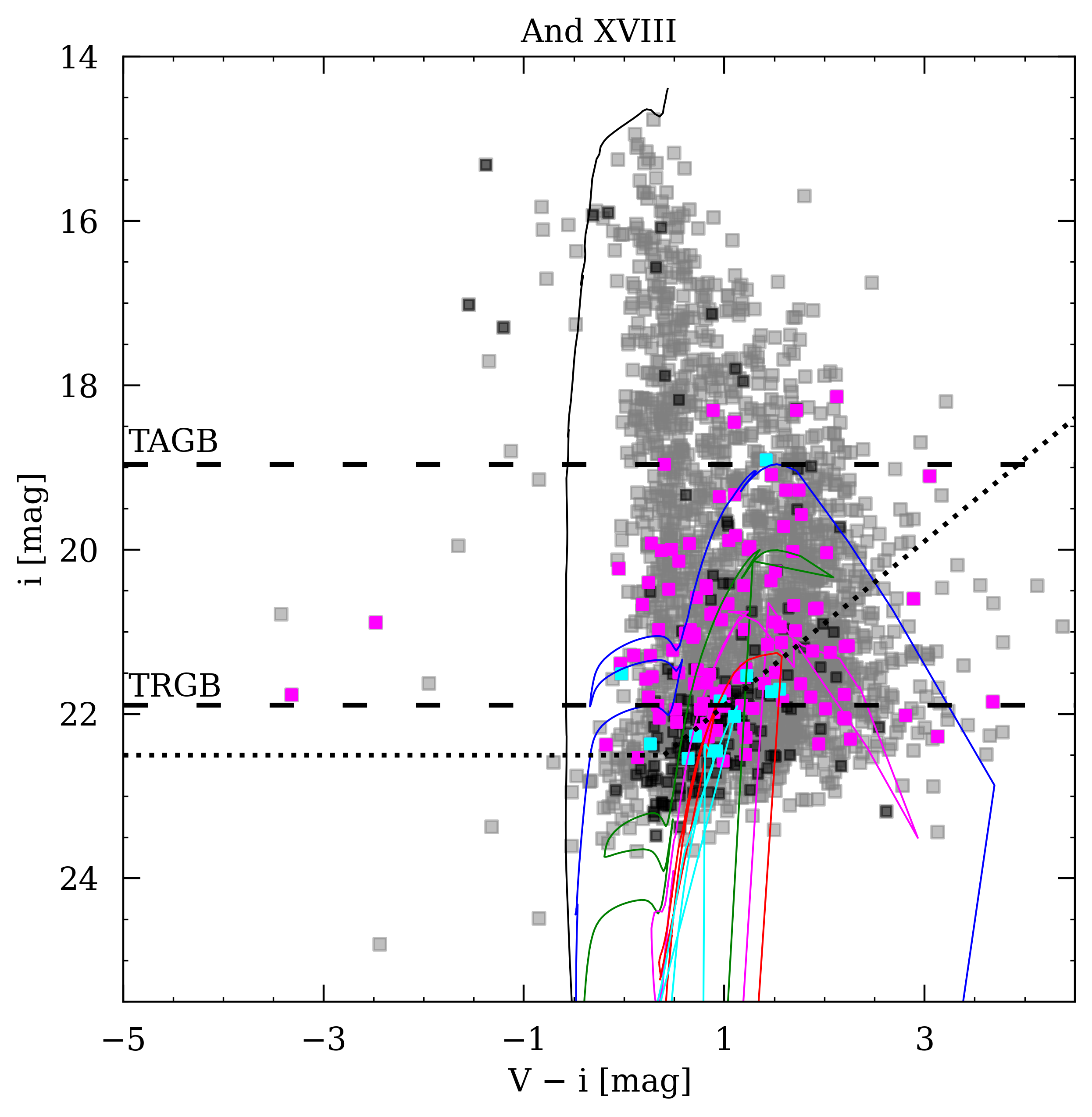}    
     \includegraphics[width=0.4\textwidth]{ 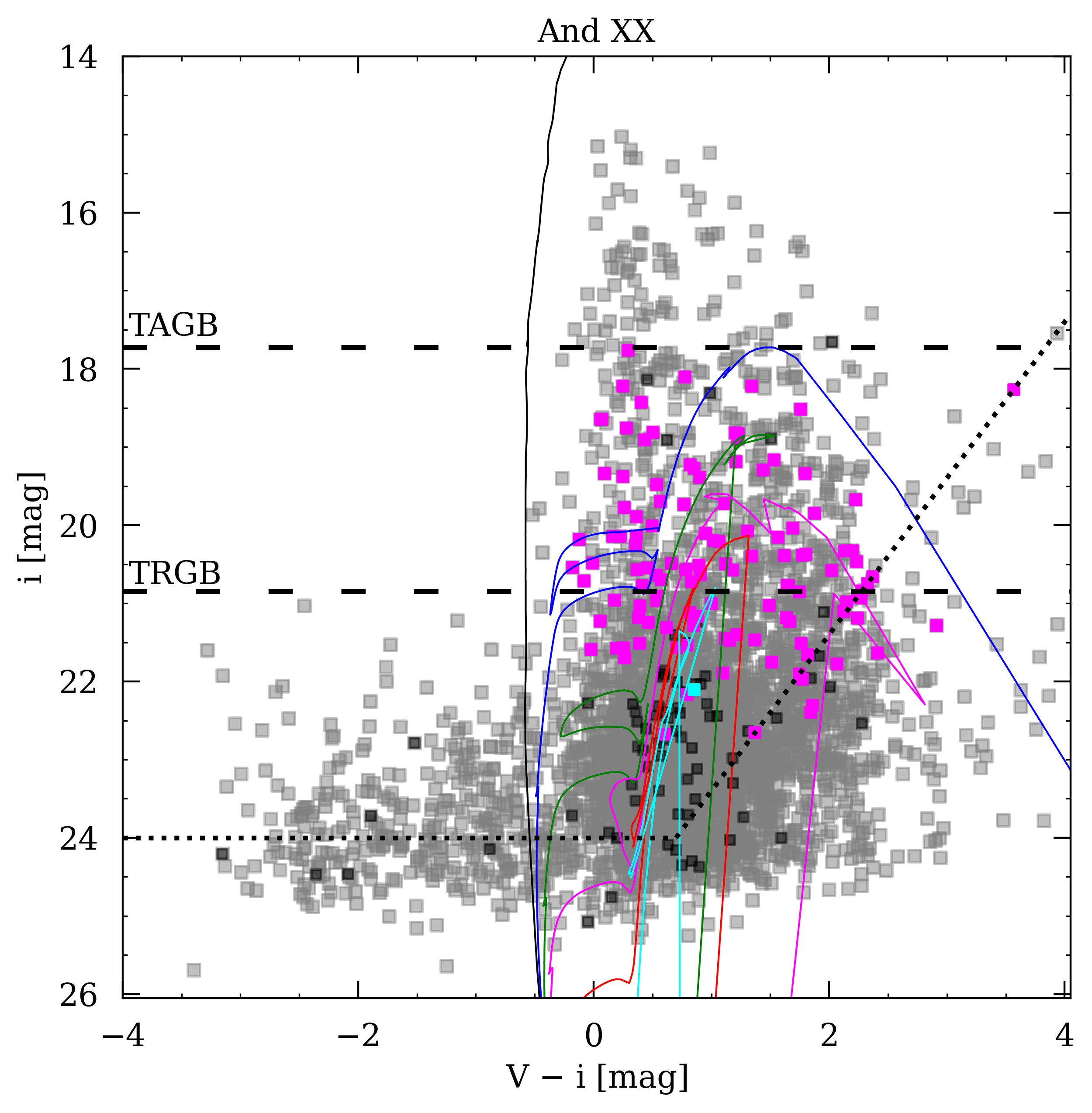} 
      \includegraphics[width=0.4\textwidth]{ 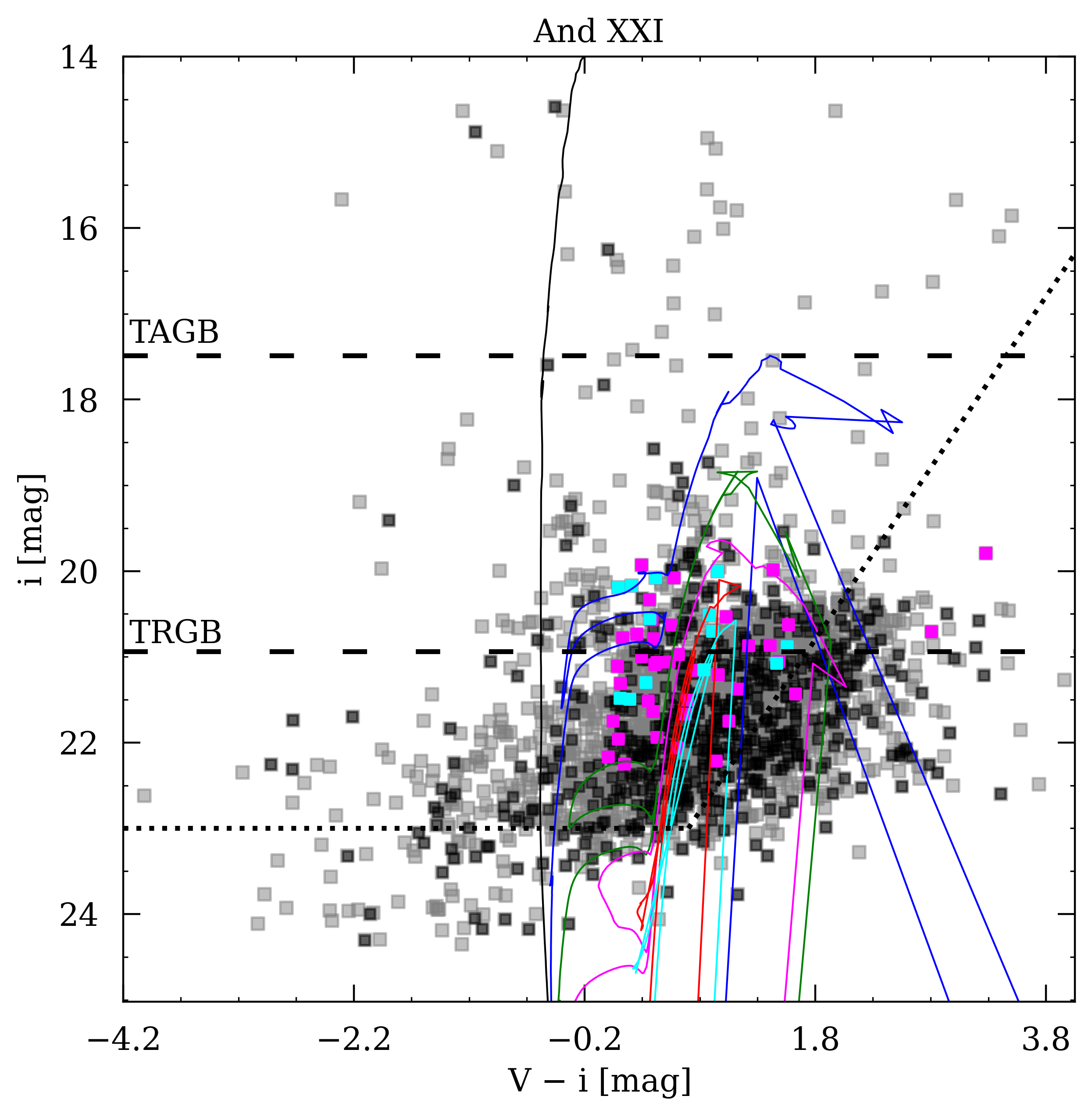} 
       \includegraphics[width=0.4\textwidth]{ 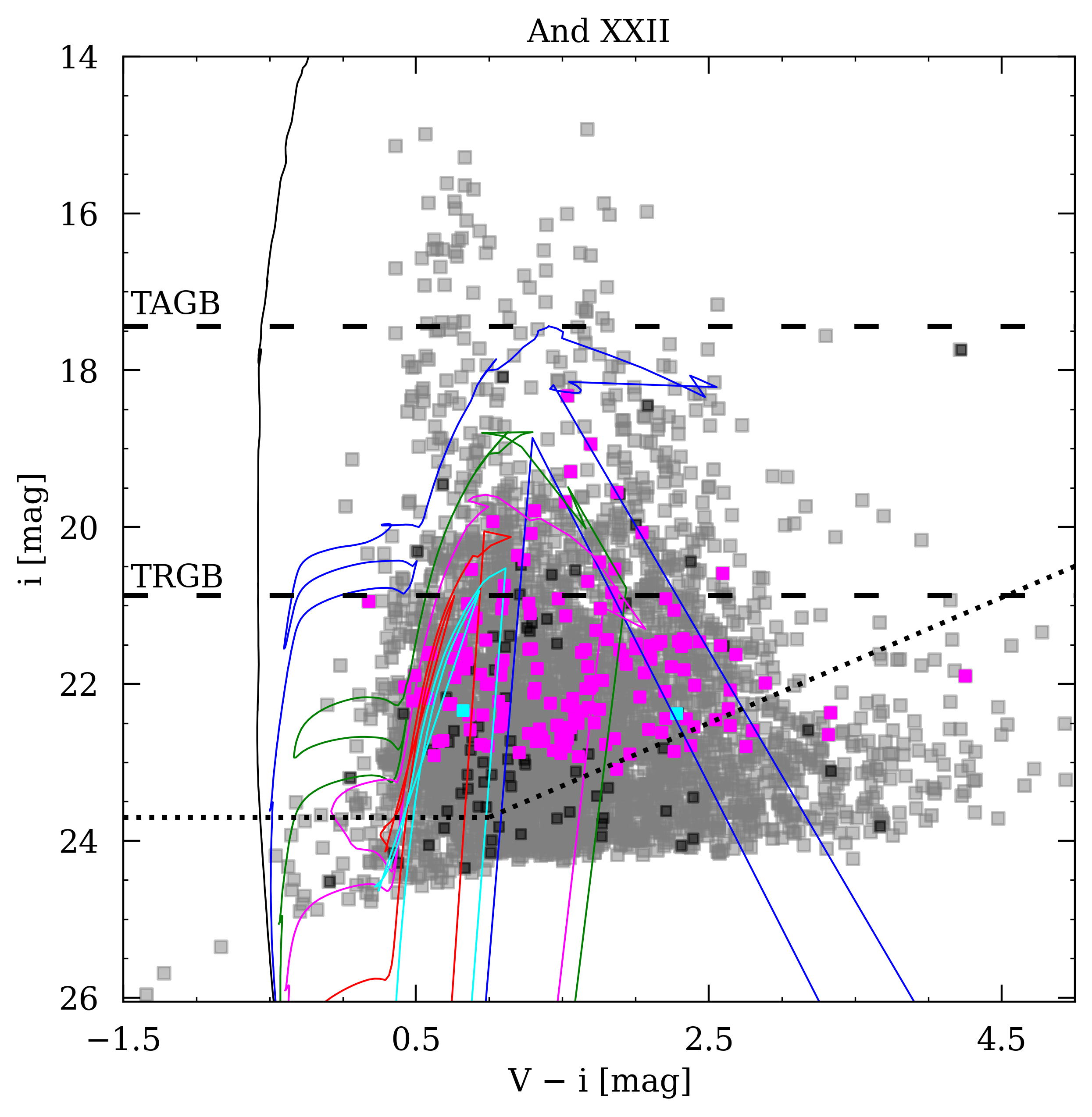} 
   \caption{CMDs of dwarf satellites.}
    \label{fig:CMDs3}
\end{figure}

Figs.~\ref{fig:LC1} and \ref{fig:LC2} show light-curve examples of a representative LPV from each dwarf galaxy. Figs.~\ref{fig:Location1}--\ref{fig:Location3} provide the spatial distribution of the LPVs within the observed fields of all dwarf galaxies. Figs.~\ref{fig:TRGB1}--\ref{fig:TRGB3} present the TRGB determination, while Figs.~\ref{fig:CMDs1}--\ref{fig:CMDs3} display the CMD of the total stellar population together with the LPV distribution, shown both across the full observed field and within two half-light radii.


\bibliography{References}{}
\bibliographystyle{aasjournal}



\end{document}